\documentclass{emulateapj}

\usepackage{enumitem}
\usepackage{color}
\usepackage{multirow}

\newcommand{\msun}{\ensuremath{\rm M_\odot}}

\newcommand{\kms}{km s$^{-1}$}

\newcommand{\lya}{Ly$\alpha$}
\newcommand{\ha}{H$\alpha$}
\newcommand{\hb}{H$\beta$}
\newcommand{\oiii}{[\ion{O}{3}]}

\newcommand{\zla}{\ensuremath{z_{\rm Ly\alpha}}}

\newcommand{\super}[1]{\ensuremath{^\textrm{\scriptsize{#1}}}}
\newcommand{\sub}[1]{\ensuremath{_\textrm{\scriptsize{#1}}}}
\newcommand{\pp}{\phantom{1}}
\newcommand{\ps}{\phantom{$-$}}
\newcommand{\arcs}{\ensuremath{''}}

\shorttitle{Spectroscopy of \lya-emitters at z$\sim$2.7}
\shortauthors{Trainor et al.}

\begin{document}

\title{The Spectroscopic Properties of Ly$\alpha$-Emitters at
  $Z\sim2.7$:\\
Escaping Gas and Photons from Faint Galaxies\altaffilmark{1}}

\author{Ryan F. Trainor\altaffilmark{2}}
\affil{Department of Astronomy, University of California, Berkeley, 501 Campbell Hall, Berkeley, CA 94720; trainor@berkeley.edu}
\author{Charles C. Steidel, Allison L. Strom}
\affil{Cahill Center for Astrophysics, MC 249-17, 1200 E California Blvd, Pasadena, CA 91125}
\and \author{Gwen C. Rudie\altaffilmark{3}}
\affil{Carnegie Observatories, 813 Santa Barbara Street, Pasadena, CA 91101}

\altaffiltext{1}{Based on data obtained at the W.M. Keck Observatory, which is operated as a scientific partnership among the California Institute of Technology, the University of California, and NASA, and was made possible by the generous financial support of the W.M. Keck Foundation.}
\altaffiltext{2}{Miller Fellow.}
\altaffiltext{2}{Carnegie-Princeton Fellow.}

\begin{abstract}
We present a spectroscopic survey of 318 faint
($\mathcal{R}\sim27$, $L\sim0.1L_{*}$), \lya-emission-selected 
galaxies (LAEs) in regions centered on the positions of hyperluminous QSOs
(HLQSOs) at $2.5<z<3$. A sample of 32 LAEs with
rest-frame optical emission line spectra from  
{\it Keck}/MOSFIRE are used to interpret the LAE spectra in the
context of their systemic redshifts. The fields are part of
the Keck Baryonic Structure Survey (KBSS), which includes substantial
ancillary multi-wavelength imaging from both the ground and space. From
a quantitative analysis of the diverse \lya\ spectral morphologies,
including line widths, asymmetries, and multi-peaked profiles, we find that
peak widths and separations are typically smaller 
than among samples of more luminous continuum-selected galaxies (LBGs) at similar
redshifts. We find tentative evidence for an association between \lya\
spectral morphology and external illumination by the nearby
HLQSO. Using the MOSFIRE subsample, we find that the peak of the
resolved ($R\approx 1300$) \lya\ line is shifted by +200 km s$^{-1}$
with respect to systemic
across a diverse set of galaxies including both LAEs and LBGs. We
also find a small number of objects with 
significantly blueshifted \lya\ emission, a potential indicator of
accreting gas. The \lya-to-\ha\ line ratios measured for the MOSFIRE
subset suggest that the LAEs in 
this sample have \lya\ escape fractions $f\sub{esc,\lya}\approx30\%$,
significantly higher than typical LBG samples. Using
redshifts calibrated by our MOSFIRE sample, we 
construct composite LAE spectra, finding the first evidence for
metal-enriched outflows in such intrinsically-faint high-redshift galaxies. These
outflows have smaller continuum covering fractions ($f_c\approx 0.3$) and
velocities ($v\sub{ave}\approx100-200$ km s$^{-1}$, $v\sub{max}\approx
500$ km s$^{-1}$) than those
associated with typical LBGs, suggesting that gas covering fraction is
a likely driver of the high \lya\ and Ly-continuum escape fractions of
LAEs with respect to LBGs. Our results suggest a similar scaling of
outflow velocity with star formation rate as is observed at lower
redshifts ($v\sub{outflow}\sim $ SFR$^{0.25}$) and indicate that a
substantial fraction of gas is ejected with $v>v\sub{esc}$. Further
observations, including deep 
spectroscopy in the observed near-IR, will further probe
the evolution and enrichment of these galaxies in the context of their
gaseous environments.
\end{abstract}

\keywords{galaxies: formation --- galaxies: high-redshift --- galaxies: intergalactic medium}

\section{Introduction} \label{laes:intro}


Stellar feedback plays a crucial role in galaxy
formation.  Massive stars enrich, heat, ionize, and expel gas
throughout their lives and deaths, and understanding these processes is key
to modeling the observational characteristics of galaxies at all
redshifts. These characteristics include the low instantaneous
efficiency of star-formation (e.g., \citealt{ken98}), the low stellar
masses and baryon content of galaxies (e.g.,
\citealt{whi91,ker09,hop14}), and the presence and evolution of the mass-metallicity
relationship of galaxies (e.g., \citealt{erb06a,and13,ste14}). Each
of these observations requires that the majority of gas associated
with galaxy halos is prohibited from forming stars, particularly in
the lowest-mass galaxies. There are many potential causes for
this inefficiency: the temperature, ionization, and/or kinematics of
the gas may suppress its conversion into 
stars. Simulations can reproduce many characteristics of
the stellar populations of galaxies simply by artificially delaying
gas cooling or implementing other ad hoc solutions specific to
these discrepancies; these naturally lack
the predictive power of a feedback model based on the physics of star
formation and supernovae themselves (see e.g., \citealt{hum13,hop14}).

In particular, many simulations struggle to reproduce properties of
the circumgalactic medium (CGM; e.g., \citealt{rud12a}) of galaxies,
which must be closely linked to the same feedback processes that 
shape their stellar populations. Studies of the gas in and around galaxy
halos conclusively show that galaxies enrich their local CGM to large
($r\gtrsim r\sub{vir}$) galactocentric radii
\citep{agu01,pet03,ade03,ade05d,ste10,mar10,tur14}, and it is
expected that the chemistry and kinematics of these halos are tied to
the properties of the enriched gaseous outflows seen ubiqitously in
the spectra of high-redshift star-forming galaxies
\citep{ste96,pet01,sha03,jon12,mar12}.

Among Lyman-break galaxies and their analogs (hereafter referred to as
LBGs), these outflows produce spectral signatures in both absorption
and emission, allowing the characterization of the chemistry and
kinematics of the outflows for 
gravitationally-lensed sources \citep{pet00,pet02}, or stacks of
galaxies \citep{sha03,ste10}. Lyman-$\alpha$ emitters
(LAEs) as a population are found to be younger, fainter, and less
reddened (e.g., \citealt{gaw06}) than LBGs on average,\footnote{Note, however,
that all LBGs emit at least low-level \lya\ emission, and all LAEs
would display Lyman breaks in sufficiently sensitive broadband images
or spectra \citep{ste11}.} so LAE surveys provide a complementary
sample to LBGs for characterizing the effects of these gas flows (and
stellar feedback in general) on galaxies over a broad range of ages
and masses. Compared to most LBGs, LAEs may also ``leak'' a relatively large
fraction of the Lyman-continuum (LyC) photons produced in their
embedded star-forming regions \citep{nes13,mos13}, which may be linked to the
observed association between higher \lya\ equivalent widths and lower
neutral gas covering fractions among LBGs
\citep{sha03,jon13}. Therefore, LAEs may be most analogous to those
galaxies that dominate the reionization of the universe (e.g., \citealt{rob13,dre14}).

Most \lya-selected galaxies are too faint for continuum absorption
features to be easily observed. At low redshifts, the \lya\
Reference Sample (LARS; \citealt{hay13,ost14}) includes 14
local galaxies with {\it HST} imaging and spectroscopy of their
\lya\ emission. \citet{riv15} analyze metal absorption and \lya\
emission from these galaxies, finding that gaseous outflows and the
covering fractions of absorbers play a strong role in modulating \lya\
escape (see also similar results by \citealt{wof13}). The few measurements of interstellar 
absorption in high-redshift LAE spectra come from especially bright examples
\citep{has13,shi14}, with continuum magnitudes ($B\sim24$, $L\sim L_{*}$) bright
enough to fall into typical samples of LBGs.\footnote{For example, 25\%
  of the 811 ``LBGs'' in \citet{sha03} have $W\sub{\lya}>20$\AA\
  and spectroscopic properties similar to these bright-LAE samples.} However, it is
the faint LAEs that provide the most interesting environment for studying the kinematics of
gas and galaxies. Assuming that the feedback processes ubiquitous in LBGs extend
to the earliest and faintest galaxies, then the effect of even faint galaxies
on their surrounding gaseous environments must be taken into account when 
modeling the enrichment of the circumgalactic and intergalactic media
(CGM and IGM; e.g., \citealt{ste10}) and its transparency to ionizing
photons (e.g., \citealt{sch13,rud13}). Furthermore, the
low masses and star formation rates of galaxies selected as LAEs may
enable the detection of the gas infall predicted by simulations
(\citealt{bir03,ker05}; and others) whereas in LBGs with stronger
star-formation rates, gaseous outflows may be more likely to dominate over infall in
determining the state of circumgalactic gas. 

Regions of the sky around bright QSOs provide particularly
advantageous environments for LAE surveys. Hyperluminous QSOs inhabit
known galaxy overdensities \citep{tra12} and thereby yield a large number of
LAEs within a single telescope pointing and narrowband-selected
redshift interval. Additionally, the ionizing fields of
the QSOs can boost the \lya\ emission of faint, gas-rich galaxies
through \lya\ ``fluorescence'' \citep{hog87,ade06,can07,can12,kol10,tra13}. These
effects make QSO fields extremely efficient for the selection and
follow-up of faint LAEs. Fluorescent \lya\ emission
can allow the characterization of the QSO radiative geometry and
history \citep{tra13}, but it also provides a means of selecting
galaxies irrespective of their intrinsic stellar luminosities. 
LAEs selected in these regions may include  ``dark galaxies''
\citep{can12}, halos of gas and dark matter in which only
negligible star formation has occurred.  Whether or not such ``dark''
objects are observable, the fluorescent boost of nearby QSOs may allow
us to probe an even greater range of galaxy properties than typical
LAE surveys alone. 

In this paper, we present rest-frame UV spectroscopy from a large (68,000
cMpc$^3$) narrowband survey conducted in eight QSO fields with
$2.55<z\sub{QSO}<2.85$. All eight fields are part of the Keck Baryonic Structure
Survey (KBSS; \citealt{rud12a}), and this \lya\ survey (designated
KBSS-\lya) is complementary to the original KBSS for probing the
interactions of QSOs, galaxies, and gas at the peak redshifts of star
formation. The authors note that the effect of the bright QSOs in
  the KBSS-\lya\ fields on their nearby galaxies is not generally
  known, but the QSO-induced contribution to the \lya\ emission is expected to be
  independent of the intrinsic star-formation properties of these
  galaxies. In this paper, we generally assume that the kinematic and
  transmissive properties of the KBSS-\lya\ LAEs are most directly related
  to their internal star formation and stellar populations. However,
  the results of our analysis that may depend strongly on QSO
  illumination and proximity are noted explicitly. Future work will
  consider the effect of the QSO on associated galaxies in greater detail.

The spectroscopic component of the KBSS-\lya\ survey is
described in Sec.~\ref{laes:obs}. In Sec.~\ref{laes:lya}, we
present the properties of the \lya\ emission lines, including analysis
of the \lya\ escape fraction, spectral morphology, and kinematics
in relation to comparison samples of continuum-bright galaxies. In
Sec.~\ref{laes:stacks}, we present 
measurements of metal absorption lines in the stacked continuum
spectra of the rest-UV spectroscopic 
sample and characterize the ionization, covering fraction, and
kinematics of star-formation driven outflows in the LAEs. A
brief discussion of these results in the context of the physical
properties of the galaxies and previous work is
given in Sec.~\ref{laes:discussion}, and our results are summarized in
Sec.~\ref{laes:conclusions}. Physical quantities are given in
observable units when possible, but a $\Lambda$CDM universe with
($\Omega\sub{m}$, $\Omega_\Lambda$, $H_0$) = (0.3, 0.7, 70 km s$^{-1}$)
is assumed when necessary.

\section{Observations} \label{laes:obs}

\subsection{LAE Sample} \label{sublaes:photobs}

\begin{figure*}
\center
\includegraphics[width=\linewidth]{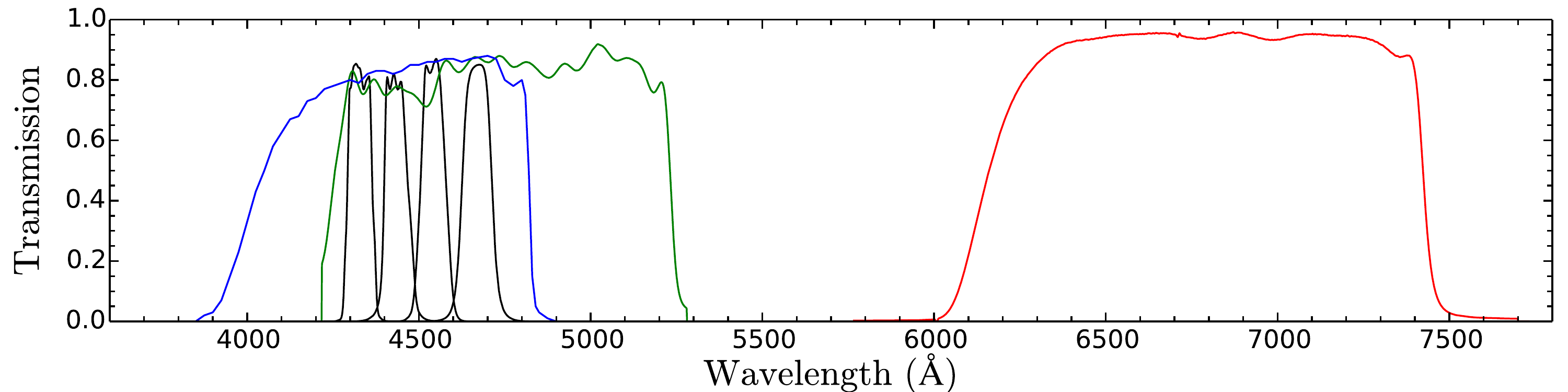}
\caption[Filters used for LAE characterization]{The bandpasses of the primary
  narrowband and broadband filters used to select and characterize the
  LAEs in this paper. Black curves are the four narrowband filters
  (from left to right: NB4325, NB4430, NB4535, NB4670, with names
  reflecting the central wavelength in Angstroms). The blue, green,
  and red curves show the bandpasses of the LRIS $B$, $G$, and
  $\mathcal{R}$ filters, respectively.}
\label{fig:filters}
\end{figure*}

\begin{figure}
\center
\includegraphics[width=\linewidth]{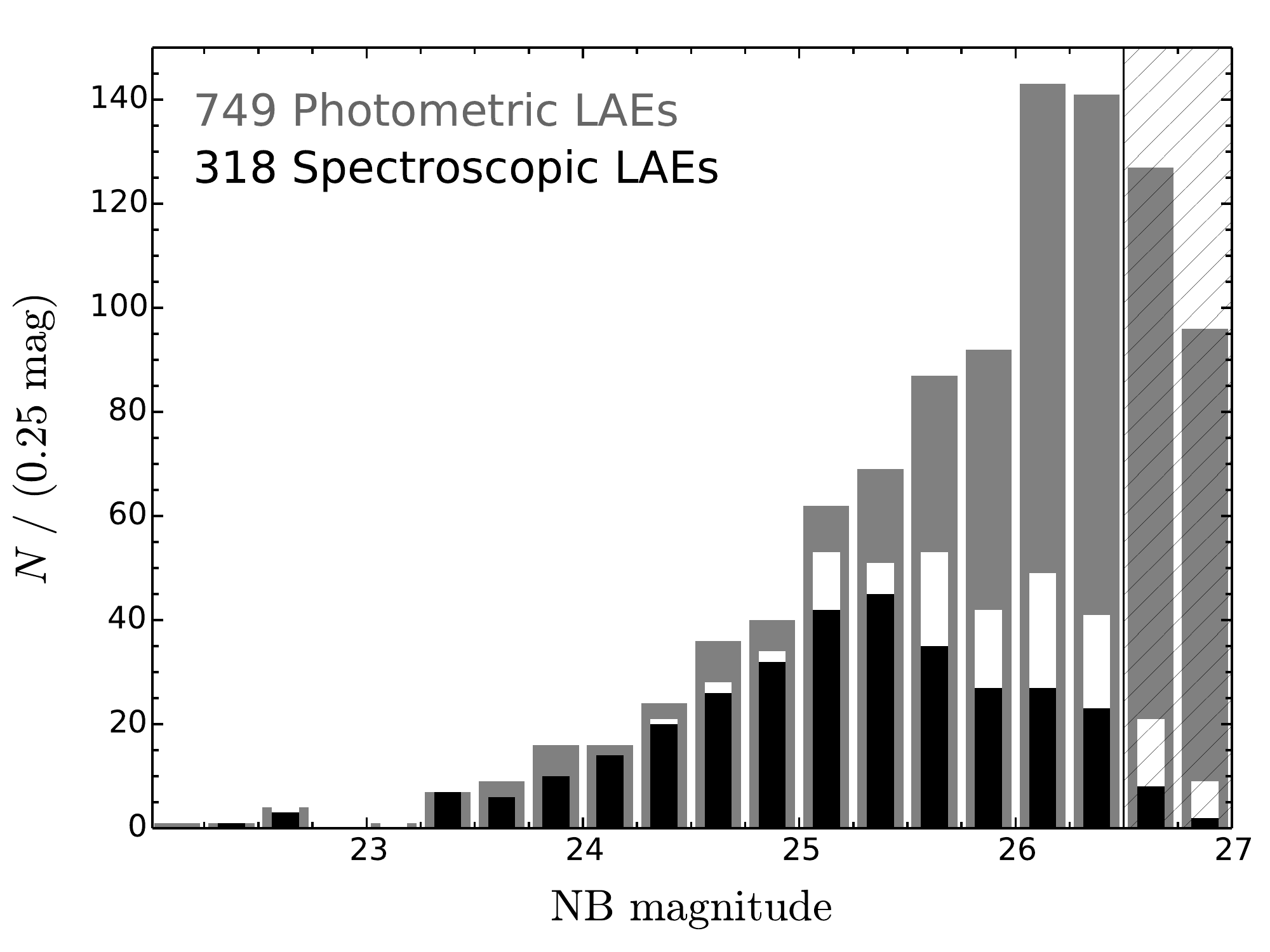}
\caption[Apparent magnitude distribution of LAEs]{The apparent
  magnitude distribution of all photometric (grey) and 
  spectroscopic (black) LAEs observed in the survey. NB magnitude denotes the
  apparent magnitude of each LAE in the narrowband (NB) filter used
  to select it. The white sections of the bars denote the number of LAEs in each bin
  observed spectroscopically for which no line was detected by our
  algorithm. These objects are excluded from the spectroscopic sample
  of LAEs. For $m\sub{NB} > 26.5$ (hatched region), the success rate of   
  spectroscopic identification of the \lya\ emission line dropped
  precipitously; for this reason, this paper primarily considers only
  those LAEs with a NB magnitude brighter than 26.5
  ($F\sub{\lya}\ge6-10\times10^{-18}$ erg s$^{-1}$ cm$^{-2}$).}
\label{fig:nbma_hist}
\end{figure}

The full parent sample of $\sim$800 photometrically-identified LAEs will be
described separately (R. Trainor et al., in prep.); this LAE sample was also
used and discussed by \citet{tra13}. LAEs were identified on the basis 
of paired narrowband and broadband observations (from Keck 1/LRIS-B)
using a custom set of narrowband filters. A total of four narrowband
filters were used, each with a bandpass constructed to select the wavelength
of \lya\ at one or more QSO redshifts (Fig.~\ref{fig:filters}). Objects
were identified and photometry 
was performed using SExtractor in one- and two-image modes to extract
narrowband and broadband magnitudes ($m\sub{NB}$ and $m\sub{BB}$)
sampling \lya. Hereafter, $m\sub{BB}$ refers to the measured LAE
magnitude in the $B$ and/or $G$ filter used to estimate the continuum
emission near \lya, whichever is better matched
at a given redshift (Table \ref{table:fields}). Because the $B/G$ filter bandpasses include the
\lya\ emission line, we use $\mathcal{R}$-band measurements to sample
the continuum without  \lya\ contamination. We conducted follow-up spectroscopy of
the subset of objects identified to have a narrowband color excess
$m\sub{BB}-m\sub{NB}>0.6$, which corresponds to a limit in rest-frame \lya\
equivalent width $W\sub{\lya}\gtrsim20$\AA. The success of our
spectroscopic follow-up observations dropped significantly for
$m\sub{NB}>26.5$, so this limit (approximately the 3-$\sigma$ depth of
our narrowband observations) was adopted as the faint cut-off for the
photometric and spectroscopic samples. This magnitude limit
corresponds to an integrated narrowband flux
$F\sub{\lya}\approx1\times10^{-17}$ erg s$^{-1}$ cm$^{-2}$ for a
continuum-free \lya\ emission line, or $F\sub{\lya}\approx6\times10^{-18}$
erg s$^{-1}$ cm$^{-2}$ for an emission line with our median value of
$\langle W\sub{\lya}\rangle\approx40$\AA. The apparent magnitude
distribution of both the parent sample and the spectroscopic subsample of
LAEs discussed in this paper are given in
Fig.~\ref{fig:nbma_hist}. Both the spectroscopic and photometric
samples have median $\langle
\mathcal{R}\sub{AB}(6930\AA)\rangle\approx 27$. According to 
the galaxy luminosity functions measured by
\citet{red08}, this continuum magnitude corresponds to
$L\sim0.1 L_{*}$ at $z\sim 2.7$. Details of the QSO fields and the spectroscopic
sample are given in Table~\ref{table:fields}.  

\begin{deluxetable}{lcccc}
\tablecaption{\lya-Emitter Field Descriptions}
\tablewidth{0pt}
\tablehead{
\multicolumn{1}{c}{\multirow{2}{*}{QSO Field}} & $z\sub{Q}$ & NB & BB 
& \multirow{2}{*}{$N\sub{spec}$} \\
& ($\pm$0.001) & Filter & Filter 
&
}

\startdata
Q0100+13 (PHL957)   &  $ 2.721 $ & NB4535 & $B$+$G$ & 20  \\
HS0105+1619         &  $ 2.652 $ & NB4430 & $B$ & 22    \\
Q0142$-$10 (UM673a) &  $ 2.743 $ & NB4535 & $B$+$G$ & 22   \\
Q1009+29 (CSO 38)   &  $ 2.652 $ & NB4430 & $B$ & 35  \\
HS1442+2931         &  $ 2.660 $ & NB4430 & $B$ & 41  \\
HS1549+1919         &  $ 2.843 $ & NB4670 & $G$ & 95    \\
HS1700+6416         &  $ 2.751 $ & NB4535 & $B$+$G$ & 20 \\
Q2343+12            &  $ 2.573 $ & NB4325 & $B$ & 63 
\enddata
\label{table:fields}
\end{deluxetable}

\subsection{UV Spectroscopic Observations} \label{sublaes:specobs}

Rest-UV spectra were obtained with Keck 1/LRIS-B in the multislit mode using the
600/4000 grism and 560nm dichroic. All objects were observed with
1\farcs2 wide slits, but the spectral resolution is limited by the
0\farcs6$-$0\farcs8 seeing disk for these (typically spatially unresolved) LAEs. The
effective resolution near the observed \lya\
wavelength for these spectra is $R\sim 1300$, corresponding to a
velocity resolution $\sigma_v\sim 100$ km s$^{-1}$. Spectroscopic
observations were performed in sets of 3$-$4 1800 s exposures, for a
total of 1.5$-$2 hours of total integration time per object. By the
completion of the survey, we had obtained spectra for 415 of the 749
photometric LAE candidates, with some candidates observed on as many as
three different masks. The observations were conducted and reduced
in essentially the same manner 
as those of the KBSS and preceding surveys, a detailed description of
which can be found in \citet{ste03} and \citet{ste04}. Specifically,
masks were constructed and targets assigned as described by
\citet{ste03}, but with a minimum slit length of 11\arcs\ in order to
provide better background subtraction for continuum-faint
($\mathcal{R} \sim 27$) sources. The data were
reduced (including flat-fielding, cosmic-ray rejection, 
background subtraction, flexure compensation, and wavelength/flux
calibration) using a custom set of IRAF scripts as described in
\citet{ste03}. The two-dimensional spectra were rectified using
flat-field observations, and one-dimensional spectra were then extracted
using the IRAF task {\it apall} with 10 pixel apertures (1\farcs35,
$\sim$2$\times$ the seeing FWHM). We detect little or no
continuum emission in individual spectra, so all pixels in the aperture received
equal weighting during the extraction and the object trace was not
allowed to vary with wavelength along the rectified 2D spectrum. Note
that the Keck 1 Cassegrain Atmospheric Dispersion Corrector (ADC)
removes the wavelength-dependent shifts in object position due to
differential atmospheric refraction. The two-dimensional spectra were
binned in two-pixel blocks along the wavelength axis during readout,
and the one-dimensional 
extracted spectra were resampled to the median resulting wavelength 
scale of the observations, 1.26 \AA/pix. 

The resulting one-dimensional spectra were then subjected to an
automated line-detection algorithm described here. First, each
spectrum was smoothed with a kernel corresponding to the instrumental
resolution ($\sim$3.5\AA\ FWHM). A detection region of the spectrum was
then isolated: in order to ensure that spectroscopically-detected
objects corresponded to their NB-selected counterparts, the
algorithm only located emission lines that would fall between the 10\%
power points of the corresponding NB filter used to select them. The
highest peak in this detection region was then used to define the
first guess at the emission line wavelength. The significance of the
detected peak was then estimated by

\begin{equation}
P(y) = 1-\left(\int_{y\sub{max}/\sigma_y}^\infty G(x)\, dx\right)^{n\sub{pix}} \,,
\end{equation}

\noindent where $y\sub{max}$ is the observed peak value (minus any
measured continuum) in the smoothed
spectrum, $\sigma_y$ is the sample standard deviation of the
smoothed spectrum in the detection region, $G(x)=e^{-x^2/2}/
\sqrt{2 \pi}$ is the normal distribution function, and
$n\sub{pix}$ is the number of pixels in the detection region of the
spectrum. In this way, $P(y)$ represents the probability of measuring
one or more pixels with value $y \geq y\sub{max}$ within the detection
region under the null hypothesis that the pixel values are normally distributed with
variance $\sigma_y^2$. Spectra with $P(y)\geq0.05$ were rejected as
non-detections, while those with $P(y)<0.05$ were designated candidate
emission lines. This threshold was chosen because it closely reproduces the
samples of significant and non-significant spectral \lya\ detections
identified by eye. The median signal-to-noise ratio (S/N) of the
spectroscopic detections is 10.1. The total flux in each
candidate emission line was then estimated both by Gaussian fitting
and by direct integration of the unsmoothed spectrum.
Lastly, the line was re-fit using an asymmetric line profile: 

\begin{equation}
  f_\lambda(\lambda) = \left\{
      \begin{array}{lr}
        A e^{-(\lambda-\mu)/2\sigma\sub{blue}^2} & \lambda < \mu \\
        A e^{-(\lambda-\mu)/2\sigma\sub{red}^2} & \lambda \geq \mu 
     \end{array}
    \right.
\label{eq:linefit}
\end{equation}

\begin{equation}
\alpha\sub{asym}\equiv\sigma\sub{red}/\sigma\sub{blue} \,\,.
\label{eq:asym}
\end{equation}
 
Here, $\sigma\sub{blue}$ represents the line-width measured from the ``blue''
side of the peak ($\lambda < \mu$), while $\sigma\sub{red}$ is the width on the ``red''
side of the peak ($\lambda > \mu$), where $\mu$ is the fitted peak of the
\lya\ profile. From these quantities, we define the
asymmetry of the line shape, $\alpha\sub{asym}$, a potential
diagnostic of \lya\ escape physics \citep{zhe13,cho13}. Many lines were
found to be multi-peaked, defined by the existence 
of a second peak in the smoothed spectrum at least 2.5$\sigma$ above
the continuum and within 1500 km s$^{-1}$ of the primary peak,
separated from the primary peak by an inter-peak minimum. As these secondary
peaks typically have lower S/N than the primary peak, 
they were fit using symmetric Gaussian profiles. No more than two
Gaussian peaks were allowed in each 
fit. When fitting line profiles, the fits were constrained such
that the fit width $\sigma$ (both $\sigma\sub{blue}$ and $\sigma\sub{red}$
for asymmetric profiles) could not fall below
the instrument resolution ($\sigma>1.5$\AA). Measured line widths,
asymmetries, and multi-peaked profiles are discussed in detail in
Sec.~\ref{sublaes:multipeak}. As demonstrated in that section,
the resolution of these spectra is sufficient to resolve the peak
width, separation, and asymmetry for the typical objects in our
survey. However, we note that these spectra may not resolve spectral
features on very small ($\delta v\lesssim100$ \kms) velocity scales,
such as those predicted by radiative transfer models and observed in
some high-resolution spectra of bright or nearby galaxies \citep{ver08,mar15}.

The observed wavelength of the \lya\ emission line was measured via two
methods: direct integration of the unsmoothed spectrum (i.e. the
flux-weighted line centroid), and the fit value of $\mu$ (i.e. the
peak of the fit asymmetric Gaussian profile). For multi-peaked lines,
the primary peak is that which 
was detected by our first-pass line detection algorithm (i.e., the
most significant spike in the smoothed spectrum). Henceforth in
this paper, $z\sub{\lya,peak}$ denotes the redshift of the
\lya\ line derived from the asymmetric Gaussian peak, while $z\sub{\lya,ave}$
refers to the redshift derived from direct integration; these
values are compared in Sec.~\ref{sublaes:lyashift}.

We used spectroscopic and photometric cuts to eliminate non-\lya\
contaminants from our sample. Because the narrowband filters
used in this survey all have central wavelengths $\lambda_c<4900$\AA,
\oiii\ emitters are excluded by design. [\ion{O}{2}] emitters
($\lambda\lambda$3726,3729) may be selected at $z\sim0.15-0.25$
depending on the specific narrowband filter used for each
field. However, the total comoving volume selected by these filters is
$\sim$50$\times$ larger for \lya\ at $z\sim2.6$ than for [\ion{O}{2}]
at $z\sim0.2$. The [\ion{O}{2}] luminosity density is also
very low at $z\approx0.2$ with respect to higher redshifts, and
objects with [\ion{O}{2}]  equivalent widths in emission greater than
our observed-frame cut off ($W\sub{obs}>72$\AA) are extremely rare
\citep{hog98,cia13}. Furthermore, the [\ion{O}{2}] doublet should be
marginally resolved in our $\sigma_v\approx 100$ \kms\ spectra, with a
line separation $\Delta \lambda\sub{OII}\approx 220$ \kms. As
shown in Sec.~\ref{sublaes:multipeak}, there is no detected
population of objects that have multi-peaked emission lines with
$\Delta v\sub{peaks}\approx 200$ \kms, so we do not expect that
low-redshift [\ion{O}{2}] emitters are a significant source of
contamination to our photometric or spectroscopic samples.

The observed spectral range for each spectrum depends on the
location of the corresponding slit on the mask, but most slits
were placed to cover a rest-wavelength range
900\AA\ $\lesssim \lambda \lesssim$ 1500\AA\ at $z \sim 2.7$. As such,
other contaminants to the LAE sample were considered by searching for
anomalous emission lines outside of the \lya\ detection window. A small
number of such objects were found. All the spectroscopically-identified
contaminants were lower-redshift AGN in which strong \ion{C}{4}
$\lambda$1550 ($z \sim 1.9$) or \ion{He}{2} $\lambda$1640 ($z \sim 1.7$)
emission was detected by our narrowband filter. In each case, 
\lya, \ion{C}{4}, and \ion{He}{2} emission are all present with 
high equivalent width. Five objects
were found where the narrowband excess was caused by \ion{C}{4}
emission, and another 5 objects were selected due to \ion{He}{2}
emission falling in the narrowband passband. All 10 objects were
removed from both the spectroscopic and 
photometric samples. These constraints define a sample of
422 spectra for 318 unique LAEs. Using the contamination results, we
estimate the contamination fraction for the
photometric sample to be $10/(318+10)=3.0\%$.

To facilitate comparison with samples of continuum-bright galaxies, we
also analyze a sample of 65 LBGs from the KBSS-MOSFIRE
\citep{ste14}. These 65 galaxies are hereafter described as the
KBSS LBG sample. Details of the object selection and data collection and 
reduction are described by \citet{ste14} and references therein. The
65 KBSS LBG galaxies discussed here are a small subsample of the entire 
KBSS-MOSFIRE catalog that meet the following requirements: 

\begin{enumerate}[leftmargin=*]
\item rest-UV spectra acquired using the same LRIS
instrument setup (including the 600-line grism) as the KBSS-\lya\ LAEs;
\item \lya\ emission spectroscopically-detected by the same
  line-detection algorithm employed for the KBSS-\lya\ LAEs (56\% of
  the above objects);
\item systemic redshift measurements acquired as part of the
  KBSS-MOSFIRE via rest-frame optical emission lines (similar to the
  KBSS-\lya\ MOSFIRE subsample described in Sec.~\ref{sublaes:mosobs});
\item a lack of broad, high-equivalent width \ion{C}{4}
  emission indicative of AGN activity (97\% of objects meet this requirement).
\end{enumerate}

The KBSS LBG sample
provides a means of comparing the spectral properties of LAEs to
objects $\sim$10$\times$ brighter in the UV continuum
using the same spectral resolution and analysis methods. Note that
$\sim$50\% of all galaxies in the full KBSS-MOSFIRE sample 
have UV redshifts measurable only in continuum absorption lines. Such
objects cannot be selected by narrowband searches, so they were
omitted from the KBSS LBG sample to ensure appropriate comparison
(criterion \#2 above). However, note also that the KBSS LAEs have no
photometric cut based on $W\sub{\lya}$, as these galaxies occupy a
large range of redshifts and do not generally have narrowband images
probing their \lya\ emission.

The line-fitting algorithm was changed slightly
for the KBSS LBG spectra to accommodate the strong continuum, which often
varies significantly on the redward and blueward sides of the \lya\
line. When fitting the LBGs, a free parameter was added to provide a linear fit to the
continuum over the line region, but the one-peak or two-peak
asymmetric Gaussian fits were otherwise conducted in an identical
manner for the LAE and LBG samples. In Sec.~\ref{sublaes:lbgoutflows},
we also consider a composite LBG spectrum from a survey of
LBGs at $z\sim3$ (C.~Steidel et al., in prep.). The data collection
and reduction of the objects in this composite spectrum are similar to the
other LAEs and LBGs described above and will be discussed in detail in
future work. 

\begin{figure*}
\center
\includegraphics[width=\linewidth]{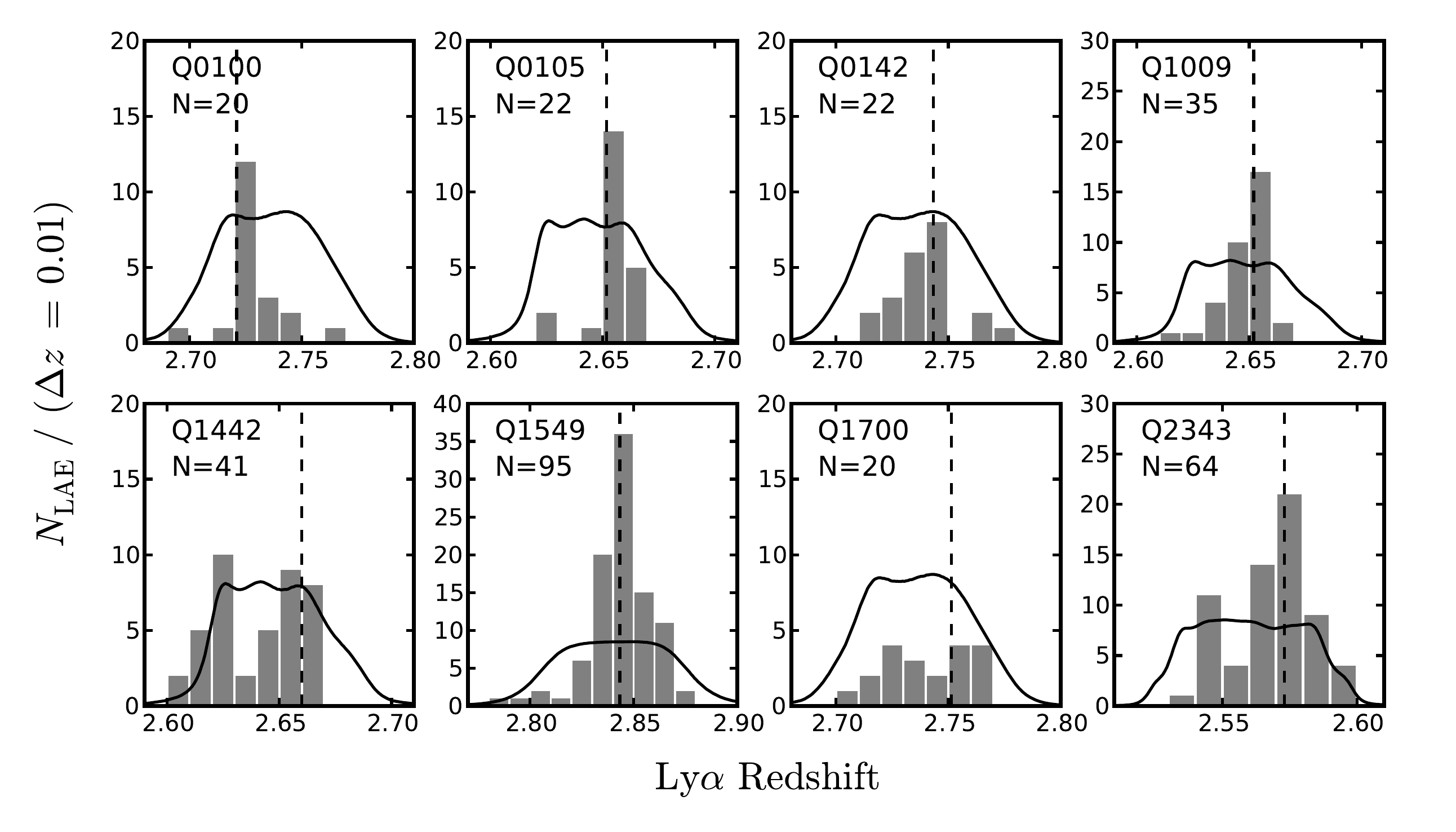}
\caption[Redshift distributions of LAEs for each field]{The redshift
  distribution of all spectroscopically-identified LAEs in each
  field (grey). For each field, the number of objects with spectra is
  given in the upper left.  Vertical dashed lines correspond to the
  HLQSO redshift in each field. Solid black lines denote the transmission
  function for the narrowband filter used to select objects in each
  field; the normalization is such that N=1 is 10\% transmission, the
  cut-off value for the automatic line-detection algorithm. Note that
  nearly all fields display a strong association of LAEs with the QSO
  redshift in addition to a more broadly-distributed component.}
\label{fig:zhists_grid}
\end{figure*}

\begin{figure}
\center
\includegraphics[width=\linewidth]{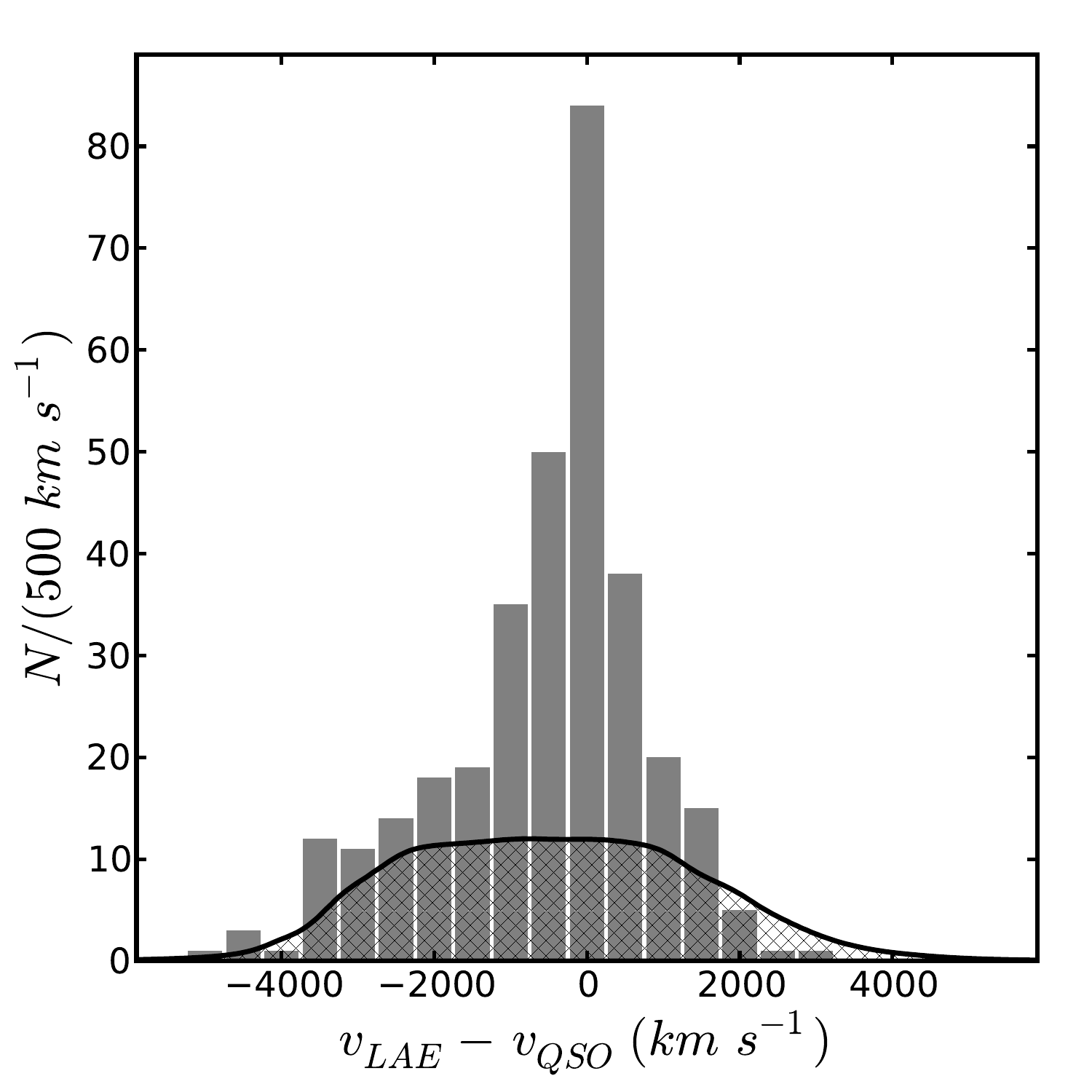}
\caption[Velocity distribution of LAEs with respect to their nearby
HLQSOs]{The velocity distribution of all spectroscopically-identified 
  LAEs with respect to their nearby hyperluminous QSO. The LAE redshifts
  are taken from the spectral detection of \lya, but are shifted by
  $-$200 km s$^{-1}$ to account for the mean \lya\ offset with
  respect to systemic. The hatched curve is the selection function 
  defined by the set of narrowband filters used to select the LAEs.}
\label{fig:vel_hist}
\end{figure}

QSO redshifts were determined as described in \citet{tra12} and have
estimated uncertainties $\sigma\sub{z,QSO}\approx 270$ km
s$^{-1}$. The redshift distribution of LAEs with respect to their
nearby QSO is shown for each field in Fig.~\ref{fig:zhists_grid}. On
small scales, the redshift distribution of LAEs with respect to the
HLQSOs is more meaningful to consider as a distribution of
velocities. The relative LAE-QSO velocities are calculated as
$v\sub{LAE} = c(z\sub{LAE}-z\sub{QSO}) / (1+z\sub{QSO})$. The
distribution of QSO-centric velocities for our entire spectroscopic
LAE sample is given in Fig.~\ref{fig:vel_hist}. Here and elsewhere in
this paper we assume that the LAEs have systemic redshifts given by 
$z\sub{LAE}=z\sub{\lya}-200$ km s$^{-1}$ to account for the typical
redshift of the \lya\ emission peak with respect to systemic derived
from our MOSFIRE data set (Sec.~\ref{sublaes:mosobs}).

\subsection{MOSFIRE Spectroscopic Sample} \label{sublaes:mosobs}

For a small subset of these \lya-selected objects, rest-frame optical spectra were also
obtained via the new Multi-Object Spectrometer For InfraRed
Exploration (MOSFIRE; \citealt{mcl10,mcl12}) on the Keck 1
telescope. These data were taken over the course of the 
KBSS-MOSFIRE survey \citep{ste14}. All the MOSFIRE data considered in
this analysis are from the field around the HLQSO Q2343+12
($z=2.573$). ${H}$-band and/or ${K}$-band spectra were obtained for 43 LAEs in
this field and reduced via the MOSFIRE-DRP; the observational
strategies employed and reduction software are described in
\citet{ste14}. Redshift fitting and one-dimensional spectral extraction were performed
using the IDL program MOSPEC (A. Strom et al., in prep.). With 0\farcs7 slits, these
spectra have $R\sim3600$, $\sigma_v\sim35$ km s$^{-1}$. Of the
43 observed LAEs, 32 yielded robust redshifts via 
measurement of \ha\ (${K}$-band, 29 objects), \hb, and/or the \oiii\
$\lambda\lambda 4959,5007$ doublet (${H}$-band, 12 objects) in
emission. Nine objects have measured redshifts in both
bands. Four of the MOSFIRE objects do not meet the \lya\ equivalent width threshold
employed here to define LAEs  in a strict sense ($W\sub{\lya}>20$\AA), but
these four are kept in the sample because they were otherwise selected
via the same photometric techniques used to define the rest of the
sample, and they all display strong \lya\ emission in their
spectra. Furthermore, the other photometric and spectroscopic
properties of these four objects are typical with respect to the remainder
of the MOSFIRE sample.

Additional information on these spectra is given in Table  
\ref{table:mosobs} and Fig.~\ref{fig:lya_stamps}. The redshift measured
from the MOSFIRE spectra, whether from the ${H}$ or ${K}$
band, is hereafter denoted by $z\sub{neb}$. In the nine cases where
a measurement was obtained in both bands, the redshift measurement from the
higher-S/N $H$ was used, but both bands generally agree quite closely:
the median redshift difference for the two bands is 17 km s$^{-1}$,
roughly half the $\sigma\sim35$ \kms\ resolution of these spectra. The fit
redshifts have typical uncertainties $\sigma_H\sim5$ \kms\
and $\sigma_K\sim 7-20$ \kms\ for the $H$ and $K$ band redshifts,
respectively, based on their Gaussian fits.

Our pre-existing $\mathcal{R}$-band image in the Q2343 field is
rotated with respect to the narrowband and $B$-band images used to select
LAEs, so $\sim$16\% of the \lya-imaged field does not currently have
$\mathcal{R}$-band coverage. Objects falling in this region are
omitted when discussing average $\mathcal{R}$-band properties of the
LAEs. In particular, 4 of the 32 MOSFIRE LAEs have no measured
$\mathcal{R}$ magnitudes (Table~\ref{table:mosobs}), but their
continuum properties in other bands ($B$ and NIR where available) are
consistent with those falling within our existing $\mathcal{R}$
coverage. The 3-$\sigma$ $\mathcal{R}$-band limit is reported in Table~\ref{table:mosobs}
for objects with measured $\mathcal{R}$ magnitudes fainter than that value.

Velocity dispersions are measured (or constrained) for each MOSFIRE LAE from the
Gaussian fit to the detected spectral line. As above, the highest-S/N
line (generally [\ion{O}{3}]
$\lambda$5007) width is used when multiple lines are detected.
The measured line widths (deconvolved from the instrumental profile)
range from $\sigma\sub{neb}=14\pm9$ km s$^{-1}$ to 
$114\pm8$ km s$^{-1}$ (Table~\ref{table:mosobs}), with a median 
$\langle\sigma\sub{neb}\rangle= 34$ km s$^{-1}$. The object with the
largest fit dispersion (Q2343-NB2807) has a two-peaked nebular line
profile that is not well-fit by a single Gaussian; with the exception
of this object, no LAE has a fit dispersion larger than 73 \kms. Ten objects have
line widths consistent with the MOSFIRE instrument profile and are given
as limits ($\sigma\sub{neb}<35$ km s$^{-1}$). The LAEs are generally
unresolved in our ground-based images, and further analysis of the
limited {\it HST}/WFC3 imaging in these fields is forthcoming. The similar
LAE sample of \citet{erb14}, however, have a median effective
half-light radius $a=0.6$ kpc in {\it HST} F814W photometry. Taking this
value as a typical size for the KBSS-\lya\ LAEs, we can estimate their
dynamical masses via the formula $M\sub{dyn}=5 a \sigma^2/G$. As in
\citet{erb14}, we assume a spherical geometry for these systems,
noting that the dynamical mass would be smaller for a disk-like
geometry. Further discussion of the likely physical morphologies of
these objects is given in Sec.~\ref{sublaes:lbgoutflows}. For the
assumed geometries and sizes, the KBSS-\lya\ LAEs would have dynamical
masses $1.4\times10^8$ $\msun<M\sub{dyn}<8.4\times10^{9}$
$\msun$.\footnote{With the exception of Q2343-NB2807 (for which a
  single Gaussian function is a poor fit to the nebular line profile),
the largest estimated dynamical mass in our LAE sample is
$3.7\times10^9$ $\msun$. Note also that the lower mass bound does not
include those objects with only an upper bound on their nebular
velocity dispersion.} The median mass is $\langle M\sub{dyn}
\rangle=8.0\times10^{8}$ $\msun$.

\begin{deluxetable*}{lccccccccccc}
\tablecaption{MOSFIRE Observations}
\tablewidth{0pt}
\tablehead{
\colhead{Object Name} & $W\sub{\lya}$& $z\sub{\lya}\super{peak}$\tablenotemark{a} & $z_{H}$\tablenotemark{a}
& $z_{K}$\tablenotemark{a} & $v\sub{\lya}\super{peak}$ & $v\sub{\lya}\super{ave}$ & $\sigma\sub{neb}$\tablenotemark{b} & $\mathcal{R}$ &
$F\sub{\lya}$\tablenotemark{c} & $F\sub{\ha}$\tablenotemark{c} & $F\sub{\lya}/$\\
& (\AA) & & & & (km s$^{-1}$)  & (km s$^{-1}$) & (km s$^{-1}$) & (mag)
& ($10^{-17}$) & ($10^{-17}$) & $F\sub{\ha}$
}

\startdata
Q2343-NB0280 &  \pp52.6 & 2.5801 & 2.5777 & $-$ &   \ps201 &   \ps364 &  \pp34$\pm$9\pp &  26.6 &  \pp6.4$\pm$0.2 &  $-$ &  \pp$-$ \\
Q2343-NB0308\tablenotemark{e} &  \pp58.8 & 2.5690 & 2.5663 & \phantom{\super{d}}(2.5666)\tablenotemark{d} &   \ps226 &   \ps360 &  $<$35 &  26.4 &  \pp4.0$\pm$0.3 &  1.3$\pm$0.3 &  \pp3.1 \\[1pt] 
Q2343-NB0345 &  \pp36.9 & 2.5917 & $-$ & 2.5876 &   \ps343 &   \ps414 &  \pp21$\pm$6\pp &  25.3 & 13.2$\pm$0.4 &  3.8$\pm$0.3 &  \pp3.5 \\
Q2343-NB0405 & 134.8 & 2.5892 & 2.5862 & \phantom{\super{d}}(2.5860)\tablenotemark{d} &   \ps247 &   \ps293 &  \pp33$\pm$8\pp &  27.1 &  \pp2.9$\pm$0.2 &  1.0$\pm$0.2 &  \pp2.9 \\[1pt] 
Q2343-NB0565 & 313.9 & 2.5661 & $-$ & 2.5635 &   \ps215 &   \ps\pp54 &  \pp34$\pm$12 &  26.8 &  \pp5.7$\pm$0.3 &  1.3$\pm$0.2 &  \pp4.4 \\[1pt] 
Q2343-NB0791 &  \pp37.4 & 2.5710 & $-$ & 2.5741 &  $-$259 &  $-$116 &  \pp39$\pm$9\pp & $-$ &  \pp3.6$\pm$0.3 &  2.7$\pm$0.3 &  \pp1.3 \\[1pt] 
Q2343-NB1154 &  \pp54.1 & 2.5773 & $-$ & 2.5765 &   \ps\pp65 &   \ps219 &  $<$35 & $-$ &  \pp5.8$\pm$0.2 &  1.0$\pm$0.3 &  \pp5.6 \\[1pt] 
Q2343-NB1174 & 236.6 & 2.5509 & $-$ & 2.5478 &   \ps262 &   \ps283 &  $<$35 & $-$ & 12.5$\pm$0.4 &  1.5$\pm$0.3 &  \pp8.6 \\
Q2343-NB1361 &  \pp90.6 & 2.5599 & 2.5587 & \phantom{\super{d}}(2.5586)\tablenotemark{d} &   \ps105 &   \ps\pp50 &  \pp36$\pm$8\pp &  27.0 &  \pp3.4$\pm$0.2 &  0.3$\pm$0.1 & 11.2 \\[1pt] 
Q2343-NB1386 &  \pp25.8 & 2.5691 & $-$ & 2.5647 &   \ps371 &   \ps\pp89 &  $<$35 & $>$27.3\phantom{$>$} &  \pp1.6$\pm$0.2 &  1.1$\pm$0.3 &  \pp1.5 \\
Q2343-NB1416 &  \pp20.7 & 2.5602 & 2.5582 & \phantom{\super{d}}(2.5594)\tablenotemark{d} &   \ps169 &   \ps275 &  \pp35$\pm$14 &  26.8 &  \pp1.8$\pm$0.2 &  0.7$\pm$0.3 &  \pp2.4 \\
Q2343-NB1501 &  \pp35.2 & 2.5614 & 2.5593 & \phantom{\super{d}}(2.5592)\tablenotemark{d} &   \ps180 &   \ps158 &  \pp71$\pm$12 & $>$27.3\phantom{$>$} &  \pp1.7$\pm$0.2 &  1.2$\pm$0.4 &  \pp1.4 \\[1pt] 
Q2343-NB1518 &  \pp22.3 & 2.5890 & $-$ & 2.5860 &   \ps251 &   \ps238 &  $<$35 &  26.2 &  \pp3.3$\pm$0.2 &  0.8$\pm$0.3 &  \pp4.0 \\
Q2343-NB1585 & 303.4 & 2.5669 & 2.5649 & \phantom{\super{d}}(2.5652)\tablenotemark{d} &   \ps169 &   \ps\pp93 &  $<$35 & $>$27.3\phantom{$>$} &  \pp5.1$\pm$0.2 &  0.6$\pm$0.2 &  \pp9.0 \\[1pt] 
Q2343-NB1692 & 359.2 & 2.5625 & $-$ & 2.5602 &   \ps194 &   \ps\pp85 &  \pp14$\pm$9\pp &  26.8 &  \pp5.6$\pm$0.2 &  1.1$\pm$0.3 &  \pp5.0 \\
Q2343-NB1783 &  \pp35.0 & 2.5784 & 2.5767 & \phantom{\super{d}}(2.5766)\tablenotemark{d} &   \ps143 &   \ps167 &  \pp51$\pm$3\pp &  25.7 &  \pp7.8$\pm$0.2 &  2.0$\pm$0.2 &  \pp3.8 \\[1pt] 
Q2343-NB1789 &  \pp88.2 & 2.5471 & $-$ & 2.5448 &   \ps196 &   \ps\pp59 &  \pp45$\pm$16 & $>$27.3\phantom{$>$} &  \pp2.3$\pm$0.2 &  1.4$\pm$0.3 &  \pp1.6 \\[1pt] 
Q2343-NB1806 &  \pp19.3 & 2.5982 & $-$ & 2.5954 &   \ps237 &   \ps204 &  \pp73$\pm$19 &  26.7 &  \pp1.2$\pm$0.1 &  1.4$\pm$0.3 &  \pp0.8 \\[1pt] 
Q2343-NB1828 &  \pp26.7 & 2.5753 & $-$ & 2.5724 &   \ps247 &   \ps216 &  \pp38$\pm$15 &  26.9 &  \pp2.7$\pm$0.3 &  1.0$\pm$0.2 &  \pp2.6 \\[1pt] 
Q2343-NB1829 &  \pp32.1 & 2.5782 & $-$ & 2.5754 &   \ps233 &   \ps271 &  \pp41$\pm$8\pp &  25.9 &  \pp5.7$\pm$0.2 &  2.6$\pm$0.3 &  \pp2.2 \\[1pt] 
Q2343-NB1860 &  \pp23.6 & 2.5740 & $-$ & 2.5741 &   \pp$-$11 &   \ps237 &  $<$35 &  25.7 &  \pp1.2$\pm$0.2 &  1.0$\pm$0.3 &  \pp1.2 \\
Q2343-NB2211\tablenotemark{e} &  \pp66.2 & 2.5776 & 2.5760 & \phantom{\super{d}}(2.5758)\tablenotemark{d} &   \ps135 &   \ps206 &  \pp30$\pm$13 &  26.9 &  \pp4.3$\pm$0.3 &  1.1$\pm$0.2 &  \pp4.0 \\[1pt] 
Q2343-NB2571 &  \pp29.4 & 2.5820 & $-$ & 2.5850 &  $-$249 &  $-$503 &  \pp38$\pm$26 & $>$27.3\phantom{$>$} &  \pp1.4$\pm$0.3 &  0.8$\pm$0.2 &  \pp1.8 \\[1pt] 
Q2343-NB2785 &  \pp14.2 & 2.5723 & $-$ & 2.5785 &  $-$517 &  $-$367 &  $<$35 & $>$27.3\phantom{$>$} &  \pp1.3$\pm$0.1 &  0.4$\pm$0.1 &  \pp3.2 \\[1pt] 
Q2343-NB2807\tablenotemark{e} &  \pp24.6 & 2.5479 & 2.5446 & $-$ &   \ps279 &   \ps184 & 110$\pm$8\pp &  23.8 & 23.4$\pm$0.3 &  $-$ &  \pp$-$ \\[1pt] 
Q2343-NB2816 &  \pp58.5 & 2.5782 & $-$ & 2.5742 &   \ps340 &   \ps243 &  \pp57$\pm$23 & $>$27.3\phantom{$>$} &  \pp2.0$\pm$0.1 &  1.5$\pm$0.4 &  \pp1.4 \\[1pt] 
Q2343-NB2821 &  \pp32.4 & 2.5800 & $-$ & 2.5779 &   \ps173 &   \ps\pp69 &  \pp24$\pm$23 & $-$ &  \pp1.3$\pm$0.2 &  1.3$\pm$0.4 &  \pp1.0 \\[1pt] 
Q2343-NB2834 &  \pp\pp7.2 & 2.5700 & $-$ & 2.5683 &   \ps142 &   \pp$-$69 &  $<$35 & $>$27.3\phantom{$>$} &  \pp0.9$\pm$0.1 &  0.5$\pm$0.2 &  \pp1.7 \\[1pt] 
Q2343-NB3061 &  \pp57.6 & 2.5788 & $-$ & 2.5764 &   \ps203 &   \ps270 &  \pp72$\pm$24 &  26.5 &  \pp2.2$\pm$0.2 &  2.0$\pm$0.5 &  \pp1.1 \\[1pt] 
Q2343-NB3170 &  \pp90.0 & 2.5488 & 2.5475 & $-$ &   \ps113 &   \ps152 &  \pp15$\pm$5\pp & $>$27.3\phantom{$>$} &  \pp2.6$\pm$0.1 &  $-$ &  \pp$-$ \\
Q2343-NB3231\tablenotemark{f} & 110.4 & 2.5737 & 2.5711 & \phantom{\super{d}}(2.5695)\tablenotemark{d} &   \ps218 &   \ps151 &  \pp43$\pm$12 &  25.9 &  \pp8.4$\pm$0.2 &  1.0$\pm$0.2 &  \pp8.7 \\[1pt] 
Q2343-NB3292 &  \pp13.4 & 2.5700 & $-$ & 2.5631 &   \ps579 &   \ps153 &  $<$35 & $>$27.3\phantom{$>$} &  \pp1.3$\pm$0.2 &  0.6$\pm$0.2 &  \pp2.2  
\enddata
\tablenotetext{a}{$z\sub{\lya}$, $z_H$, and $z_K$ denote the redshifts
  measured from fitting the \lya\ emission line, $H$ band emission lines (primarily
  [\ion{O}{3}] $\lambda$5007), and $K$ band emission lines (\ha\ exclusively).} 
\tablenotetext{b}{$\sigma\sub{neb}$ is the velocity width of the
  nebular emission line used to define $z\sub{neb}$. Both the line widths
  and their associated uncertainties reflect the deconvolution of the
  $\sigma\sub{inst} = 35$ km s$^{-1}$ instrumental profile.} 
\tablenotetext{c}{Line fluxes (in cgs units) are determined from a combination of
  spectroscopic and photometric measurements as described in Sec.~\ref{sublaes:lyashift}.} 
\tablenotetext{d}{In the nine cases where both the ${H}$ and
${K}$ band spectra yielded redshifts, we adopted the
higher-S/N ${H}$ band redshifts to define $z\sub{neb}$.} 
\tablenotetext{e}{The nebular redshifts of Q2343-NB0308, Q2343-NB2211,
  and Q2343-NB2807 were measured from the [\ion{O}{3}]
  $\lambda$4959\AA\ line because the $\lambda$5007\AA\ 
  line did not fall on the detector. }
\tablenotetext{f}{The nebular redshift of Q2343-NB3231 was measured from 
  the \hb\ line (and cross-checked with \ha) because the [\ion{O}{3}]
  $\lambda$5007\AA\ and $\lambda$4959\AA\ lines did not fall on the detector. }
\label{table:mosobs}
\end{deluxetable*}


\section{Lyman-$\alpha$ spectral morphology}\label{laes:lya}

The observed luminosity and spectral profile
of the \lya\ line is the result of both the input distribution of
\lya\ photons produced directly from recombination and the resonant
scattering that impedes and shapes their escape. The net
result of these processes in star-forming galaxies is generally seen
to be the broadening of the \lya\ line and an overall shift redward
with respect to the systemic redshift (e.g., \citealt{sha03}), an
observation that can be simply explained by outflowing gas that
drives \lya\ photons to significant shifts in both space and velocity
\citep{mas03,ste10}. The scattering and absorption of these \lya\ photons
likewise leads to severely attenuated luminosities with
respect to early predictions of star-formation-associated \lya\
emission (e.g., \citealt{par67,gia96}). The details of these effects,
however, depend on many factors, including the dust content, column
density, covering fraction, and velocity distribution of the scattering medium. 

As the \lya\ line profile bears the imprint of these diverse galaxy
and gas properties, it is an important tool for studying galaxy
evolution. However, the degenerate effects of these processes require
care in their interpretation. For instance, \citet{has13},
\citet{sch13}, and \citet{shi14} show that galaxies selected via their \lya\ emission
have a smaller velocity shift of \lya\ with respect to systemic
compared to continuum-selected galaxies (e.g.,those of
\citealt{ste10}). \citet{erb14} extend this study to a large sample of
LAEs and LBGs with 
systemic redshifts, demonstrating that the \lya\ velocity shift shows
a significant inverse trend with $W\sub{\lya}$ for the combined population of
galaxies selected by either technique. The physical basis of this
relationship remains unclear, 
however. In particular, it is not clear whether the velocity offset of
\lya\ is predominantly driven by outflow velocity (as suggested by
\citealt{has13}) or \ion{H}{1} optical depth (as suggested by
\citealt{cho13}). 

Furthermore, many LAEs do not exhibit simple \lya\
shifts, but rather show complex, multi-peaked profiles that may
correspond to either internal star formation \citep{kul12} or
externally-illuminated fluorescent processes \citep{kol10}. In
particular, \lya\ lines that are {\it blueshifted} with respect to systemic
are predicted by simple models of emission from inflowing \ion{H}{1}
\citep{ver06} whose low star-formation and intrinsic \lya-production
would likely hide them from view in non-fluorescing populations. The
distribution of \lya\ spectral properties and their relation to other
physical properties of the galaxies (including the \lya\ escape
fraction) are thus an important window into 
the processes that drive gas into and out of galaxies throughout
their evolution.

\subsection{\lya\ escape fraction}\label{sublaes:lyaescape}

\begin{figure*}
\center
\includegraphics[width=\linewidth]{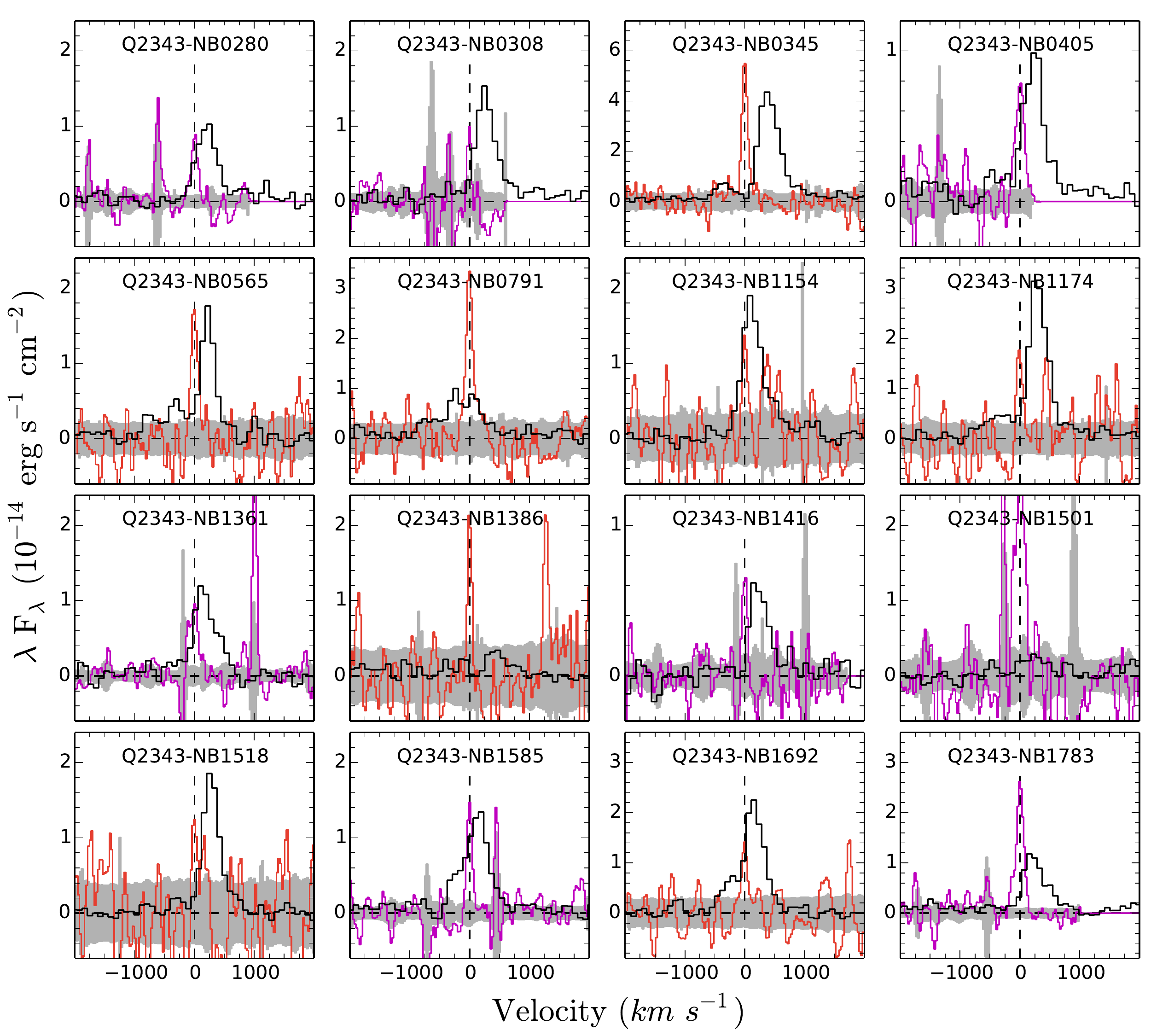}
\caption[\lya\ and nebular line spectra for MOSFIRE sample]{\lya\
 and nebular line spectra for the subsample of objects with MOSFIRE
 redshifts. \lya\ spectra (black) are from LRIS, with resolution
 $R\sim1300$. Nebular spectra are for the \ha\ line ($K$ band, red) or \hb\ or
 [\ion{O}{3}] ($H$ band, purple), whichever was used to estimate the systemic
 redshift (see Table~\ref{table:mosobs} and notes below). MOSFIRE
 spectra have been smoothed by a three-pixel boxcar to reduce
 noise. Grey regions denote the $\pm1\sigma$ uncertainties on the
 MOSFIRE spectra. Both spectra are shifted to the redshift frame
 estimated from $z\sub{neb}$. The fluxes are in $\lambda F_\lambda$
 units to facilitate comparison of their integrated  line
 luminosities. (Figure is continued below)}
\label{fig:lya_stamps}
\end{figure*}

Postage stamps of each \lya\ and nebular line spectrum in the MOSFIRE
sample are displayed in Fig.~\ref{fig:lya_stamps}. The majority of the
nebular redshifts are derived from \ha, for which the intrinsic flux 
ratio is $F\sub{\lya}/F\sub{\ha} \approx 8.7$ under the typical assumption
of case-B (i.e., ionization-bounded) recombination and $T_e=10^4$
K.\footnote{The assumed ratio $F\sub{\lya}/F\sub{\ha} \approx 8.7$
  is commonly adopted in the literature, often with reference to
  \citet{bro71}. However, \citet{bro71} does not explicitly
  calculate $F\sub{\lya}/F\sub{\ha}$, but only line ratios including
  states at energy levels $n\ge2$ (e.g., \ha/\hb; see note in
  \citealt{hen15}). While there is some variation in the calculated
  \lya/\hb\ and \ha/\hb\ ratios with temperature and density (e.g., \citealt{dop03}), values
  in the range $F\sub{\lya}/F\sub{\ha} = 8-10$ are typical and do
  not change our results appreciably. We adopt $F\sub{\lya}/F\sub{\ha}
  \approx 8.7$ for consistency with previous studies at high and low
  redshifts (e.g., \citealt{hay10,hen15}).}

The LAE \lya\ line fluxes ($F\sub{\lya}$)
are measured photometrically from their continuum-subtracted
narrowband detections. Directly integrating the rest-UV spectrum
yields a \lya\ flux $\sim$2$\times$ smaller on average. This
difference may be partially due to uncertainties in the absolute
flux calibration of the rest-UV spectra, but the consistency between
our stacked continuum spectra and broadband flux measurements
(Sec.~\ref{sublaes:abs}) suggests that this effect is
minimal. Similarly, the photometric errors in the measured narrowband
\lya\ fluxes are estimated to be $\lesssim$15\%. Rather,
it appears that the differential slit losses between \lya\ and
continuum photons are significant within our 1\farcs2 wide slits.

For the MOSFIRE subsample of LAEs with $K$-band spectra, \ha\ fluxes
($F\sub{\ha}$; Table~\ref{table:mosobs}) are estimated by a Gaussian 
fit to the emission line. Slit losses are difficult to estimate for
individual faint objects, so the measured values have been multiplied by a
factor 1.5$-$2.1 (depending on the mask) to account for the average
slit loss measured for bright point sources on the same slitmasks
(A. Strom et al., in prep.). Most LAEs in our sample do not have
  robust detections of both \ha\ and \hb, so we have
not corrected the \ha\ fluxes for dust attenuation, but the four
objects with \ha\ and \hb\ SNR $>2$ have a mean ratio
$\langle F\sub{\ha}/F\sub{\hb}\rangle =3.3\pm1.1$, consistent with
the canonical (non-attenuated) case-B value
$F\sub{\ha}/F\sub{\hb}=2.86$ \citep{bro71}. 

The measured line ratios in Fig.~\ref{fig:lya_stamps} vary
considerably ($F\sub{\lya}/F\sub{\ha}\sim 1 - 11$,
Table~\ref{table:mosobs}) with respect to that predicted by case-B
recombination. If the intrinsic
$F\sub{\lya}/F\sub{\ha}$ flux ratio is assumed to be 8.7 (and the \ha\ flux is
not significantly affected by extinction), then the median ratio
$\langle F\sub{\lya}/F\sub{\ha}\rangle = 2.7$ suggests a typical \lya\
escape fraction $\langle f\sub{esc,\lya}\rangle =\langle F\sub{\lya}\super{observed}/
F\sub{\lya}\super{intrinsic}\rangle \approx 30\%$. This
  $F\sub{\lya}/F\sub{\ha}$ ratio is consistent with that observed in the 4
  objects with \ha\ line detections in the similar faint-LAE sample of
\citet{erb14}. If the \ha\
emission is attenuated by interstellar dust, then the intrinsic values of
$f\sub{esc,\lya}$ may be lower, but the measured values are considerably
higher than previous estimates of $f\sub{esc,\lya}\approx5\%$ for typical
high-redshift star-forming galaxies \citep{hay10,ste11,cia14}. Such high
\lya\ escape fractions are suggestive of the high LyC escape
fractions observed in samples of LAEs \citep{nes13,mos13} and may
indicate low \ion{H}{1} optical depths for photons escaping from
star-forming regions in these galaxies. Given the proximity of these
LAEs to hyperluminous QSOs, however, it may be that $f\sub{esc,\lya}$
is elevated by the ``fluorescent'' boosting of the total \lya\ flux
from these objects.

For comparison, the KBSS LBGs have a median flux ratio $\langle
F\sub{\lya}/F\sub{\ha}\rangle \approx 0.5$, or a corresponding escape
fraction $f\sub{esc,\lya}\approx 6\%$.  The LBG \ha\ slit losses were corrected
in a similar manner to the LAEs, and further analysis of the
rest-frame optical spectra of the KBSS LBGs will be presented
in A. Strom et al., (in prep.). The LBGs lack narrowband \lya\
photometry, so the \lya\ fluxes were measured by direct integration of
the rest-UV spectrum (after flux-calibrating the spectrum based on the
broadband $G$-band flux). \citet{ste11} measure \lya\ aperture
corrections for LBG spectra showing \lya\ in net emission or
absorption in comparison to LAEs, finding that LBGs with detected \lya\
emission show similar differential slit losses to LAEs in our 1\farcs2
slits. In this case the \lya\ luminosities of the KBSS LBGs may be
underestimated by a factor of $\sim 2\times$, and the escape
fractions may be  $f\sub{esc,\lya}\approx 12\%$ in a larger
aperture. This \lya\ escape fraction is similar to that derived for
\lya-emitting LBGs ($f\sub{esc,\lya}= 14.4\%$) by \citet{ste11}. It thus appears that 
our sample of \lya-emitting, continuum-color selected KBSS 
LBGs have a broadly similar \lya\ escape fraction with respect to
previously studied samples of LBGs, and significantly smaller values than the
KBSS-\lya\ LAEs.

Given the small number of LAEs with \ha\ detections, it is difficult to
investigate trends within the MOSFIRE LAE sample. In particular, we
find no association between $f\sub{esc}$ and the \lya\ offset, peak
multiplicity, or peak separation ($v\sub{\lya}$, $f\sub{mult}$, and
$v\sub{peaks}$; see discussion in Secs.~\ref{sublaes:lyashift}$-$\ref{sublaes:multipeak}). However, we 
find a significant trend between $f\sub{esc,\lya}$ and $W\sub{\lya}$,
which are positively correlated with $p<4\times10^{-3}$ for a Pearson
correlation test. This result suggests that $W\sub{\lya}$ is
primarily an indicator of \lya\ {\it escape}, not \lya\ {\it
  production} among these faint galaxies. In fact, the relation
between \ha\ luminosity (a standard indicator of star-formation rate)
and $W\sub{\lya}$ is weakly negative in the sample. Given that a higher
$W\sub{\lya}$ at a given observed \lya\ luminosity corresponds to a
fainter UV continuum luminosity (another proxy for star-formation rate),
we suggest that $W\sub{\lya}$ is likely {\it anti-correlated} with
star-formation rate among these faint LAEs. This effect is similar to
the anti-correlation observed between $W\sub{\lya}$ and UV
star-formation rate (or luminosity) seen in several previous studies (e.g.,
\citealt{and06,ouc08,sta10,kor10,ate14}).  We also find a 
tentative, negative association between $W\sub{\lya}$ and
$\sigma\sub{neb}$ (the width of the nebular
\ha\ and/or [\ion{O}{3}] emission). Because $\sigma\sub{neb}$ traces the stellar velocity
dispersion and thus the dynamical mass of the galaxy, this may suggest
that higher $W\sub{\lya}$ is associated with lower dynamical masses. However,
further study of the stellar populations and star-formation rates of
the LAEs (including our ongoing photometry and rest-frame
optical spectroscopy of these objects) will be necessary to confirm
the physical basis of these trends.

\renewcommand{\thefigure}{\arabic{figure} (Cont.)}
\addtocounter{figure}{-1}
\begin{figure*}
\center
\includegraphics[width=\linewidth]{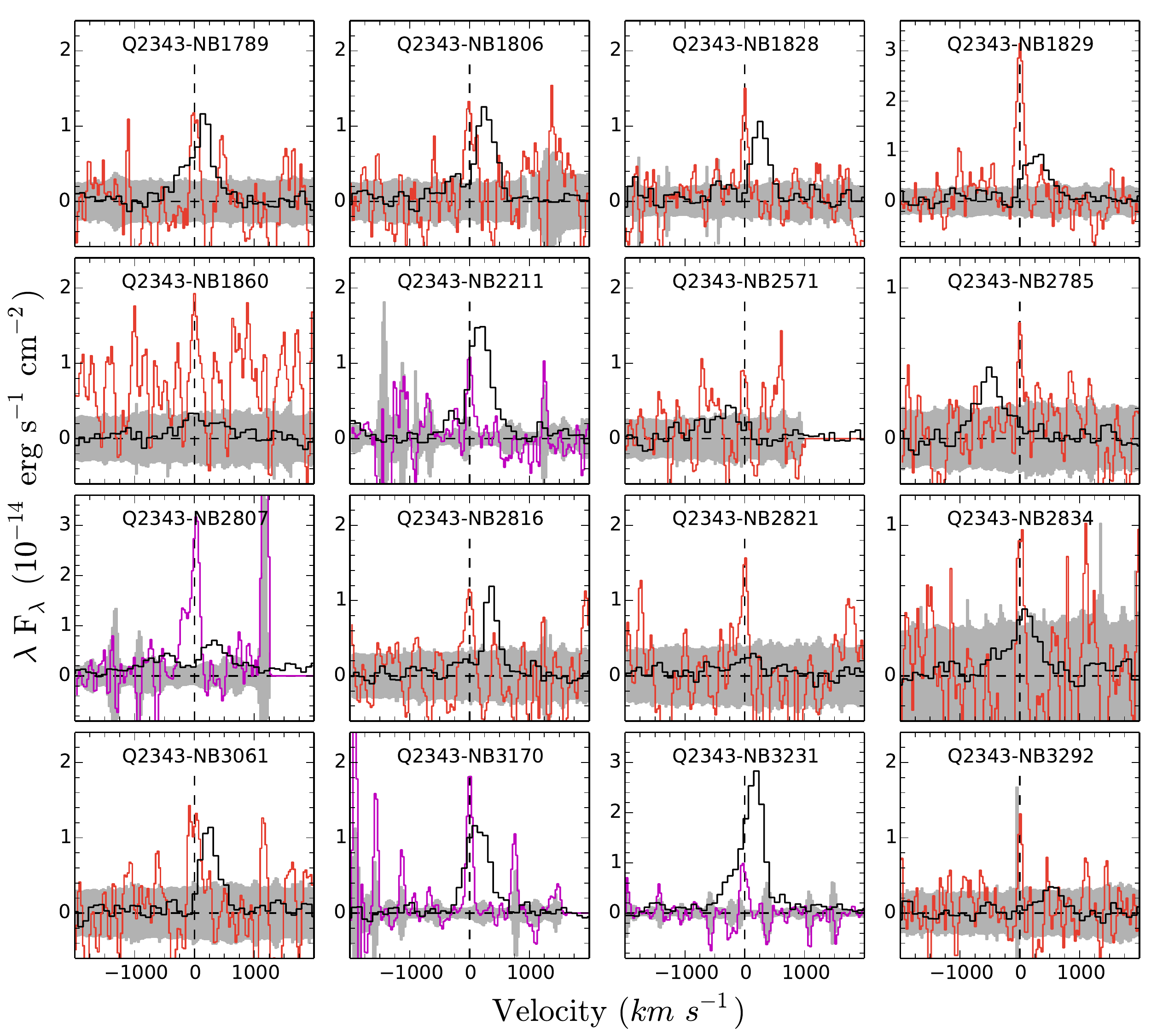}
\caption{Spectra for the remainder of the MOSFIRE LAE sample.}
\end{figure*}
\renewcommand{\thefigure}{\arabic{figure}}

\subsection{Velocity shift with respect to systemic redshift}\label{sublaes:lyashift}
 
The offset of the \lya\ line with respect to the systemic redshift was
measured for each of the 32 LAEs in the MOSFIRE sample using the
$z\sub{\lya}$ and $z\sub{neb}$ values estimated as described in
Sec.~\ref{sublaes:specobs} and \ref{sublaes:mosobs}. The velocity
offset was then inferred from the measured redshifts as

\begin{equation}
v\sub{\lya}=\frac{z\sub{\lya}-z\sub{neb}}{1+z\sub{neb}} c\,\, .
\label{eq:vlya}
\end{equation}

\begin{figure}
\center
\includegraphics[width=\linewidth]{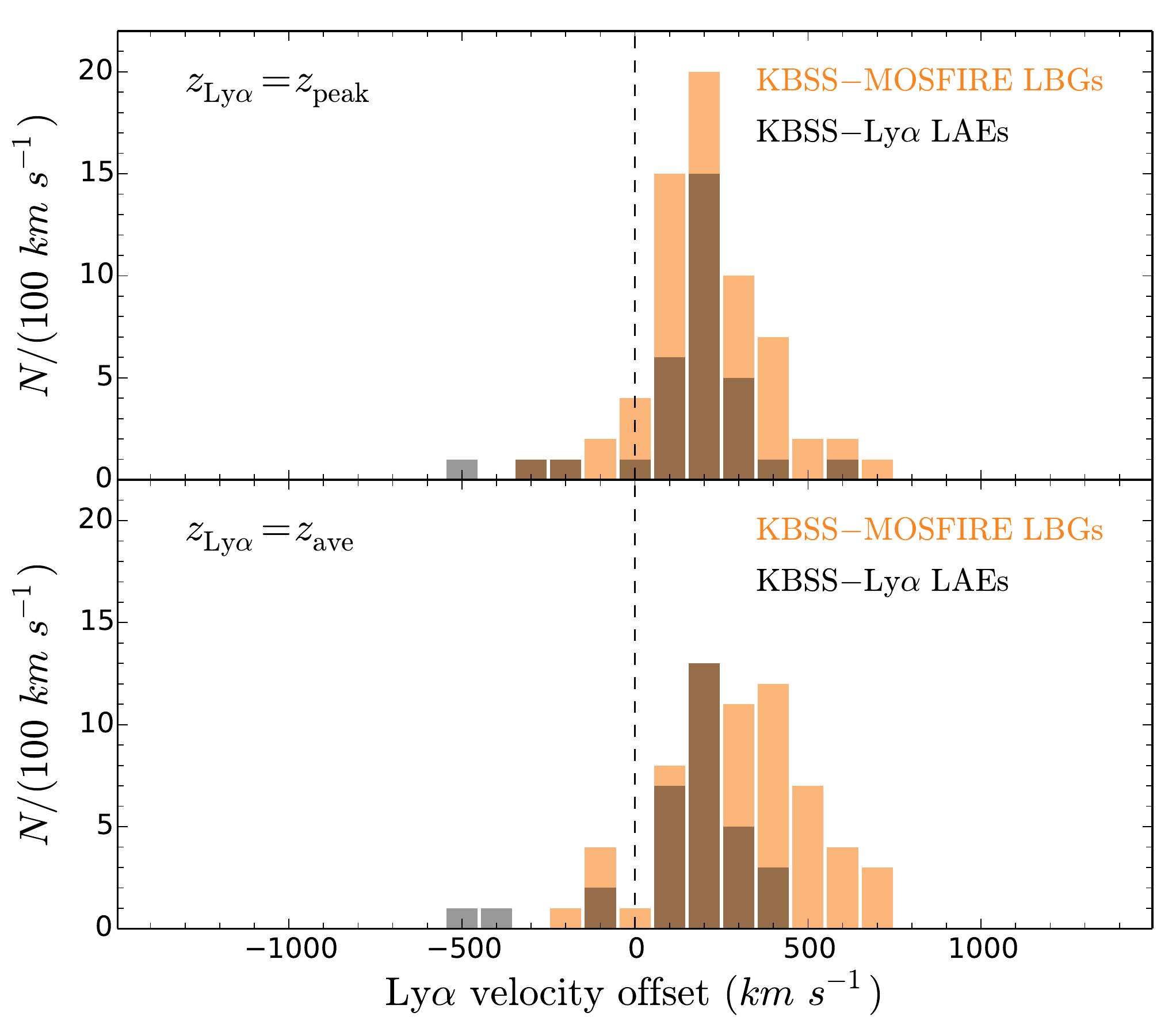}
\caption[Distribution of \lya\ velocity offsets for the MOSFIRE
sample]{Distribution of \lya\ velocity offsets with respect to
  systemic for LAEs (grey) and KBSS LBGs (orange) with
  nebular redshifts measured with MOSFIRE. In each panel, the vertical
  dashed line denotes $v\sub{\lya}=0$. {\it Top:} The
  distributions when $z\sub{\lya}$ is defined by the peak of the fit
  line profile ($z\sub{\lya,peak}$). Most of the galaxies have offsets in a 
  tight range around $v\sub{\lya}=200$ km s$^{-1}$, but a few LAEs have
  \lya\ lines blueshifted with respect to the systemic redshift
  ($v\sub{\lya}<0$). {\it Bottom:} The distributions when
  $z\sub{\lya}$ is defined by the line centroid
  ($z\sub{\lya,ave}$). Most LAEs have $v\sub{\lya}\sim200$ 
  km s$^{-1}$, but with a broader distribution compared to the
  $z\sub{\lya}=z\sub{peak}$ distribution. The LBG centroid velocities
  occupy a broader, more highly-redshifted range than either the LAEs or the
  corresponding LBG \lya\ peaks.}
\label{fig:hist_dv}
\end{figure}

The velocity shift of the \lya\ line with respect to the systemic
redshift is clear from the panels in Fig.~\ref{fig:lya_stamps}. The
majority (28/32; 88\%) of spectra display the redshifted \lya\ profile typical of
star-forming galaxies, but there are several objects with minimal
velocity shift, or even a dominant peak blueward of the systemic
redshift. The distribution of measured velocity shifts (Eq.~\ref{eq:vlya})
is given in Fig.~\ref{fig:hist_dv}. The typical \lya\ shift of $+200$
km s$^{-1}$ is clearly visible. This shift is consistent with that
found for the few samples of LAEs with systemic redshifts in the
literature (e.g., \citealt{mcl11,has13,cho13,shi14}), and significantly
less than that previously measured in samples LBGs (300$-$400 km
s$^{-1}$, \citealt{ste10}). However, we demonstrate below that the detailed
velocity distribution depends on the method by which the \lya\
emission redshift is assigned. 

The two panels of Fig.~\ref{fig:hist_dv}
show the effect of defining \zla\ by either the fitted peak
($z\sub{\lya,peak}$) or the flux-weighted centroid  of the line profile
($z\sub{\lya,ave}$; see Sec.~\ref{sublaes:specobs}). Both
methods yield qualitatively similar distributions, with median values
$\langle v\sub{\lya,peak}\rangle=198$ km s$^{-1}$ and $\langle
v\sub{\lya,ave}\rangle=176$ km s$^{-1}$. However, $z\sub{\lya,peak}$
displays a tighter correlation with the redshift derived from
nebular emission lines than $z\sub{\lya,ave}$.
The median absolute deviations (a scale estimator that is insensitive
to outliers) of the two
distributions of $v\sub{\lya}$ are MAD($v\sub{\lya,peak}$) $= 54$ km $s^{-1}$ and
MAD($v\sub{\lya,ave}$) $= 93$ km $s^{-1}$.
The similarities between the
two distributions suggest that either quantity is a reasonable proxy for
the systemic redshift after correcting for the typical 200 km s$^{-1}$
offset, and even the individual velocities computed by both
measurements are consistent within 100 km s$^{-1}$ for the majority of
objects (Table~\ref{table:mosobs}). Because $z\sub{\lya,peak}$ tracks
$z\sub{neb}$ more tightly, 
however, we adopt $z\sub{sys,\lya}=z\sub{\lya,peak}-200$ as the best
estimate of the systemic redshift in the absence of nebular emission
line measurements.

The distribution of \lya\ line offsets from the comparison KBSS 
LBG sample is also displayed in Fig.~\ref{fig:hist_dv}, and
there the differences between the two \lya\ redshift estimators are
more apparent. While the distribution of fit peak velocities
$v\sub{\lya,peak}$ is again sharply peaked around 
$v\sub{\lya}\approx 200$ km s$^{-1}$, the centroid velocities $v\sub{\lya,ave}$
are more extended, particularly toward higher (more strongly
redshifted) velocities. The median velocity offsets for the LBG sample are
$\langle v\sub{\lya,peak}\rangle=205$ km s$^{-1}$ and $\langle
v\sub{\lya,ave}\rangle=306$ km s$^{-1}$. Note that our measurement of
$v\sub{\lya,peak}$ in particular suggests a significantly smaller
\lya\ offset than previous studies of LBGs. Qualitatively, the difference
between the two distributions is due to the fact that the LBG \lya\
emission lines have extended red tails (often combined with
continuum flux levels that are higher on the red side of \lya\ than
the blue side) that draw the velocity centroid 
to higher/redder values. Fitting the \lya\ emission line with a symmetric
Gaussian model will likewise cause the fit emission-line center to
be redder than the \lya\ emission peak. While the LAE emission lines
tend to be asymmetric as well, they typically have more symmetric
profiles and/or stronger blue secondary peaks that compensate for
the extended red wing. The LAEs also have less continuum emission that
is more symmetric about the wavelength of \lya\ than typical LBGs
(e.g., the observed ratio of $f_{1125}$ to $f_{1325}$; see
Sec.~\ref{sublaes:lbgoutflows}). The distributions of emission line widths,
asymmetry, and peak multiplicity for each sample are discussed in
Sec.~\ref{sublaes:multipeak} below.

Despite their differing line
morphologies and photometric properties, it is significant that both the
LAE and LBG samples have \lya\ emission line peaks that are
well-described by a uniform offset of $\sim$200 km s$^{-1}$ redward of
the systemic redshifts. While the measured velocity of the \lya\ peak
may be strongly dependent on spectral resolution, its constancy across a
broad range of galaxy luminosities suggests that it is less sensitive
to galaxy properties than the \lya\ emission centroid. For a sample of
galaxies observed {\it with the same spectral resolution}, therefore,
the \lya\ emission peak appears to be a robust redshift indicator across
a diverse range of galaxy types (with the caveat that $\sim$50\% of $z\sim 2.4$
LBGs have no net \lya\ emission in their spectra; \citealt{ste11}). We use this redshift
calibration to create composite LAE spectra from our large sample of
LAEs without nebular redshift measurements in Sec.~\ref{laes:stacks}.

\begin{figure}
\center
\includegraphics[width=\linewidth]{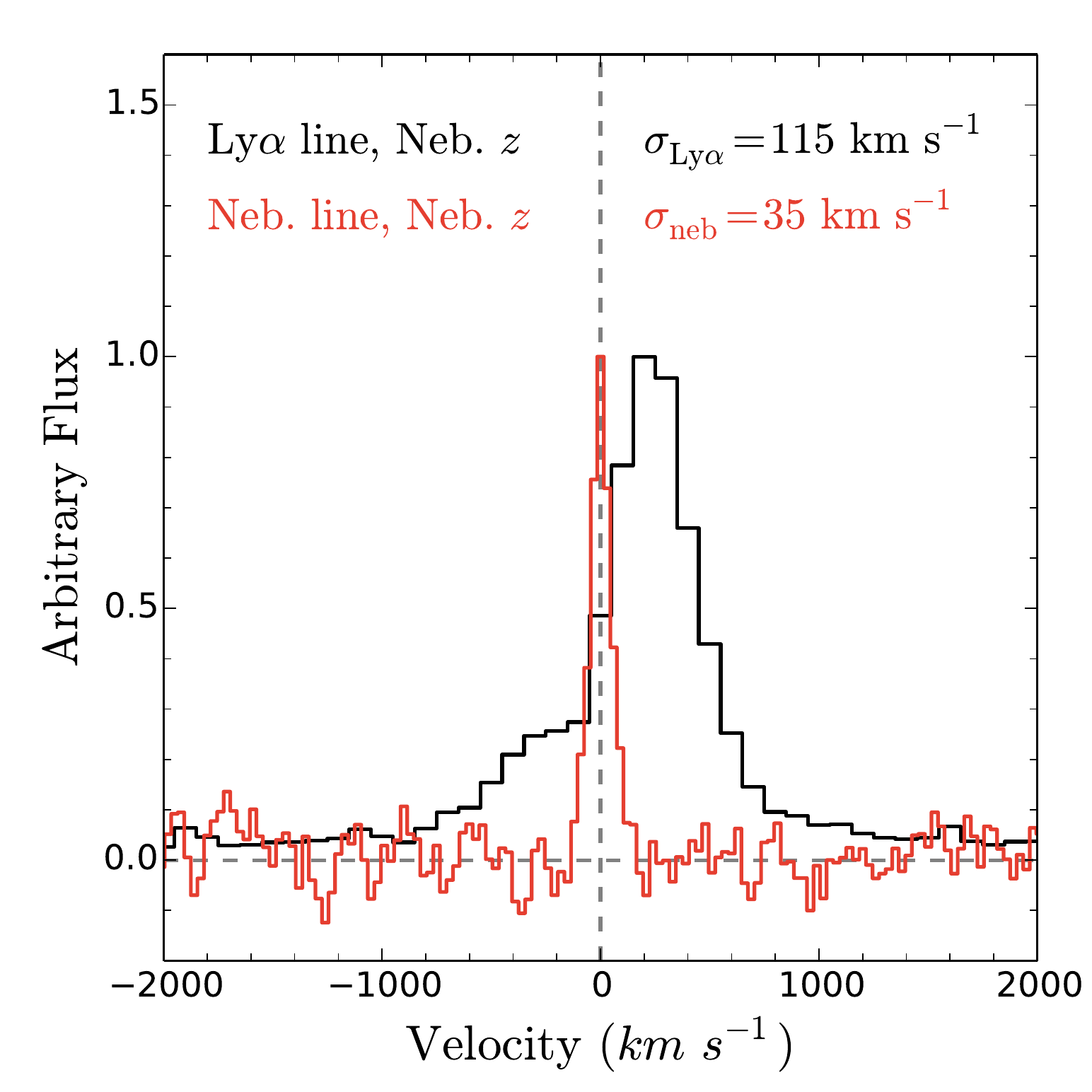}
\includegraphics[width=\linewidth]{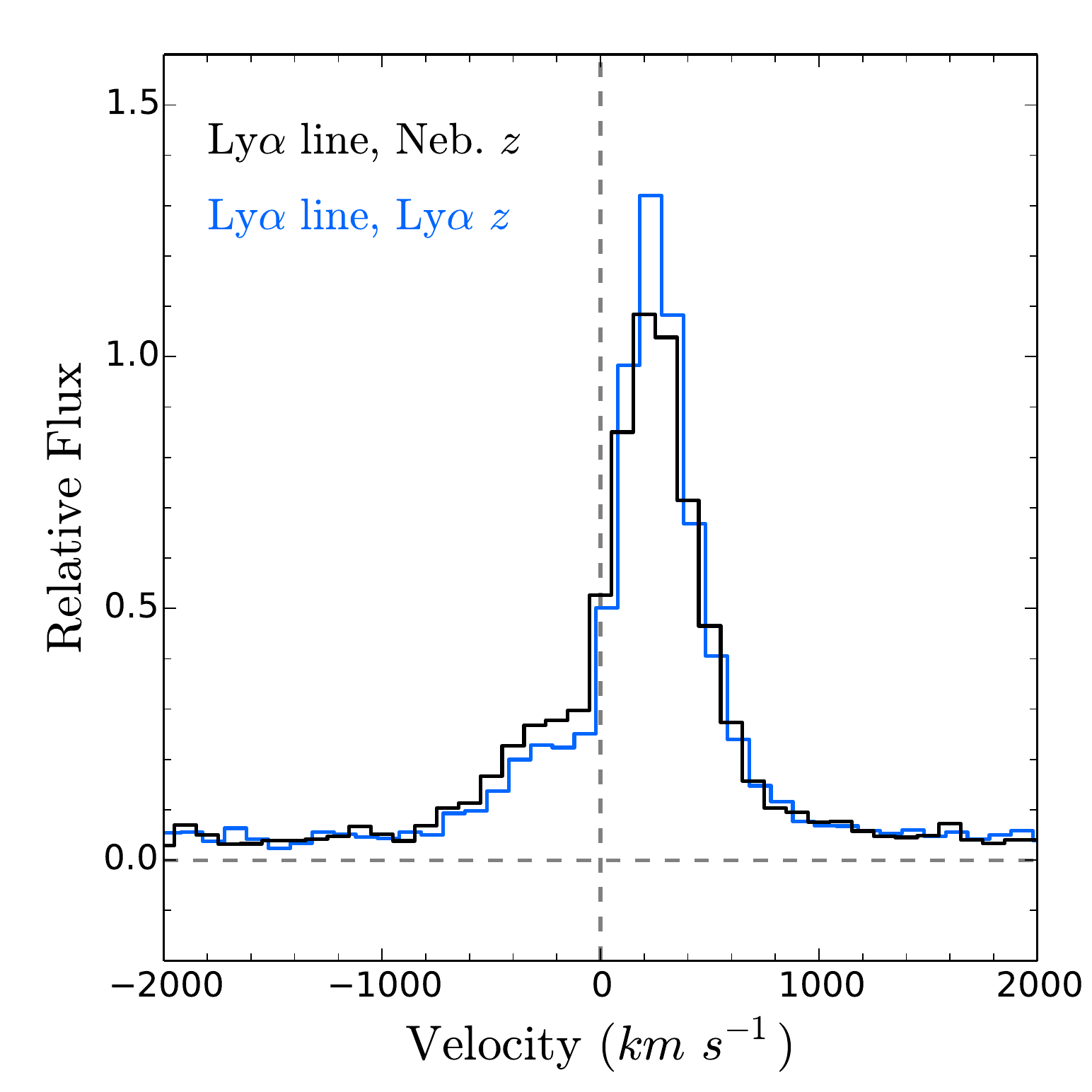}
\caption[Stacked profiles of the \lya\ emission line for the MOSFIRE
sample]{Stacked spectral profiles for the 32 LAEs with systemic 
  (nebular) redshift measurements. {\it Top:} Comparision of the
  stacked nebular (red) and \lya\ (black) line profiles for these
  LAEs. The nebular stack consists of all measured nebular lines (\ha,
  \hb, and/or \oiii\ $\lambda \lambda$4959,5007) for each LAE. The lines are stacked according
  to their corresponding nebular 
  redshifts, and the effect of resonant scattering on escaping \lya\
  photons is clearly visible, as is the typical \lya\ velocity offset
  of $\sim +200$ km s$^{-1}$ relative to systemic. {\it Bottom:}
  Comparison of the average \lya\ profiles when spectra are stacked
  according to their nebular, systemic redshift (black, as above) and
  a redshift derived directly from the \lya\ line peak (blue, shifted $+30$
  km s$^{-1}$ for clarity), wherein the ``systemic'' redshift is estimated as
  $v\sub{sys,\lya}=v\sub{\lya}-200$ km s$^{-1}$. Stacking via
  the \lya\ redshift distorts the \lya\ profile and diminishes
  the measured flux blueward of the systemic redshift.}
\label{fig:lya_stack}
\end{figure}

In addition to the distribution of average or peak \lya\ line
velocities, it is interesting to compare the full velocity distributions of
\lya\ and nebular emission. Fig.~\ref{fig:lya_stack} displays the stacked
\lya\ and nebular line profiles of the 32 MOSFIRE LAEs. Of particular note is
the breadth of the \lya\ line with respect to the nebular line
profile; assuming that the \lya\ photons are generated by
recombination processes in the same \ion{H}{2} regions that generate
the nebular emission, the top panel of Fig.~\ref{fig:lya_stack}
clearly displays the diffusion of the \lya\ photons in velocity as
they resonantly scatter through the surrounding \ion{H}{1}
gas. Fitting a Gaussian function to the stack of nebular lines yields
a velocity width of $\sigma\sub{neb}=35\pm3$ km s$^{-1}$. The stacked \lya\
profile has a width of $\sigma\sub{\lya}=115\pm8$ km s$^{-1}$ for the
primary (red) peak. These figures reflect the deconvolution of the
instrumental resolution of MOSFIRE ($\sigma\sub{inst} = 35$ km s$^{-1}$) and LRIS
($\sigma\sub{inst} = 100$ km s$^{-1}$) in their observed modes.\footnote{Note that the fits to the
composite line profiles yield similar (though not identical) values to
the averages of the individually-fit lines
(Table~\ref{table:laelbgem}).} The 
peak of the stacked \lya\ spectrum is again quite close to $+200$ km
s$^{-1}$, but there is a distinct component blueward of the systemic
redshift as well. After subtracting the faint continuum, the
fraction of \lya\ line flux emitted with $v<0$ is 16\%. For
comparison, we show the effective profile that would result from
stacking all the \lya\ lines at the redshifts derived from the \lya\
line peak and then making a $+200$ km s$^{-1}$ shift in the bottom panel of
Fig.~\ref{fig:lya_stack}. The resulting profile is narrower
($\sigma\sub{\lya}=85\pm8$ km s$^{-1}$),
with only 11\% of the line flux emitted at $v<0$, roughly 2/3 the flux
measured using the true systemic redshifts.

The four MOSFIRE LAEs with $v\sub{\lya}<0$ compose an extremely
interesting, though small, group of objects. Further analysis of this
sample is ongoing, including spectroscopy and photometry to constrain
the stellar contribution to the LAE luminosity, enrichment, and
dynamics. While only three of these blueshifted LAEs currently have $\mathcal{R}$
photometry, their average flux in that band is $\sim$1/2$\times$ that of
the LAEs with redshifted \lya\ lines, tentatively suggesting that
blueshifted \lya\ emitters have physical properties that are distinct
from typical LAEs.  Other properties associated with redshifted or
blueshifted \lya\ emission are discussed in the context of
multi-peaked \lya\ emission below. 

\subsection{Multi-peaked and asymmetric \lya\ profiles}\label{sublaes:multipeak}

\begin{figure*}
\center
\includegraphics[width=\linewidth]{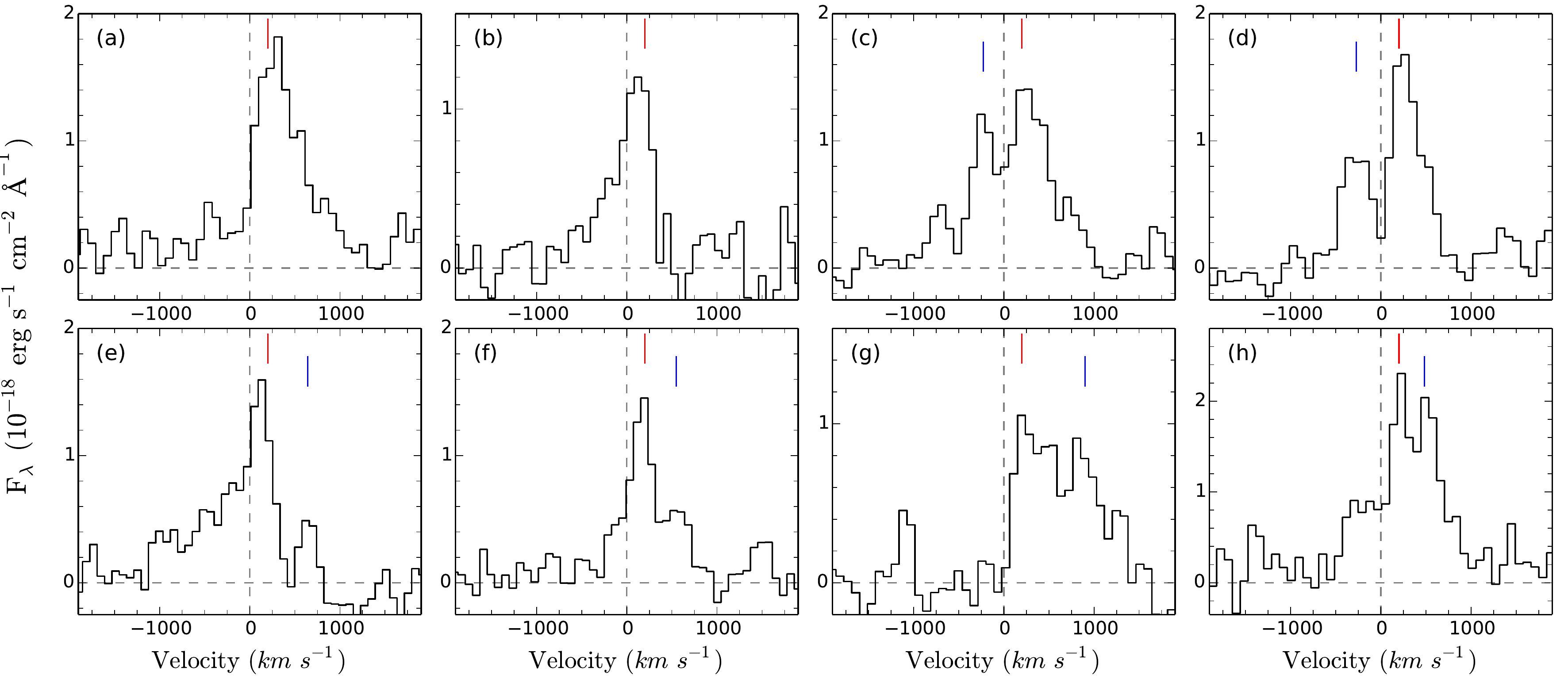}
\caption{A sample of the diverse \lya\ spectral morphologies represented
within our spectroscopic sample of LAEs. Red vertical lines denote the
redshift fit to the primary \lya\ peak by our algorithm, while blue vertical
lines denote the fit position of the secondary peak
(when detected with significance). Redshifts are assigned to be $-$200
\kms\ with respect to the primary \lya\ peak. The top row (panels $a$$-$$d$) includes
examples of common morphological types: single-peaked lines with red
$(a)$ or blue $(b)$ asymmetric tails, or double-peaked lines with dominant red
peaks $(c$, $d)$. The bottom row (panels $e$$-$$f$) shows examples of the
less common blue-dominant double-peaked lines. Despite their similar
peak ratios, panels $(e$ \& $f)$ have extended blue tails, whereas panels
$(g$ \& $h)$ have extended red tails.}
\label{fig:lya-diverse-thumbs}
\end{figure*}


As discussed above, star-forming galaxies are often associated with
multi-peaked \lya\ emission. In a systematic study of \lya\ emission
among $\sim$1500 star-forming galaxies at $z\sim2-3$, 
\citet{kul12} found that $\sim$30\% are multi-peaked, with some
dependence on the spectral resolution and S/N of the
observations. While most of these galaxies were observed at lower
resolution, a subsample of their spectra were obtained with the same
600-line grism of Keck/LRIS used for our LAE sample and have similar integration
times ($\sim$1.5 hours). This subsample included 44 multi-peaked
spectra, a multiplicity rate of 27\%.

\citet{yam12b} studied the \lya\ peak morphologies of 91 LAEs at $z\sim
3$ at similar resolution ($R\sim 1700$), finding a $\sim$50\%
multiplicity rate. This heterogeneous sample contains 12 \lya-blobs
from \citet{mat04}, as well as compact, faint LAEs similar to those of
the KBSS-\lya\ sample (although with a limiting \lya\ flux
$\sim$2$\times$ brighter, according to \citealt{yam12a}).

For the KBSS-\lya\ sample of this paper, multi-peaked systems were identified as
described in Sec.~\ref{sublaes:specobs}. Of the 318 unique
spectroscopically-identified LAEs, 129 are found to be multi-peaked,
for a multiplicity fraction of 40\%. The criteria for identifying a
secondary peak were not identical to that of \citet{kul12}; in
particular, the significance threshold was not dependent on the
magnitude or S/N of the primary peak, and no minimum peak separation
was enforced. However, we found that adjusting the multiplicity criteria
to match those of \citeauthor{kul12} as closely as possible did not
significantly affect the overall rate of selection nor the
spectral/physical properties of the selected sample.

\begin{figure}
\center
\includegraphics[width=\linewidth]{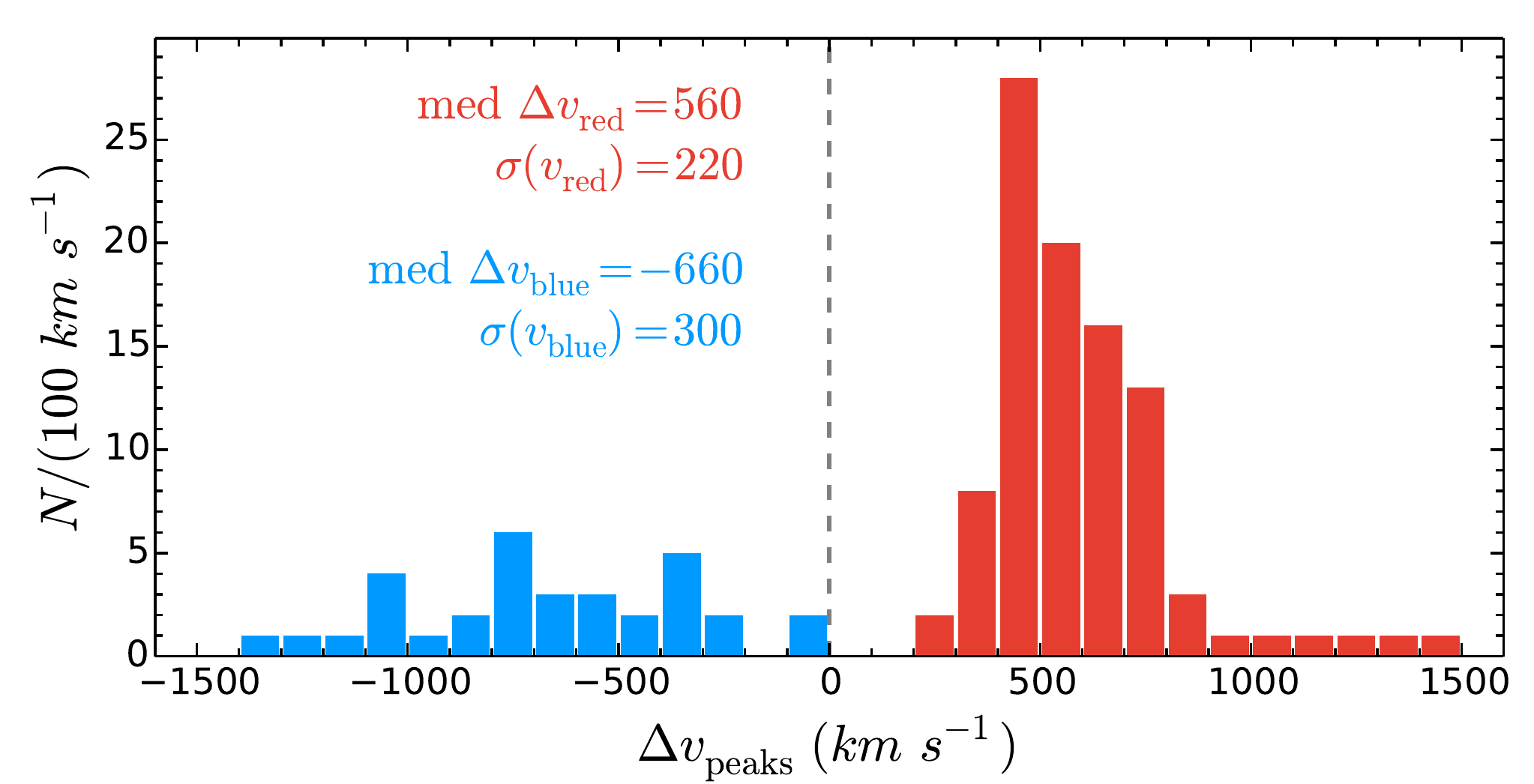}
\includegraphics[width=\linewidth]{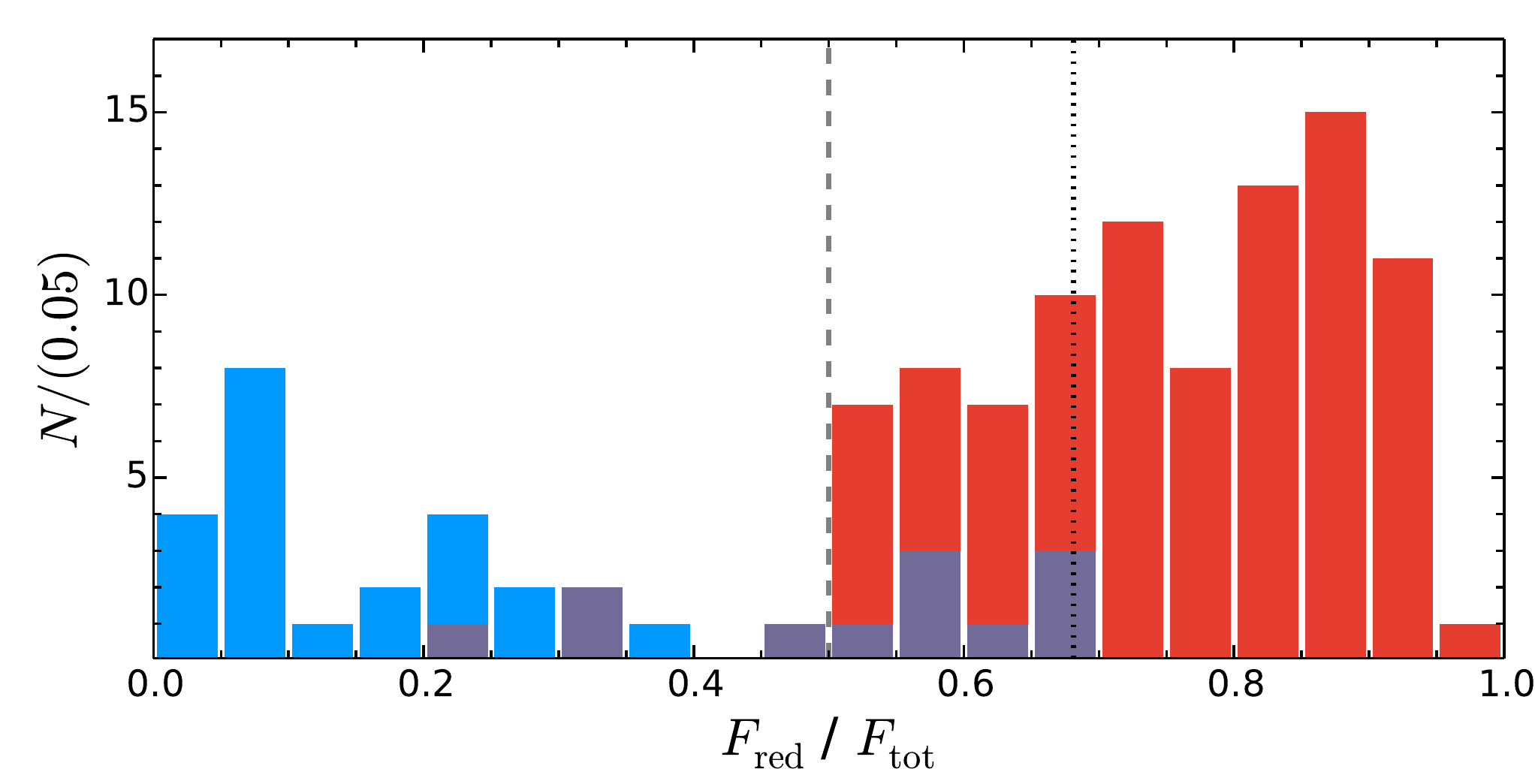}
\caption{{\it Top:} Distribution of peak separations among the 129
  LAEs with identified multi-peaked \lya\ emission lines (40\% of the
  318 LAEs with \lya\ 
  spectra). The majority (96/129) have dominant red peaks ($\Delta
  v>0$, red), but a substantial minority (33/129) have dominant blue
  peaks ($\Delta v<0$, blue). The dashed grey line denotes $\Delta v =
  0$. The red-peak-dominant systems have a smaller median peak
  separation (560 km s$^{-1}$ vs. 660 km s$^{-1}$) than the
  blue-peak-dominant systems, as well as a smaller scatter about this
  value (220 km s$^{-1}$ vs. 300 km s$^{-1}$). 
  {\it Bottom:} The distribution (stacked histogram) of flux in the red
  peak ($F\sub{red}$) as a fraction of 
  total \lya\ flux ($F\sub{tot}$) for the same 129 multi-peaked
  LAEs, with colors matched to the top panel. Red-peak-dominant \lya\
  lines (red, $\Delta v >0$ in the top panel) generally have
  $F\sub{red}/F\sub{tot}>0.5$, and the median flux ratio among all
  systems is $F\sub{red}/F\sub{tot}=0.68$ (dotted black line).}
\label{fig:multipeak}
\end{figure}

Previous studies of \lya\ kinematic multiplicity (among samples with or without
systemic redshifts) typically group these lines into blue-peak
dominant or red-peak dominant lines, motivated by the association of
blue (red) peak dominance with inflowing (outflowing) gas in radiation
transfer models (e.g., \citealt{zhe02,ver06,dij06}). \citet{kul12}
found that 67\% of their 239 multi-peaked objects (the majority of
which had lower spectral resolution than the KBSS-\lya\ sample) have
dominant red peaks. A selection of examples of the diverse \lya\
spectral morphologies represented among the KBSS-\lya\ LAEs is displayed in
Fig.~\ref{fig:lya-diverse-thumbs}, and the distribution of red/blue
peak dominance in our sample is displayed in Fig.~\ref{fig:multipeak}. In the
top panel of Fig.~\ref{fig:multipeak}, we show the distribution of inter-peak
velocities, defined as $\Delta
v\sub{peaks}=c(z\sub{peak,pri}-z\sub{peak,sec})/(1+(z\sub{peak,pri})$.
$z\sub{peak,pri}$ is the redshift of the first peak identified by the
line detection algorithm discussed in Sec.~\ref{sublaes:specobs}
(i.e., the peak of the asymmetric Gaussian profile fit to the highest
peak in the smoothed spectrum), and $z\sub{peak,sec}$ is the redshift
corresponding to the mean of the second fit Gaussian component. For
this definition, $\Delta v\sub{peaks}>0$ corresponds to red-dominant
systems, such that the primary peak is redward of the inter-peak
trough.

Of the 129 multi-peak LAEs, 96 (74\%) have dominant red peaks ($\Delta
v\sub{peaks}>0$), while 33 (26\%) are blue dominant ($\Delta
v\sub{peaks}<0$; Table~\ref{table:laelbgem}). The red-dominant LAEs
have a quite narrow distribution of peak separations, with a median value
$\langle \Delta v\sub{peaks}\rangle=560$ km s$^{-1}$ and a standard
deviation $\sigma(\Delta v\sub{peaks})=220$ km s$^{-1}$. Conversely,
the blue-dominant LAEs have a larger velocity shift over a broader
range: $\langle \Delta v\sub{peaks}\rangle=-660$ km s$^{-1}$ and 
$\sigma(\Delta v\sub{peaks})=300$ km s$^{-1}$. A similar contrast can
be seen in the flux distribution within these multi-peaked LAEs
(Fig.~\ref{fig:multipeak}, bottom panel). The fraction of total \lya\
flux in the red peak, $F\sub{red}/F\sub{tot}$,  is defined as the
integral of the redder of the two Gaussian components divided by
the integral of the sum of the two components. There are 101 LAEs (78\%) with
$F\sub{red}/F\sub{tot}>0.5$, an alternative definition of red-peak
dominance. The sets of LAEs with $\Delta v\sub{peaks}>0$ and
$F\sub{red}/F\sub{tot}>0.5$ overlap almost entirely,\footnote{The
  small number of objects with 
  $F\sub{red}/F\sub{tot}>0.5$ and $\Delta v\sub{peaks}<0$ (or {\it vice
    versa}) generally have a secondary peak that was fit with a broad
  Gaussian profile, thereby including more total flux despite having a lower
  S/N.} but we use the prior definition (based on the significance of
peaks in the smoothed spectrum) for consistency with \citet{kul12} and
other surveys that assume different model profiles. 


Due to the presence of nearby hyperluminous QSOs in these LAE fields,
it is important to discern what role the external illumination of the
QSO (i.e., ``fluorescence'') may play in shaping the observed spectral
morphology of the LAEs. Clearly, the fluorescing LAEs must inhabit the 
region of the environment illuminated by the QSO in order for the
ionizing field to have an observable effect. As argued in
\citet{tra13}, the combination of geometry and the finite QSO lifetime
(measured to be $\lesssim$20 Myr) cause this fluorescence to be observable
only when the LAE is in the foreground of the QSO, or less than
$\sim$$10^7$ lightyears (3.2 proper Mpc) distant in the background.

\begin{figure}
\center
\includegraphics[width=\linewidth]{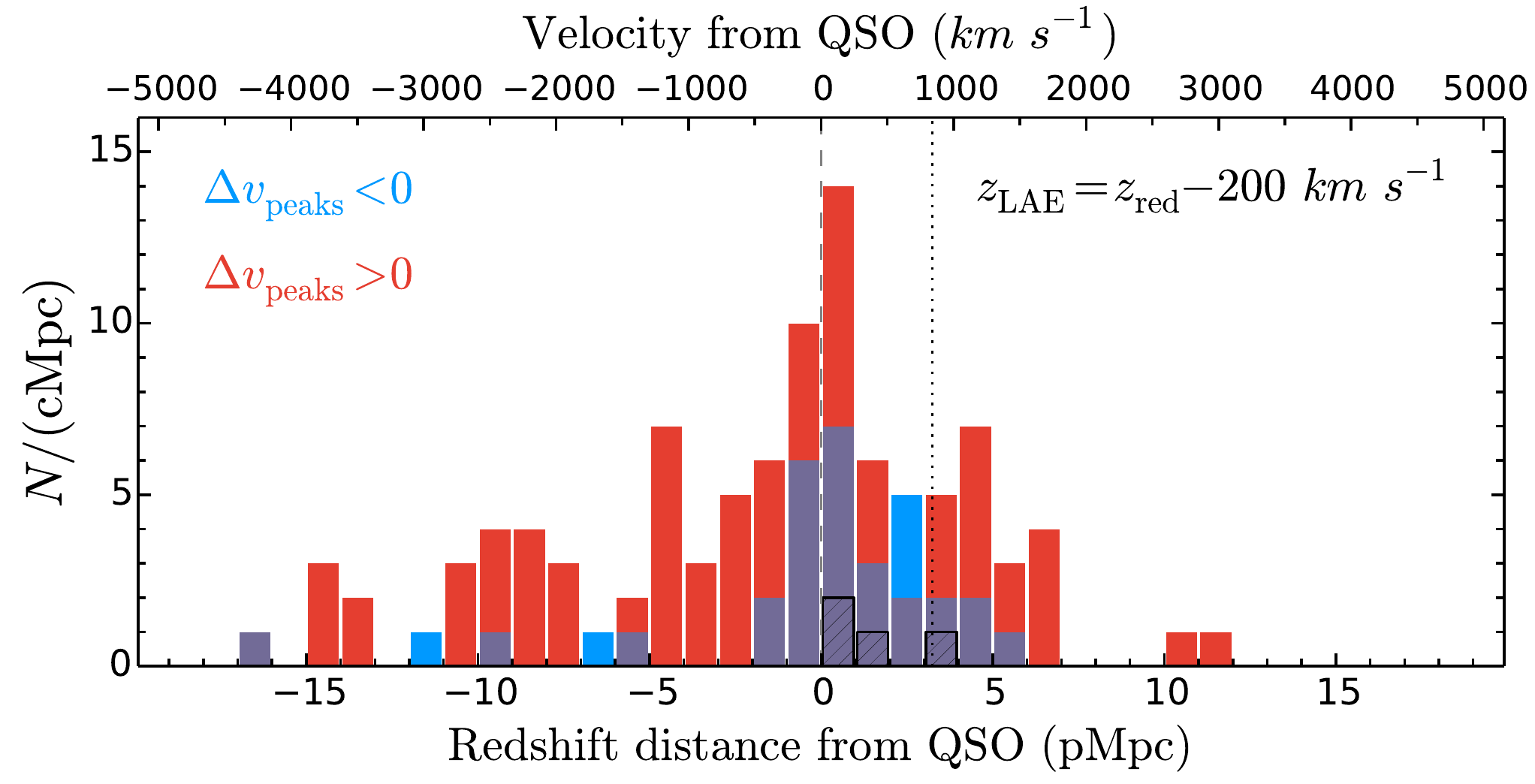}
\caption{The redshift distribution of multi-peaked \lya\ lines with
  respect to their corresponding HLQSOs. Color-coding of the stacked
  histogram matches Fig.~\ref{fig:multipeak} (top panel) and
  Fig.~\ref{fig:asym}. The dashed grey line corresponds to LAEs at the
  HLQSO redshift. The dotted black line corresponds a distance 10
  Mlyr behind the HLQSO, the upper bound on the region that could
  emit detectable QSO-generated fluorescent emission for a 20 Myr QSO
  lifetime (as argued in \citealt{tra13}). Here, the LAE systemic redshifts
  are assumed to be $-$200 km s$^{-1}$ from the {\it red} \lya\
  peak. Under this assumption, the blue-peaked LAEs are highly biased
  toward the rear of the nearby QSOs. Black hatched bars denote the
  {\it systemic} redshifts measured for the 4 LAEs with 
  blueshifted \lya\ peaks ($v\sub{\lya,peak}<0$ km s$^{-1}$), which
  appear to occupy a similar physical volume as the blue-dominant
  multipeaked LAEs.} 
\label{fig:multipeak-z}
\end{figure}

 The redshift distribution relative to the QSOs of all the multi-peaked LAEs is shown in
Fig.~\ref{fig:multipeak-z}.
Care must be taken in
comparing the redshift distributions of red- and blue-dominant LAEs
based on the redshifts of their \lya\
lines. Fig.~\ref{fig:multipeak-z} shows the redshift distribution when
the systemic redshift of each LAE is taken to be 200 
km s$^{-1}$ blueward of the red \lya\ peak, {\it regardless of whether
  the redder peak is primary or secondary}. The redshift distribution of
red-dominant \lya\ profiles is fairly symmetric, with a slight bias toward the QSO
foreground. The blue-dominant LAEs, however, show a strong association
with the region in the immediate background of the QSO. A two-sample
KS test finds only a small ($p\approx0.03$) likelihood of both samples
being drawn from the same distribution.

Given the uncertainty in assigning redshifts to those sources without
accurate systemic redshifts from optically-thin nebular emission
lines, it is unclear whether the 
blue-dominant LAEs are localized within the QSO field, or
whether they differ from the red-dominant distribution at
all. As noted above, blueshifted \lya\ emission is a possible 
observational signature of infalling gas illuminated by a central
source. Given the proximity of the HLQSOs, however, it is also
possible that the detected emission has a significant fluorescent
contribution. In this case, the blueshifted \lya\ emission could
conceivably correspond to outflowing gas illuminated externally by the
nearby QSO, a situation that could explain the association of
blue-dominant LAEs with the background of the QSO, where the geometry
would naturally predict blueshifted emission from the QSO-facing side
of a gaseous outflow. Only three of the blue-dominant LAEs currently
have MOSFIRE spectra, but their systemic redshifts place them 0.4$-$1.2
pMpc in the background of the QSO (consistent with the \lya-only
measurements). Similarly, the 4 LAEs with MOSFIRE redshifts and
$v\sub{\lya,peak}<0$ km s$^{-1}$ (that is, objects with blueshifted \lya\
with respect to systemic) lie in the immediate background of the QSO
(0.4$-$4 pMpc; hatched region in Fig.~\ref{fig:multipeak-z}), suggesting that
external illumination may play a role in their spectral morphology as well.

Detailed predictions for fluorescent emission from gas with large bulk
velocities and a realistic spatial distribution are unclear (although
see work by \citealt{kol10}), and it is
difficult to draw robust conclusions without access to the systemic
redshifts of these LAEs. Given the diversity of line morphologies
and asymmetries even within the set of blue-dominant LAEs
(Fig.~\ref{fig:lya-diverse-thumbs}), it is unclear whether these
atypical profiles are generated by a single mechanism. Further study
of these ``blue'' LAEs, including additional rest-frame optical
spectroscopy, is required to understand the physics of their \lya\
emission and transmission.   

\begin{figure}
\center
\includegraphics[width=\linewidth]{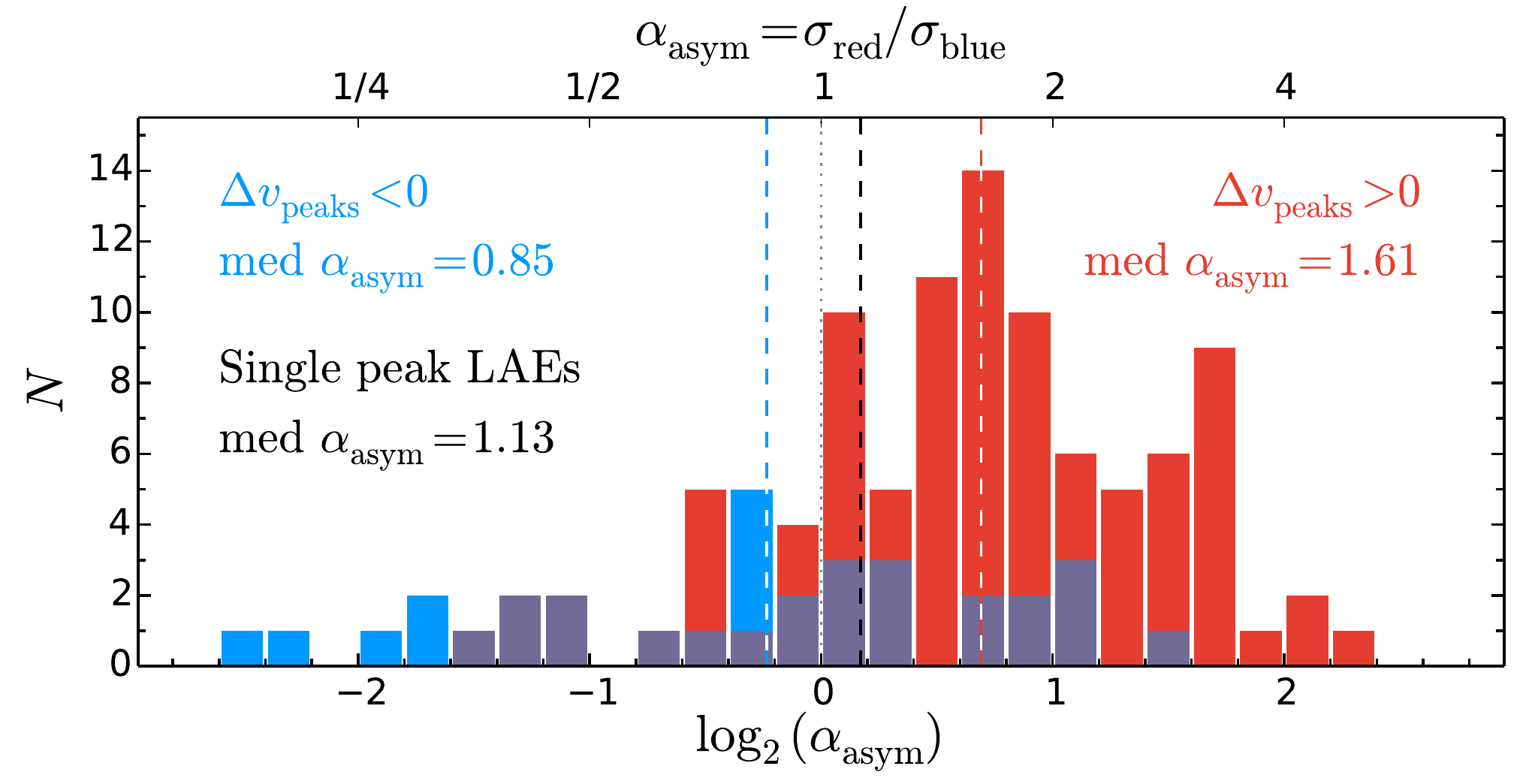}
\caption{\lya\ line asymmetries for LAEs with multi-peaked
  \lya\ detections. The reported asymmetry corresponds to the
  primary peak of the multi-peaked line. Red bars give the asymmetry
  distribution of red-peak-dominant systems ($\Delta v\sub{peaks}>0$),
  while blue bars correspond to blue-peak-dominant systems ($\Delta
  v\sub{peaks}<0$) according to the top panel of
  Fig.~\ref{fig:multipeak}. The blue-white dashed line is the median
  asymmetry of the $\Delta v\sub{peaks}<0$ sample and the red-white
  dashed line is that of the $\Delta v\sub{peaks}>0$ sample. For
  comparison, the black dashed line gives the median asymmetry of the
  sample of spectral \lya\ lines without multiple detected peaks.}
\label{fig:asym}
\end{figure}

Lastly, in the absence of systemic redshift measurements and/or
multiple emission peaks, the \lya\ line asymmetry ($\alpha\sub{asym}$;
Sec.~\ref{sublaes:specobs}) may also trace gas 
kinematics among these LAEs (see, e.g., \citealt{zhe13}). However, 
line asymmetry is often not evident 
in medium-resolution, low-S/N spectra. Not only will the 
measured line asymmetry be systematically diminished by the smoothing
effect of the convolved line spread function, but noise peaks and
unresolved emission line features may be fit by an extended tail by
our automatic fitting algorithm.

Despite these issues, some trends can be seen in the relative
distributions of \lya\ line asymmetries. Fig.~\ref{fig:asym} shows
the distribution of $\alpha\sub{asym}$ for blue-dominant and
red-dominant multi-peaked LAEs. While both subsamples have broad
distributions of $\alpha\sub{asym}$ because of the effects described
above, there is a clear tendency of the blue-dominant \lya\ lines
to have lower values of $\alpha\sub{asym}$ (that is, an extended blue
wing in the primary peak), while red-dominant LAEs typically have
$\alpha\sub{asym}>1$ 
(extended red wings). The two distributions have a Kolmogorov-Smirnov
probability of being drawn from the same underlying distribution
$p<4\times10^{-5}$. This result agrees with the findings of
\citet{yam12b}, who found a similar relation (albeit with lower
significance) in their smaller sample. 

\subsection{LBG \lya\ line morphologies}\label{sublaes:lbgemission}

\begin{figure}
\center
\includegraphics[width=\linewidth]{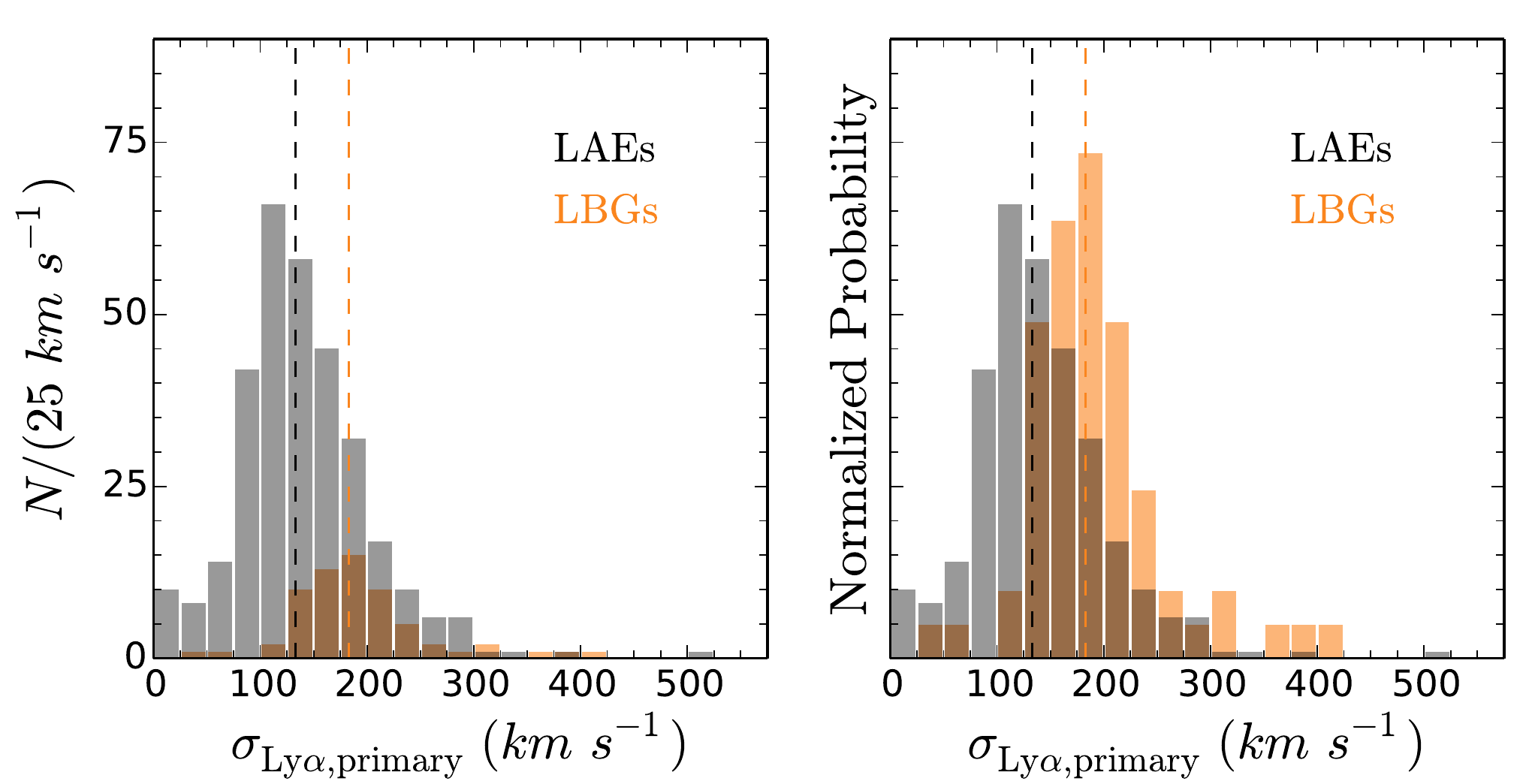}
\caption{Distribution of fit widths (Eq.~\ref{eq:linewidth}) for the
  primary \lya\ peak among the 318 
  KBSS-\lya\ LAEs (grey) and the 65 LBGs in the
  KBSS LBG (orange) sample lines. The LRIS instrumental resolution of
  $\sigma\sub{inst} = 100$ \kms\ has been subtracted in quadrature; fit widths less than
  100 \kms\ are plotted as zero. The left
  panel shows the number distribution of object line widths, while the right
  panel presents the same data normalized as a probability
  distribution for ease of comparison. The median width for each
  sample is marked with a dashed line.}
\label{fig:widthlbg}
\end{figure}

\begin{figure}
\center
\includegraphics[width=\linewidth]{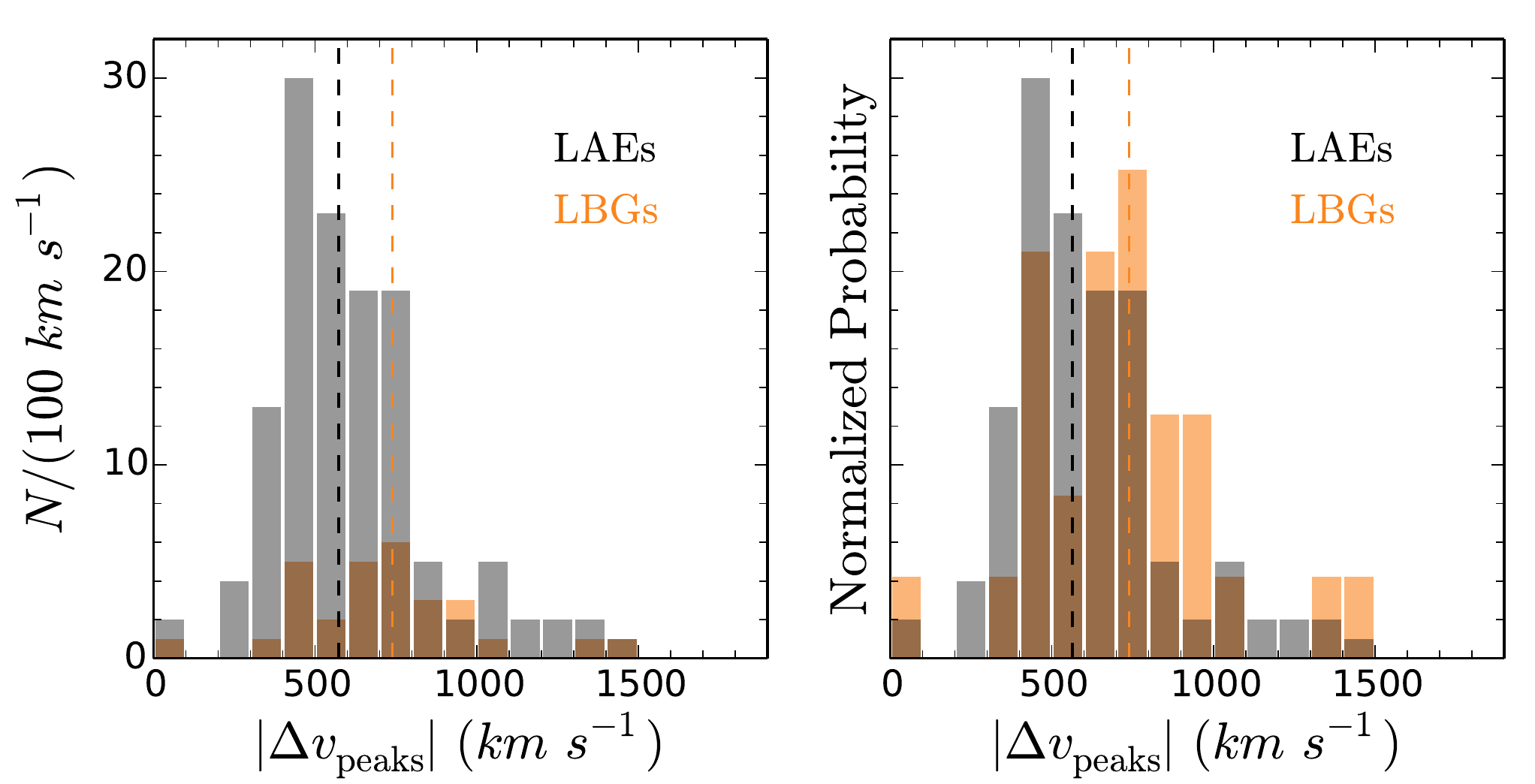}
\caption{Distribution of \lya\ peak separations among the 129
  LAEs in Fig.~\ref{fig:multipeak} (grey) and the 29 LBGs in the
  KBSS LBG (orange) sample with multi-peaked \lya\ emission
  lines. Given the smaller size of the LBG sample, positive and 
  negative separations of the same size are grouped together (unlike Fig.~\ref{fig:multipeak}). The left
  panel shows the number distribution of of objects, while the right
  panel presents the same data normalized as a probability
  distribution for ease of comparison. The median separation for each
  sample is marked with a dashed line.}
\label{fig:multipeaklbg}
\end{figure}

\begin{deluxetable}{lccc}
\tablecaption{LAE and LBG Emission-Line Properties}
\tablewidth{0pt}
\tablehead{
& LAEs & MOSFIRE & KBSS\\
& & LAEs\tablenotemark{a} & LBGs
}

\startdata
$N\sub{obj}$ & 318 &  32 &  65 \\ 
$\langle F\sub{\lya}\ \rangle$\tablenotemark{b} ($10^{-17}$ cgs) &  4.5$\pm$0.8 &  4.6$\pm$0.8 &  7.3$\pm$1.6 \\ 
$\langle \sigma\sub{\lya} \rangle$ (km s$^{-1}$) & 144$\pm$3 & 138$\pm$8 & 195$\pm$7 \\ 
$N\sub{blue}$\tablenotemark{c} &   $-$ &   4 &   4 \\ 
$f\sub{blue}$\tablenotemark{c} &  $-$ & 13\% &  6\% \\ 
$N\sub{mult}$\tablenotemark{d} & 129 &  13 &  29 \\ 
$f\sub{mult}$\tablenotemark{d} & 41\% & 41\% & 45\% \\ 
$N\sub{mult,blue}$\tablenotemark{e} &  33 &   3 &   6 \\ 
$f\sub{mult,blue}$\tablenotemark{e} & 10\% &  9\% &  9\% \\ 
$\langle |\Delta v\sub{peaks}| \rangle$ (km s$^{-1}$) & 615$\pm$14 & 608$\pm$49 & 716$\pm$35 \\ 
$\langle F\sub{\ha} \rangle$ ($10^{-17}$ cgs) &  $-$ &  1.3$\pm$0.1 &  7.2$\pm$0.7 \\ 
$\langle \sigma\sub{neb} \rangle$ (km s$^{-1}$) &   $-$ &  34$\pm$4 &  84$\pm$4
\enddata
\tablenotetext{a}{Parameters for the subset of LAEs with MOSFIRE nebular (systemic) redshifts} 
\tablenotetext{b}{Averages are given as mean values, with uncertainties
  denoting the standard error of the mean.}
\tablenotetext{c}{Number (or fraction) of objects with a blueshifted
  \lya\ line ($v\sub{\lya,peak}<0$)}
\tablenotetext{d}{Number (or fraction) of objects with multiple detected \lya\ peaks}
\tablenotetext{e}{Number (or fraction) with multiple detected \lya\
  peaks that have a lower-velocity (``blue'') dominant peak}
\label{table:laelbgem}
\end{deluxetable}

We compare the distribution of \lya\ morphologies between the
KBSS-\lya\ LAEs and the KBSS LBGs in
Figs.~\ref{fig:widthlbg}~\&~\ref{fig:multipeaklbg}, as well as in
Table~\ref{table:laelbgem}. In the LBG sample, 
29/65 (45\%) have a detected secondary peak. Of these objects, 23/29
(79\%)  are red-peak dominant. Note that these rates are 40\% and
74\%, respectively, among the LAEs (Table~\ref{table:lbgsamples}). As
described above, measured rates of \lya\ 
peak multiplicity are dependent on both the resolution and S/N of the
spectra. The KBSS LBGs considered in this paper were observed
using the same instrument setup as the KBSS-\lya\ LAEs and were
analyzed using the same line-detection software (see
Sec.~\ref{sublaes:specobs}); however, the typical LBG \lya\
fluxes are significantly brighter (Table~\ref{table:lbgsamples}) and the
spectral S/N is correspondingly higher. Secondary
peaks of low relative significance are therefore more likely to be detected in 
the LBG sample than in our LAE sample. In particular, we find that
the ratio of primary-peak flux to secondary-peak flux is larger among
the LBGs than the LAEs, which may be caused by missing similarly faint
secondary \lya\ peaks in the LAE spectra. However, radiative transfer
models of \lya\ escape also predict that the primary/secondary flux
ratio increases with outflow velocity (e.g., \citealt{ver06}). In
either case, the frequency of \lya\ peak multiplicity and blue-dominant \lya\ peaks
are broadly consistent between our LBG and LAE samples.

However, the velocity distribution of \lya\ flux varies significantly
between the LAE and LBG samples. Fig.~\ref{fig:widthlbg} displays the
distribution of fit \lya\ emission line widths for both samples of
objects. The measured line width is that of the primary \lya\ peak,
and is given by the average of the red and blue scale parameters of the
asymmetric Gaussian fit (Eq.~\ref{eq:linefit}):

\begin{equation}
\sigma\sub{\lya,primary}=\frac{\sigma\sub{red}+\sigma\sub{blue}}{2} \,\,.
\label{eq:linewidth}
\end{equation}

The LAE sample has a median line width of 166 km s$^{-1}$, with 33\%
of spectra having a primary-peak line width less than 150 km
s$^{-1}$. The LBG sample has a median line width of 210 km s$^{-1}$,
with only 3\% (two spectra) having a primary-peak line width less than
150 km s$^{-1}$. The LRIS spectral resolution $\sigma\sub{inst} = 100$ km
s$^{-1}$ (which is the same for both the LAE and LBG samples) has been
subtracted in quadrature. Note that nearly all the measured lines are
well-resolved with respect to the resolution limit. A Kolmogorov-Smirnov test gives a
probability $p\approx10^{-10}$ of both samples being drawn from the same
distribution. 

A similar contrast is seen in the distribution of peak separations
among the LAEs and LBGs with multiple detected emission peaks
(Fig.~\ref{fig:multipeaklbg}). Among the 129 LAEs with multiple
detected peaks, the median peak separation is $\langle|\Delta
v\sub{peaks}|\rangle=565$ km s$^{-1}$, while the 29 LBGs with multiple
peaks have $\langle|\Delta v\sub{peaks}|\rangle=780$ km s$^{-1}$. Note
that unlike Fig.~\ref{fig:multipeak}, only the absolute value of the
separation is considered here. A Kolmogorov-Smirnov test gives a
probability $p\approx10^{-3}$ of both samples being drawn from the
same distribution. As
above, the finite resolution of our spectra censors the distribution
of peaks with small separations, but the typical separations observed
exceed our resolution limit by a large factor, suggesting that we are
able to resolve the intrinsic distribution of peak separations for
each sample. Furthermore, our use of the same instrument
setup and line-analysis algorithms for both the LAE and LBG samples
allows us to demonstrate an 
instrinsic difference between the two populations.

In contrast to the results of \citet{yam12b}, we find no significant
correlation between the \lya\ peak width and peak separation among
multi-peaked profiles for our LAE sample, and only a moderate
association among the LBGs (Pearson correlation $r\approx0.3$,
$p\approx0.05$). However, these parameters are highly correlated
(although with significant scatter) in
the combined LBG+LAE sample, for which a Pearson correlation test
yields $p\approx10^{-4}$ ($r\approx0.3$).

Here we have defined the peak separation by the velocity difference between the
peaks of the two most dominant components of the \lya\ emission line,
rather than the flux-weighted centroids of those components. Because
the typical \lya\ emission lines have redshifted primary peaks with
asymmetric extended red tails (particularly for the LBGs;
Fig.~\ref{fig:hist_dv}), using flux centroids or symmetric Gaussian
fits to define the peak separation would lead to 1) a larger median
peak separation for both samples, and 2) a larger difference between
the LAE and LBG distributions.

\begin{figure}
\center
\includegraphics[width=0.8\linewidth]{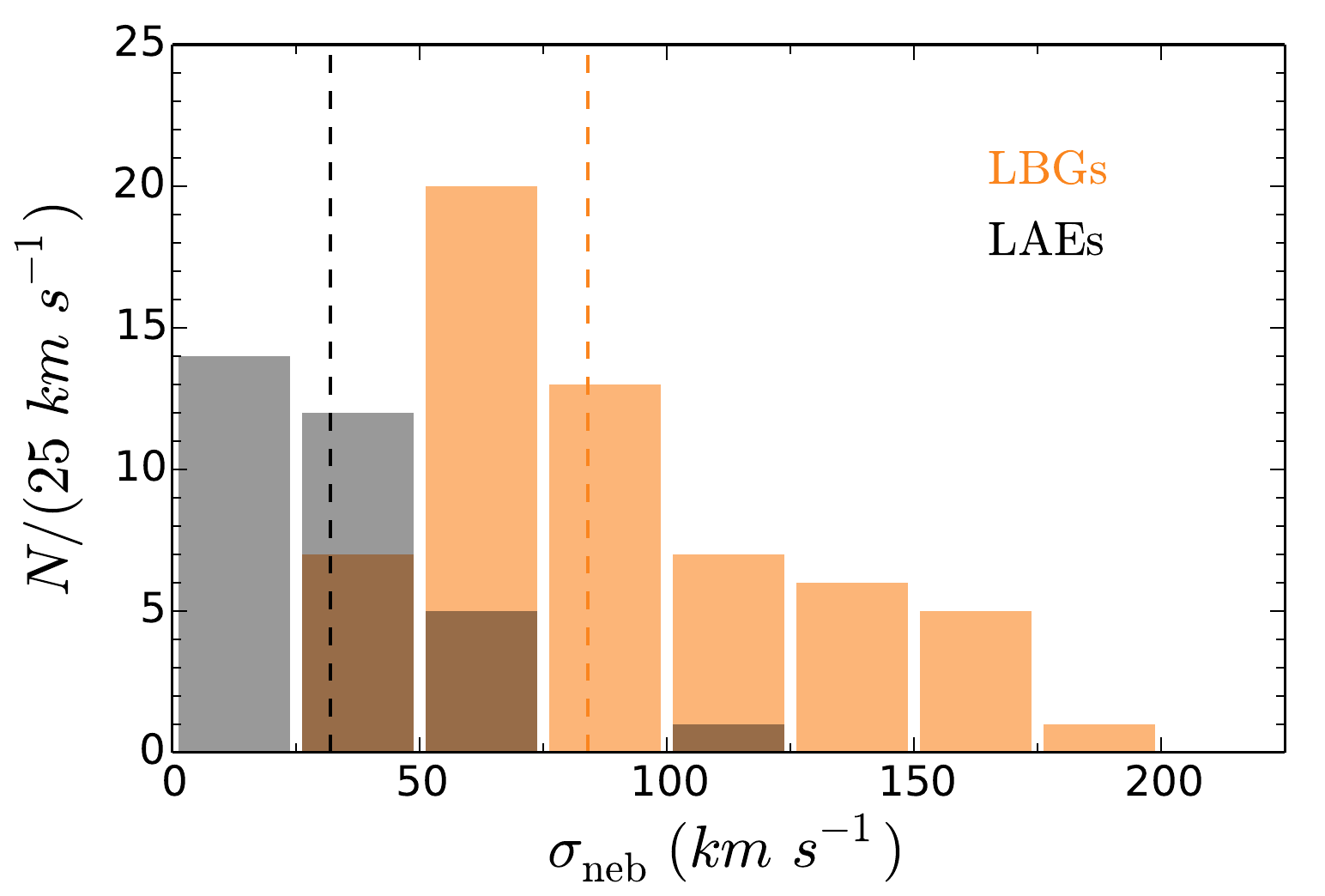}
\caption[Distribution of nebular line widths for LAEs and
LBGs]{Distribution of nebular line widths for LAEs (grey) and KBSS LBGs
  (orange) for the subset of objects with MOSFIRE rest-optical
  spectra. Dashed lines denote the average line width for each
  sample. The instrumental resolution of 35 km s$^{-1}$ has been
  removed from each value; objects consistent with this resolution
  have been assigned $\sigma\sub{neb}=0$. A typical LBG has a
  nebular line width $\sim$2$\times$ that of a typical LAE.}
\label{fig:signeb}
\end{figure}

For comparison, the distribution of nebular line widths for both the
LAEs and LBGs is given in Fig.~\ref{fig:signeb}. As discussed in
Sec.~\ref{sublaes:lyashift}, the nebular line widths trace the
velocity dispersion of the stars and \ion{H}{2} regions within the
galaxy. The larger nebular line widths of LBGs with respect to LAEs
are therefore likely to reflect the larger masses of continuum-bright
galaxies with respect to faint LAEs. Further discussion of the
mass-dependence of the interstellar and circumgalactic properties of
these galaxies is reserved for Sec.~\ref{laes:discussion}.

In contrast, the velocity distribution of emergent \lya\ flux
depends most sensitively on the velocity distribution and the optical
depth of the scattering \ion{H}{1} gas. The observed spectral
morphologies therefore suggest that LBGs are associated with higher
velocity and/or more optically-thick outflows than their LAE
analogs. It is thus evident that the observational characteristics
that govern a galaxy's selection by the narrowband-excess and/or
continuum-color techniques (namely, the observed \lya\ and continuum
emission) must be linked to properties of the outflowing gas -- whether
directly (by the coupling of \lya\ photons to interstellar and
circumgalactic gas) or indirectly (through the star-forming regions
that produce \lya, far-UV, and ionizing photons and presumably provide
the energy/momentum for large-scale galactic outflows). In the
remainder of this paper, we separate these effects of gas velocity
and optical depth on the emitted spectrum by considering the 
absorption and emission profiles in composite galaxy spectra.

\section{Signatures of outflows in stacked spectra}\label{laes:stacks}

\subsection{Absorption signatures of outflows}\label{sublaes:abs}


While the resonance of the \lya\ transition makes it a sensitive tracer
of gas in and around high-redshift galaxies, this strong coupling can
obscure the details of the physical processes driving \lya\ absorption
and emission. In particular, we noted in Sec.~\ref{laes:lya} that \lya\
transmission is sensitive to a variety of factors, including the
optical depth, covering fraction, dust content, and kinematics of the
gas distribution. Because they produce degenerate effects on the
\lya\ profile, \lya\ emission alone is insufficient to fully
characterize the physical conditions among LAEs or high-redshift
galaxies generally. 

The measurement of absorption lines in the rest-UV continuum spectra of
galaxies is thus a crucial tool for disentangling these
effects. \citet{kun98} analyzed the spectra of nearby \lya-emitting
galaxies, finding blueshifted metal absorption features from enriched,
outflowing gas. This study noted the correspondence of the \lya\
emission and metal absorption in tracing the velocity structure and
porosity of the interstellar gas. At higher redshifts, the analysis of
the continuum spectra of typical star-forming galaxies began with spectra of the 
gravitationally-lensed $z \sim 2.7$ galaxy MS 1512-cB58
\citep{pet00,pet02} and was extended to non-lensed galaxies through
stacking the spectra of many ($\sim$1000) $z\sim3$ LBGs by
\citet{sha03}. These studies utilized the high S/N of the lensed or
stacked spectra to extract metal abundances, ion-specific covering
fractions, outflow velocities, and systemic redshifts for the galaxies
while placing these properties in the context of their star-formation
rates, masses, and \lya\ emission properties. \citet{ste10} presented
both interstellar absorption line features and \lya\ emission in the
context of a unified kinematic model for gas outflows in LBGs (Sec.~5
of that paper). For faint LAEs such as
those in our sample, constraints on the metal content and outflow
velocities are especially interesting because they reveal the extent
to which past and ongoing star formation are already
having an effect on the chemistry and kinematics of these
particularly young, low-mass galaxies.

LAEs are by selection generally faint in the continuum, so absorption
measurements of comparable fidelity to those of \citet{sha03} present
a significant challege, particularly for objects as faint as those
considered here. \citet{has13} studied the absorption profiles in a
stack of four LAEs with systemic redshift measurements to measure outflow
velocities, and \citet{shi14} conducted a similar analysis for four individual
bright LAEs, finding typical outflow velocities $v\sub{abs}\sim
100-200$ km s$^{-1}$. As noted in Sec.~\ref{laes:intro}, however, the
objects in these studies have 
${B}$-band magnitudes $\langle m\sub{B}\rangle \sim 24$ ($L\sim L_*$), meaning that
they easily fall within typical selection criteria for
continuum-selected star forming galaxies.
While useful for studying 
the correlation between outflow velocity $v\sub{abs}$ and $W\sub{\lya}$, these
samples are clearly far removed from the faint LAEs considered in this
paper, with median continuum magnitude $\langle m\sub{B}\rangle =
26.8$ ($L\sim0.1L_*$, $\sim$10$\times$ fainter than these previous samples).

For such faint objects, large samples are needed to obtain even a
low-fidelity measurement of metal absorption. To make such a
measurement, we stack the rest-UV spectra of
all 318 spectroscopic KBSS-\lya\ LAEs using a rest frame defined
by $z\sub{sys,\lya}=z\sub{\lya,peak}-200$ km s$^{-1}$ based on the
redshift calibration of Sec.~\ref{sublaes:lyashift}. As many of our
objects have multiple LRIS spectra of their rest-UV continuum, we
include all spectra for each object such that each spectrum receives
equal weighting, effectively weighting each object by total exposure
time.\footnote{The results of the continuum stacking are insensitive
  to whether the objects are weighted equally, weighted by number of
  spectra, or weighted by total exposure time.} Furthermore, two spectra
with individually-significant continua (possibly AGN) were removed from
the sample in order to ensure that they did not dominate the stack; all
the remaining spectra have S/N $\ll$ 1 per pixel in the continuum individually. Lastly, the
portion of each spectrum corresponding to $\lambda \approx 5577 \pm
15$\AA\ in the observed frame was omitted from the stack in order to
remove contamination from a bright sky line. In total, 422 LAE
spectra were stacked, producing a final spectrum with an effective
exposure time of $\sim$675 hours (all from the Keck 1 10m telescope).

An error spectrum was then generated for the stack via a bootstrapping
procedure. 1000 iterations were performed in which 422 spectra were
drawn (with replacement) from our sample and stacked to make a set of
1000 randomized spectral stacks. The standard deviation of values
at each pixel was then used to define the error spectrum of the true
stacked data.\footnote{The error spectrum thus includes the
  variance in continuum level at each pixel, so it is a conservative
  estimate of the error in the mean spectrum.} From this error
spectrum, we estimate that the stack has 
a median S/N $\sim$ 10 per pixel in the range 1230\AA\ $\lesssim\lambda\sub{rest}
\lesssim$ 1420\AA, the range of interest for the absorption lines discussed
below.\footnote{Unfortunately, we are unable to obtain robust
  measurements of the \ion{C}{4} doublet 1549,1551\AA, generally one
  of the strongest absorption lines in high-redshift galaxies and a
  key indicator of AGN activity when seen in emission. Because we
  favor the resolution of the 600-line grism over the increased
  spectral range of the 300 or 400-line grisms, the \ion{C}{4} doublet
  falls off the spectrograph for many of our objects. Furthermore, the
  particular redshift range of our LAEs ($z \sim 2.6$) causes 
  significant contamination near $\lambda\sub{rest}\sim1550$\AA\ due
  to the bright sky line at $\lambda\sub{obs}\sim 5577$\AA.}

\begin{figure*}
\includegraphics[width=\textwidth]{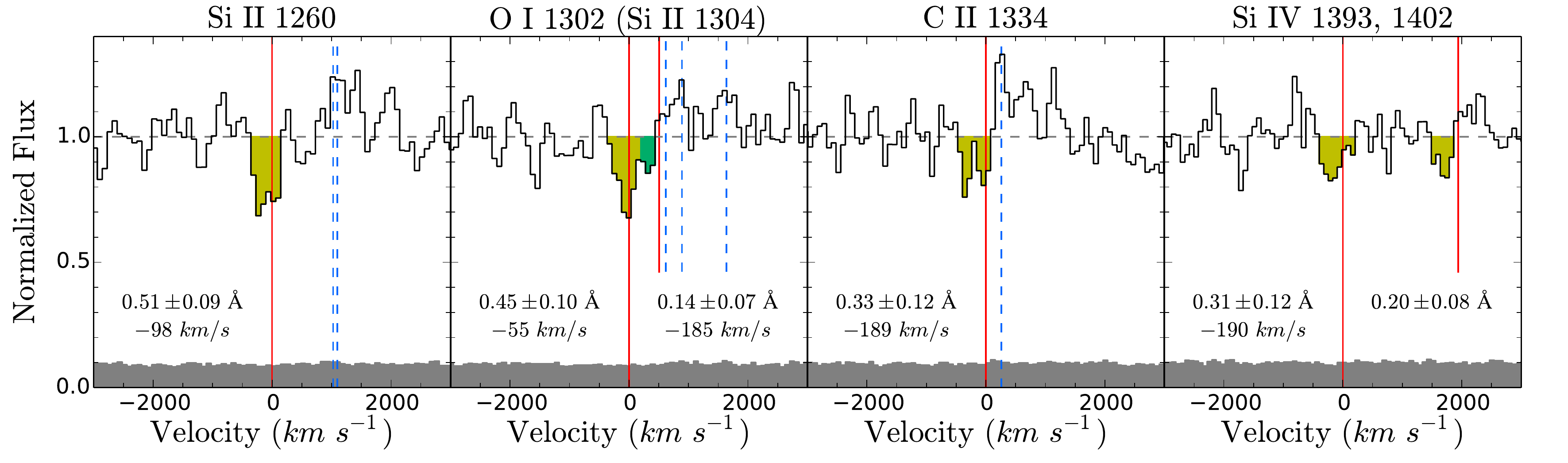}
\caption[Stacked LAE absorption-line spectrum]{Interstellar absorption
  signatures of metal lines in the 
  stacked spectrum of all 318 spectroscopic LAEs. Vertical red lines
  are the rest-wavelengths of metal ions seen in absorption, while
  dashed blue lines denote the wavelengths of fine-structure and
  nebular emission features. A detailed list of these lines is given
  in Table~\ref{table:lines}. Yellow shaded regions are the areas used
  to calculate the equivalent width and absorption-weighted velocity
  of each species; the absorption estimated for \ion{Si}{2}
  $\lambda$1304 is shaded green to differentiate it from the
  partially-blended \ion{O}{1} $\lambda$1302 absorption. Uncertainties
for the equivalent width measurements are estimated via a
bootstrapping procedure described in the text, and the bootstrapped error
spectrum is displayed in grey at the bottom of each panel.}
\label{fig:metal_abs}
\end{figure*}

\begin{deluxetable}{lllll}
\tablecaption{LAE Continuum Absorption Features}
\tablewidth{0pt}
\tablehead{
Ion & $\lambda\sub{lab}$\tablenotemark{a} & $f\sub{osc}$\tablenotemark{b} &
$W\sub{ion}$\tablenotemark{c} & $A_{ki}$\tablenotemark{d} \\
& (\AA) & & (\AA) & $(10^8$ s$^{-1})$
}

\startdata
\sidehead{Absorption features}
\ion{Si}{2}\tablenotemark{e} & 1190.416 & 0.575 & $0.53\pm0.15$ & ~~~~$-$ \\
\ion{Si}{2}\tablenotemark{e} & 1193.290 & 0.277 & $0.24\pm0.11$ & ~~~~$-$ \\[5pt]
\ion{Si}{2} & 1260.422 & 1.22 & $0.53\pm0.10$ & ~~~~$-$ \\
\ion{O}{1} & 1302.169 & 0.0520 & $0.43\pm0.10$\tablenotemark{f} & ~~~~$-$ \\
\ion{Si}{2} & 1304.370 & 0.0928 & $0.17\pm0.08$\tablenotemark{f} & ~~~~$-$ \\
\ion{C}{2} & 1334.532 & 0.129 & $0.32\pm0.11$ & ~~~~$-$ \\
\ion{Si}{4} & 1393.760 & 0.513 & $0.30\pm0.12$ & ~~~~$-$ \\
\ion{Si}{4} & 1402.773 & 0.255 & $0.21\pm0.08$ & ~~~~$-$ \\

\sidehead{Emission features}
\ion{Si}{2}* & 1264.738 & ~~~~$-$ & ~~~~$-$ & 30.4 \\
\ion{Si}{2}* & 1265.002 & ~~~~$-$ & ~~~~$-$ & \pp4.73 \\
\ion{O}{1}* & 1304.858 & ~~~~$-$ & ~~~~$-$ & \pp2.03 \\
\ion{O}{1}* & 1306.029 & ~~~~$-$ & ~~~~$-$ & \pp0.676 \\
\ion{Si}{2}* & 1309.276 & ~~~~$-$ & ~~~~$-$ & \pp6.23 \\
\ion{C}{2}* & 1335.663 & ~~~~$-$ & ~~~~$-$ & \pp0.476 \\
\ion{C}{2}* & 1335.708 & ~~~~$-$ & ~~~~$-$ & \pp2.88
\enddata
\tablenotetext{a}{Vacuum wavelength of transition}
\tablenotetext{b}{Oscillator strength from the NIST Atomic Spectra
Database (www.nist.gov/pml/data/asd.cfm)}
\tablenotetext{c}{Equivalent width of absorption in stacked LAE spectrum
  (Fig.~\ref{fig:metal_abs})}
\tablenotetext{d}{Einstein A-coefficients from the NIST Atomic Spectra
Database}
\tablenotetext{e}{The \ion{Si}{2} $\lambda\lambda$ 1190,1193 doublet
  is in the continuum blueward of \lya, which suffers from lower S/N
  and a high density of absorption and emission lines. As such, these features
  are only used here to constrain the saturation of \ion{Si}{2} $\lambda$1260.}
\tablenotetext{f}{The \ion{O}{1}
$\lambda$1302 and \ion{Si}{2} $\lambda$1304 absorption lines 
are partially blended, so the division between them is somewhat uncertain.}
\label{table:lines}
\end{deluxetable}

\begin{figure}
\includegraphics[width=\linewidth]{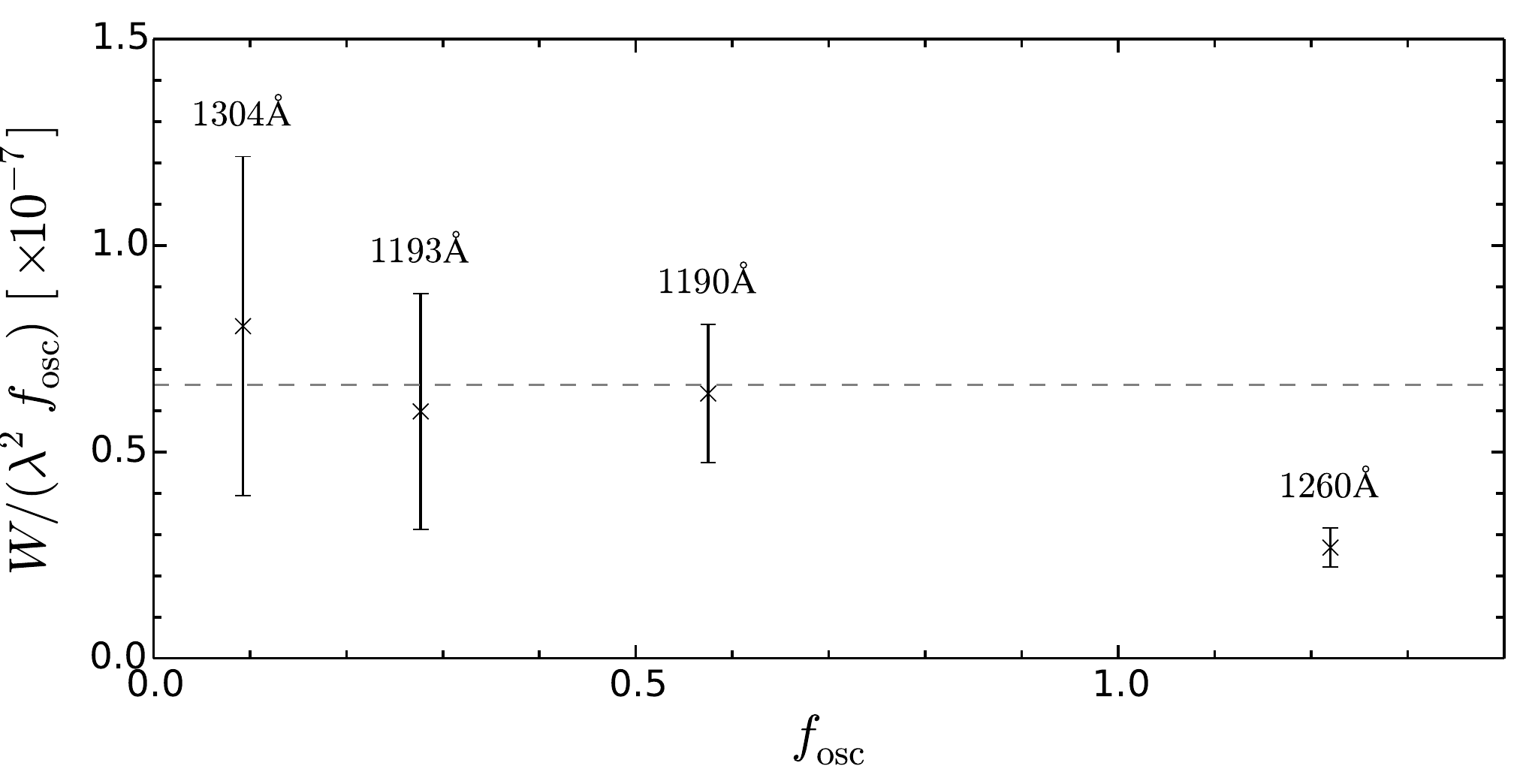}
\caption{Curve-of-growth analysis for the 4 detected \ion{Si}{2} lines
in the stacked LAE spectrum. The scaled equivalent width of absorption
for each transition $i$ is given by $W_i/\lambda_i^2 f\sub{osc,$i$}$,
which is proportional to the column density of the ion (and thus equal
for all transitions) if the transition is unsaturated. If transitions
are saturated, $W_i/\lambda_i^2 f\sub{osc,$i$}$ will decrease with
increasing $f\sub{osc,$i$}$. Errorbars are 
the 1-$\sigma$ uncertainties on $W_i$ from the bootstrap analysis. The
dashed horizontal line is the best-fit constant value of $W/\lambda^2
f\sub{osc}$ for the three transitions with the lowest values of
$f\sub{osc}$ (the least likely to be saturated).}
\label{fig:si2_sat}
\end{figure}

Measurements were made of the absorption of six metal lines,
corresponding to four spectral regions (Fig.~\ref{fig:metal_abs}, Table
\ref{table:lines}). For 
each spectral region, the continuum level was independently estimated
using the local median value, but the continuum estimates among all four
regions are consistent to $\lesssim 2$\% in $\lambda f_\lambda$
units. The measured continuum was also checked for consistency with the
broadband photometry. At the median wavelength of the absorption
lines ($\lambda\sub{obs}\sim5000$\AA), the stacked continuum flux density is
$7.5\times10^{-8}$ Jy ($m\sub{5000} =26.72$), comparable to the
mean \lya-subtracted broadband $B/G$ flux ($7.9\times10^{-8}$ Jy;
$\langle m\sub{BB}\rangle=26.66$) measured photometrically among the
318 objects in the spectral stack. Absorption measurements were then made for
the low-ionization lines \ion{Si}{2} $\lambda$1260, \ion{O}{1}
$\lambda$1302, \ion{Si}{2} $\lambda$1304, and \ion{C}{2}
$\lambda$1334, as well as the higher-ionization doublet \ion{Si}{4}
$\lambda\lambda$1393,1402. The equivalent width of absorption is
estimated in each case by numerically integrating the set of
contiguous pixels with $f\sub{norm}<1$ nearest the rest-frame
metal-line wavelength in the normalized stack. The \ion{O}{1}
$\lambda$1302 and \ion{Si}{2} $\lambda$1304 absorption lines
are partially blended, so the division between them was determined by visual
inspection of the stacked spectrum. This division was
  cross-checked in bootstrapped composite spectra resampled at a range
  of wavelength scales. Uncertainties in the equivalent
widths were computed by performing the same measurement for each line
in each of the 1000 bootstrap spectra; the quoted error refers to the
standard deviation of these bootstrap measurements. An
absorption-weighted velocity was then measured for each line via
numerical integration of the line profile: 
$v\sub{abs}=\int_{\Delta v} [1-f\sub{norm}(v)]vdv/\Delta v$. While the equivalent widths in
absorption of the two lines in the \ion{Si}{4} doublet were measured
separately, they were constrained to have the same total
absorption-weighted velocity.

 All the measured absorption lines have mean velocities
$-100\lesssim \langle v\sub{abs}\rangle\lesssim-200$ km s$^{-1}$ with absorption
extending to $v\sub{abs}\sim-500$ km s$^{-1}$ (where negative
velocities indicate blueshifted absorption), in good agreement with
the brighter LAE samples of \citet{has13} and \citet{shi14}. We
therefore find clear evidence for blueshifted metal absorption
signatures in these $L\sim0.1 L^*$ LAEs, a strong indicator of the presence of
metal-enriched outflowing gas.

Because the 
systemic redshifts are not precisely known for the LAEs in our stack,
there may be some additional uncertainty in our measured velocities
compared to those measured in samples with complete catalogs of
systemic redshifts. However, our stacked spectra also show several
fine-structure and nebular emission features
(dashed vertical lines in Fig.~\ref{fig:metal_abs}) that are more likely to trace the systemic galaxy
redshifts; the correspondence of these peaks to the rest-frame
wavelengths of these features further suggests that our corrected \lya\ line
measurements are a good proxy for the redshifts in these LAEs. Notably,
we have not corrected the line strengths or velocities for
emission from nearby fine-structure transitions; the effects of this
contamination likely causes an underestimate of intrinsic
absorption and overestimate of the velocity of \ion{Si}{2} $\lambda$1304
and \ion{C}{2} $\lambda$1334. 

\begin{deluxetable*}{crcccccc}
\tablecaption{\lya\ Spectroscopic Subsamples and Composite \lya\
  Emission Profiles}
\tablewidth{0pt}
\tablehead{
Subsample & $\langle W\sub{\lya} \rangle$ & $\langle\mathcal{R}\rangle$ & $N\sub{obj}$ &
$N\sub{spec}$\tablenotemark{a} & $\sigma\sub{red}$ (km s$^{-1}$)\tablenotemark{b} &
$\sigma\sub{blue}$ (km s$^{-1}$)\tablenotemark{b}  & $\Delta v$ (km s$^{-1}$)\tablenotemark{b}
}

\startdata
\sidehead{\lya\ redshifts\tablenotemark{c}}
$W\sub{\lya} \geq 60$\AA & 110\AA~ & 27.6 & 158 & 211 & 115$\pm$5\pp &
191$\pm$10 & 447$\pm$7 \\
$20 \leq W\sub{\lya} < 60$\AA & 37\AA~ & 26.8 & 160 & 211 & 116$\pm$5\pp & 276$\pm$15 & 468$\pm$7 \\
$W\sub{\lya} < 20$\AA\tablenotemark{d} & 11\AA~ & 26.6 & 82 & 94 &
119$\pm$7\pp & 353$\pm$37 & \pp475$\pm$10\\
KBSS LBGs & 27\AA~ & 24.4 & 65 & 65& 178$\pm$10 &
226$\pm$8\pp & \pp646$\pm$20 \\

\sidehead{Systemic redshifts}
MOSFIRE LAEs & 44\AA~ & 26.9 & 32 & 57& 131$\pm$9\pp &
234$\pm$14 & \pp490$\pm$23 \\
KBSS LBGs & 27\AA~ & 24.4 & 65 & 65& 215$\pm$20 & 246$\pm$11 & \pp568$\pm$25 
\enddata
\tablenotetext{a}{Because some objects were observed multiple times, and
  all available spectra were combined in each continuum stack, $N\sub{spec}>N\sub{obj}$.}
\tablenotetext{b}{The parameters $\sigma\sub{red}$,
  $\sigma\sub{blue}$, and $\Delta v$ are the velocity widths of the
  red peak, blue peak, and peak separation in the fit to the composite
  \lya\ spectrum (Figs.~\ref{fig:abs_ions}~\&~\ref{fig:abs_ions_lbg}). The peak widths
  reflect the subtraction in quadrature of the LRIS instrumental
  resolution of $\sigma\sub{inst} = 100$ km s$^{-1}$. }
\tablenotetext{c}{Composite spectra constructed using \lya\ redshifts,
shifted by $-$200 km s$^{-1}$}
\tablenotetext{d}{Objects observed on the basis of a narrowband excess
and spectroscopically-detected \lya\ emission line, but which have
$W\sub{\lya}$ smaller than the adopted threshold of $W\sub{\lya}\leq20\AA$}
\label{table:eqwsamples}
\end{deluxetable*}

Galaxy outflows may also be characterized by the covering fraction of
the UV continuum by absorbing materials and its
optical depth. While these properties are
typically degenerate for individual transitions (except at much higher
S/N and resolution), they may be constrained
independently when multiple transitions of the same species are
present. In particular, the expected ratio of the equivalent widths of
\ion{Si}{2} $\lambda$1304 and \ion{Si}{2} $\lambda$1260 is
$W_{1260}/W_{1304}=12.3$ when both transitions are optically thin. The
measured ratio of these lines in the LAE composite spectrum is
$W_{1260}/W_{1304}=3.6$, consistent with a saturated \ion{Si}{2}
$\lambda$1260 transition, with the caveat that the measured \ion{Si}{2}
$\lambda$1304 absorption has small S/N and is partially blended with
the \ion{O}{1} $\lambda$1302 line. To clarify this measurement, we
consider the absorption due to the \ion{Si}{2} $\lambda
\lambda$1190,1193 doublet in the rest-EUV
($\lambda\sub{rest}\lesssim1200$\AA) spectrum. The high density 
of absorption and emission lines and lower continuum level of the 
spectrum shortward of \lya\ (including foreground \lya\ absorption) result in larger
uncertainties in these line strengths, but both lines of the doublet
are detected with high significance.

The measured absorption equivalent widths ($W$) for each
of the four \ion{Si}{2} transitions are given in
Table~\ref{table:lines}, along with the corresponding vacuum
wavelengths ($\lambda$) and oscillator strengths ($f\sub{osc}$). In the
optically-thin regime, these quantities are related to the column
density of absorbing material by $N\sub{ion}\propto W_i/\lambda_i^2
f\sub{osc,$i$}$ for each transition of index $i$. Because the physical
column density of a single ionic species must be single-valued, the
quantity $W/\lambda^2 f\sub{osc}$ will likewise be constant for
all ground-state transitions of the same ion unless one or more
transitions have saturated. In this latter case, $W/\lambda^2
f\sub{osc}$ will decrease with increasing $f\sub{osc}$.

The quantity $W/\lambda^2 f\sub{osc}$ is compared for the four
\ion{Si}{2} transitions in Fig.~\ref{fig:si2_sat}. The \ion{Si}{2}
$\lambda$1190, $\lambda$1193, and $\lambda$1304 transitions are consistent with a
single value of $W/\lambda^2 f\sub{osc}$, but the \ion{Si}{2}
$\lambda$1260 absorption falls significantly below the weighted
average of the other three lines, indicating saturation. While the
other three transitions appear to be optically thin, both \ion{Si}{2}
$\lambda$1193 and \ion{Si}{2} $\lambda$1304 have nearby excited fine-structure
emission transitions that may bias our measurements toward lower
equivalent widths, as discussed above. If the \ion{Si}{2}
$\lambda$1304 line were saturated, however, the bottom of its
absorption trough would reach a similar depth as the saturated
\ion{Si}{2} $\lambda$1260 transition with respect to the local
continuum (as is seen in some LBG samples; see
Sec.~\ref{sublaes:lbgoutflows}), which is clearly not the case in the
LAE composite spectrum (Fig.~\ref{fig:metal_abs}). This suggests that any filling-in of the
absorption by excited fine-structure emission has a small effect on the
total measured absorption. The analysis of the composite spectrum thus
strongly suggests that $\tau_{1260}>1$, $\tau_{1304}\ll1$, and
$\tau_{1190}\sim \tau_{1193}\lesssim1$ based on the consistency of the
curve-of-growth analysis in Fig.~\ref{fig:si2_sat} (where $\tau_\lambda$ is
the central optical depth of the \ion{Si}{2} transition at wavelength $\lambda$).

Given the saturation of \ion{Si}{2} $\lambda$1260, the
maximum depth of the line profile with respect to the
continuum indicates a covering fraction of \ion{Si}{2}-enriched
outflowing gas $f_c\approx0.3$ among the LAEs. While multiple
unambiguous absorption features are not available for other
low-ionization species in the composite, the consistency of the
absorption profile depths and widths for the \ion{O}{1} $\lambda$1302 and
\ion{C}{2} $\lambda$1334 with the saturated \ion{Si}{2} $\lambda$1260
line suggests that those species are co-spatial with \ion{Si}{2} and 
have a similar covering fraction. We therefore take the covering
fraction of low-ionization gas outflows to be $f_c\approx0.3$ for the
LAE sample, noting the possible existence of narrow absorption
components unresolved by our $R\sim1300$ spectra.

The two lines of the \ion{Si}{4} doublet have a measured ratio
$W\sub{1393}/W\sub{1402}=1.4$, suggesting that the doublet is also
saturated ($W\sub{1393}/W\sub{1402}=2$ on the linear part of the curve
of growth). However, the weaker line has S/N $\sim 2$, and $\sim$25\%
of the bootstrap spectra have $W\sub{1393}/W\sub{1402}\gtrsim2$, so
the transition may in fact be optically thin. The higher ionization potential
of \ion{Si}{4} (33.5 eV) requires radiation from hot stars and/or
collisional ionization in $T\gtrsim10^4$K gas, and therefore traces
more strongly ionized gas in or around the galaxy. The depth of the two absorption
lines thus suggests the presence of an ionized medium with an average
covering fraction $f_c\sim20$\% (or greater if the lines are optically
thin). It is unclear whether this gas is co-spatial with the medium giving rise
to the lower-ionization lines described previously; this relationship
is probed further by comparing the velocity profiles of their
absorption below.

It is important to note that all of the above constraints on the
covering fraction of metal-enriched gas rely on the assumption that
the metal-line photons scattered out of the galaxy spectrum are
emitted away from our line of sight to the galaxy. In the limit that
the scattering medium is entirely co-spatial with the observed light
distribution from the galaxy (and that the scattering is isotropic and
photon-conserving), we would expect that the net emission and
absorption from scattering would be zero. In reality, many photons
will be scattered at large radii with respect to the light profile of
the galaxy, particularly for transitions where
excited fine-structure states provide an alternative escape pathway
with low optical depth. Given the spatial compactness of the LAEs,
however, this ``emission-filling'' may reduce the observed
absorption, particularly for those transitions without nearby excited
fine-structure states such as \ion{Si}{4} $\lambda
\lambda$1393,1402. This effect may partially explain the lower
measured value of  $f_c$ for the \ion{Si}{4} transitions, in addition
to their unknown optical depth.

The velocity shift of \lya\ emission shows a strong inverse trend with
\lya\ equivalent width in the collective populations of high-redshift
LAEs and LBGs \citep{erb14}, but there are many mechanisms
that could underly this relationship. If a change in outflow velocity
is the driver, then we should expect the velocity profile of the metal
absorption lines tracing the outflows to vary with $W\sub{\lya}$ as
well. Unfortunately, the S/N in stacked subsamples is too low to probe
this effect in individual lines. However, we can boost the S/N by
combining the line profiles for multiple lines that we expect to trace
the same gas. In particular, we can stack the low-ionization lines
(corresponding to more neutral \ion{H}{1} gas) and the high-ionization lines
(corresponding to more highly ionized \ion{H}{2} gas) and thereby extract enough signal to
compare the variation of the gas velocity distribution with $W\sub{\lya}$.

\begin{figure*}
\includegraphics[width=0.33\textwidth]{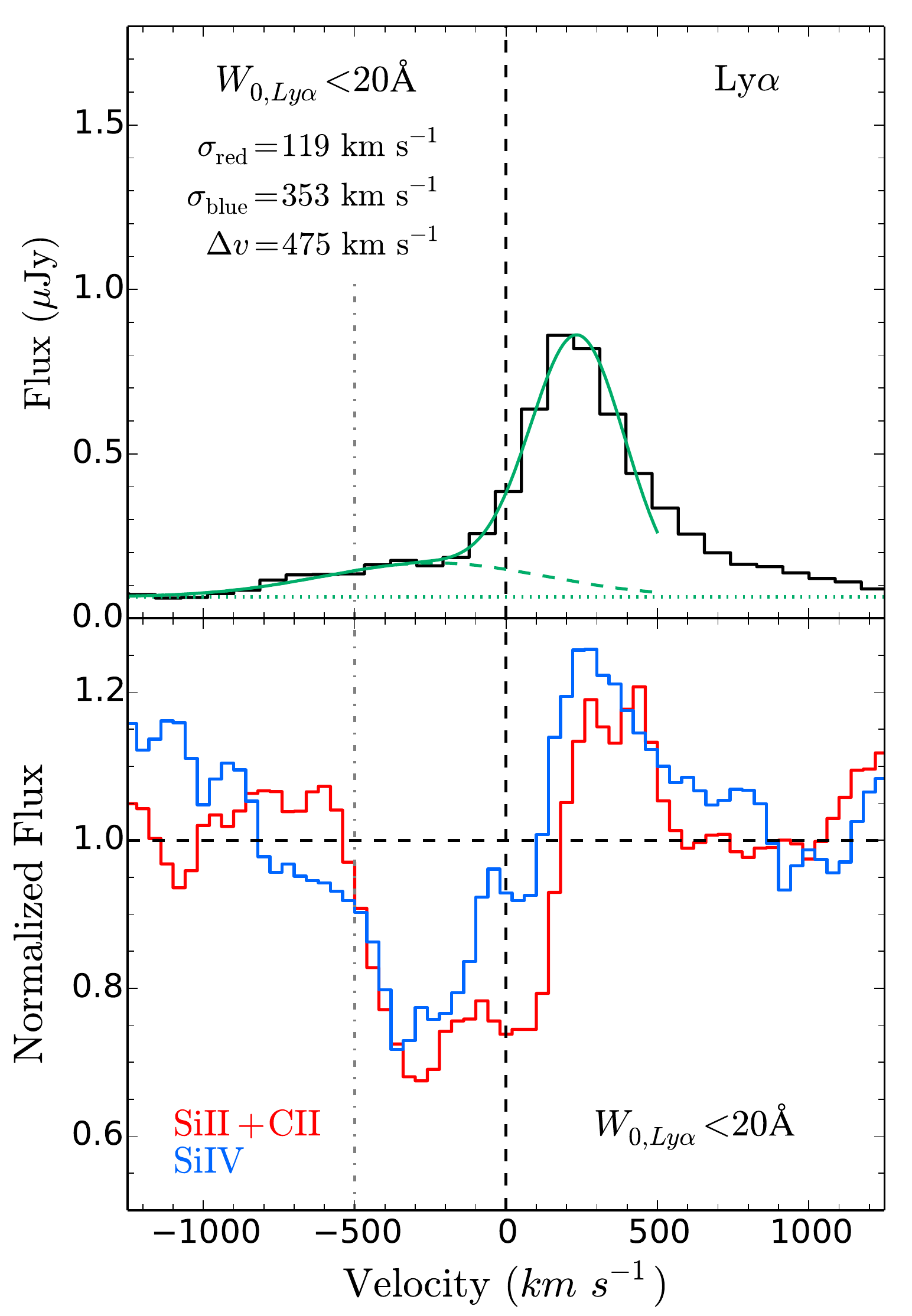}
\includegraphics[width=0.33\textwidth]{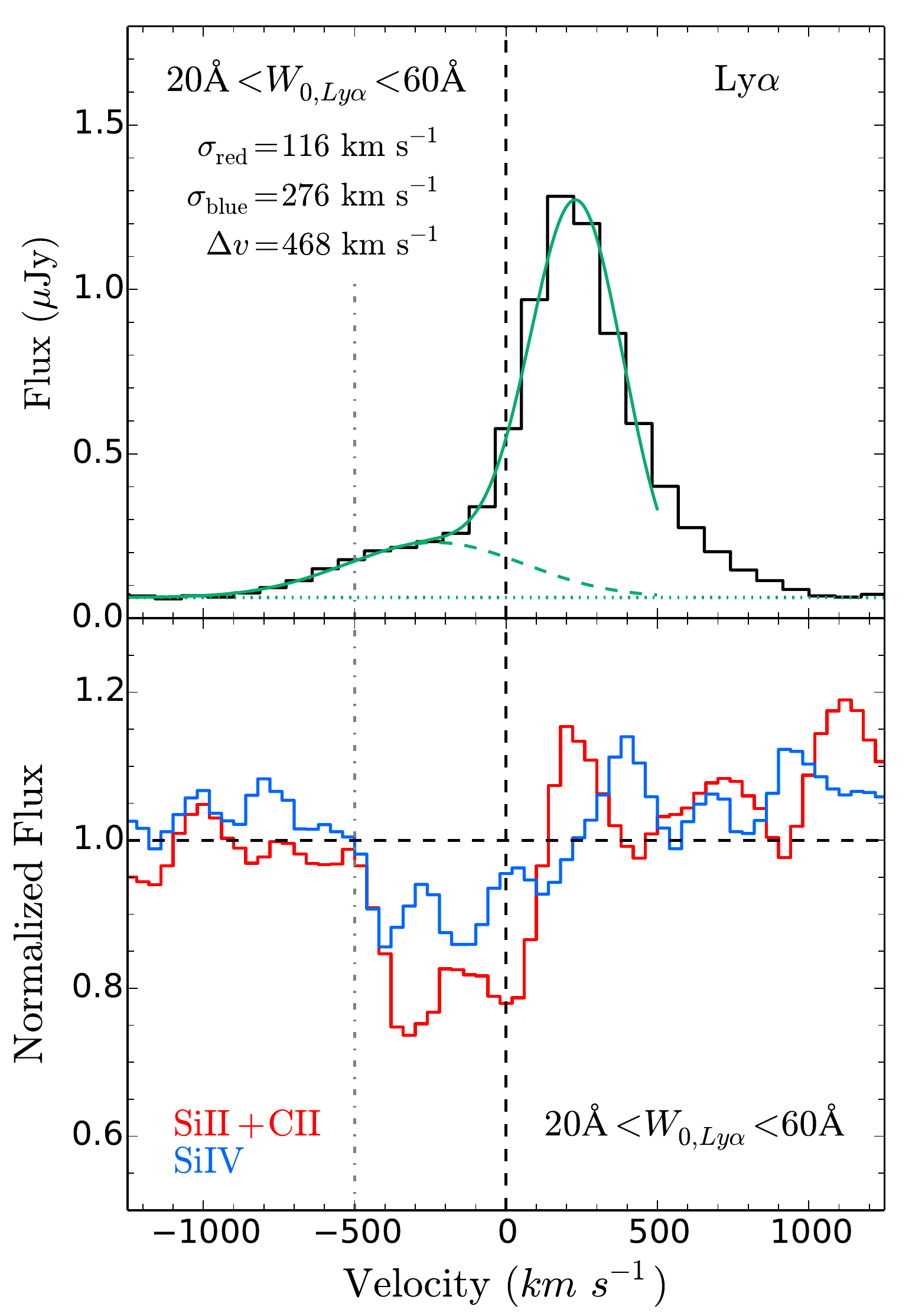}
\includegraphics[width=0.33\textwidth]{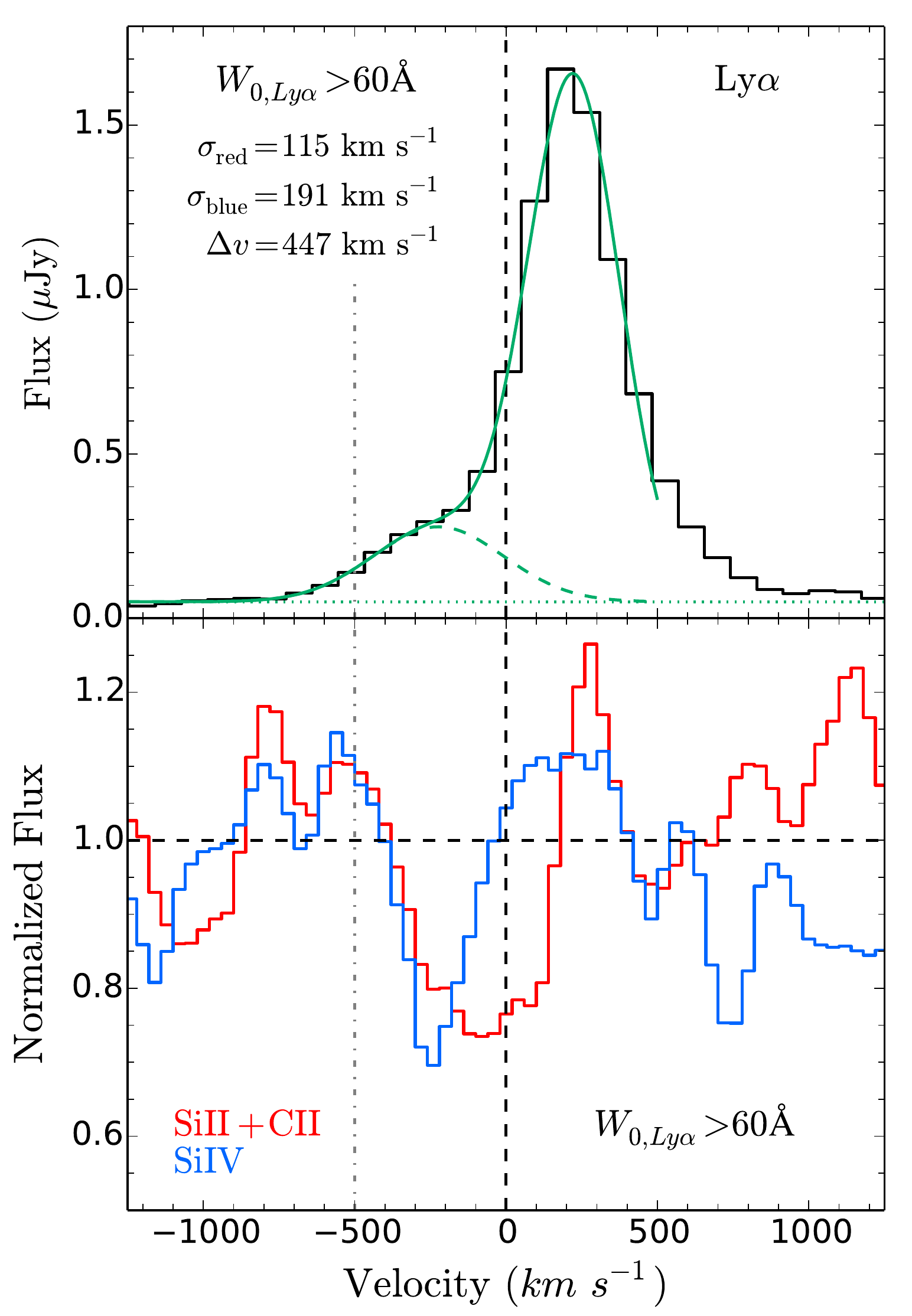}
\caption[Stacked \lya\ emission and absorption for several bins in
$W\sub{\lya}$]{\lya\ emission and interstellar absorption 
  signatures in stacked subsets of spectra grouped by their
  photometric \lya\ equivalent width, $W\sub{\lya}$. Details of each
  subsample are given in Table~\ref{table:eqwsamples}.  {\it Top panels:}
  \lya\ emission spectra (black), and profile fit (green) for each
  subsample. The solid green line is the full fit, the dotted green
  line is the fit continuum level, and the dashed green line is the
  blueshifted Gaussian component of the fit. The fits do
  not extend to $v>+500$ km s$^{-1}$, as this portion of the spectrum
  was omitted during fitting. Dashed and dot-dashed vertical lines
  denote $v=0$ and $v=-500$ km s$^{-1}$, respectively. Note that the
  specific flux in $\mu$Jy is measured in the observed frame. {\it Bottom panels:}
  Red curves are the average absorption profiles of the
  low-ionization species \ion{Si}{2} $\lambda$1260 and
  \ion{C}{2} $\lambda$1334, while blue curves are the absorption
  profiles of the more highly ionized \ion{Si}{4} ($\lambda$1393 and
  $\lambda$1402). \ion{O}{1} $\lambda$1302 and \ion{Si}{2} $\lambda$1304
  are omitted due to blending. As above, dashed and dot-dashed vertical lines
  denote $v=0$ and $v=-500$ km s$^{-1}$, while the dashed horizontal
  line denotes the local continuum level.}\label{fig:abs_ions}
\center
\includegraphics[width=0.33\textwidth]{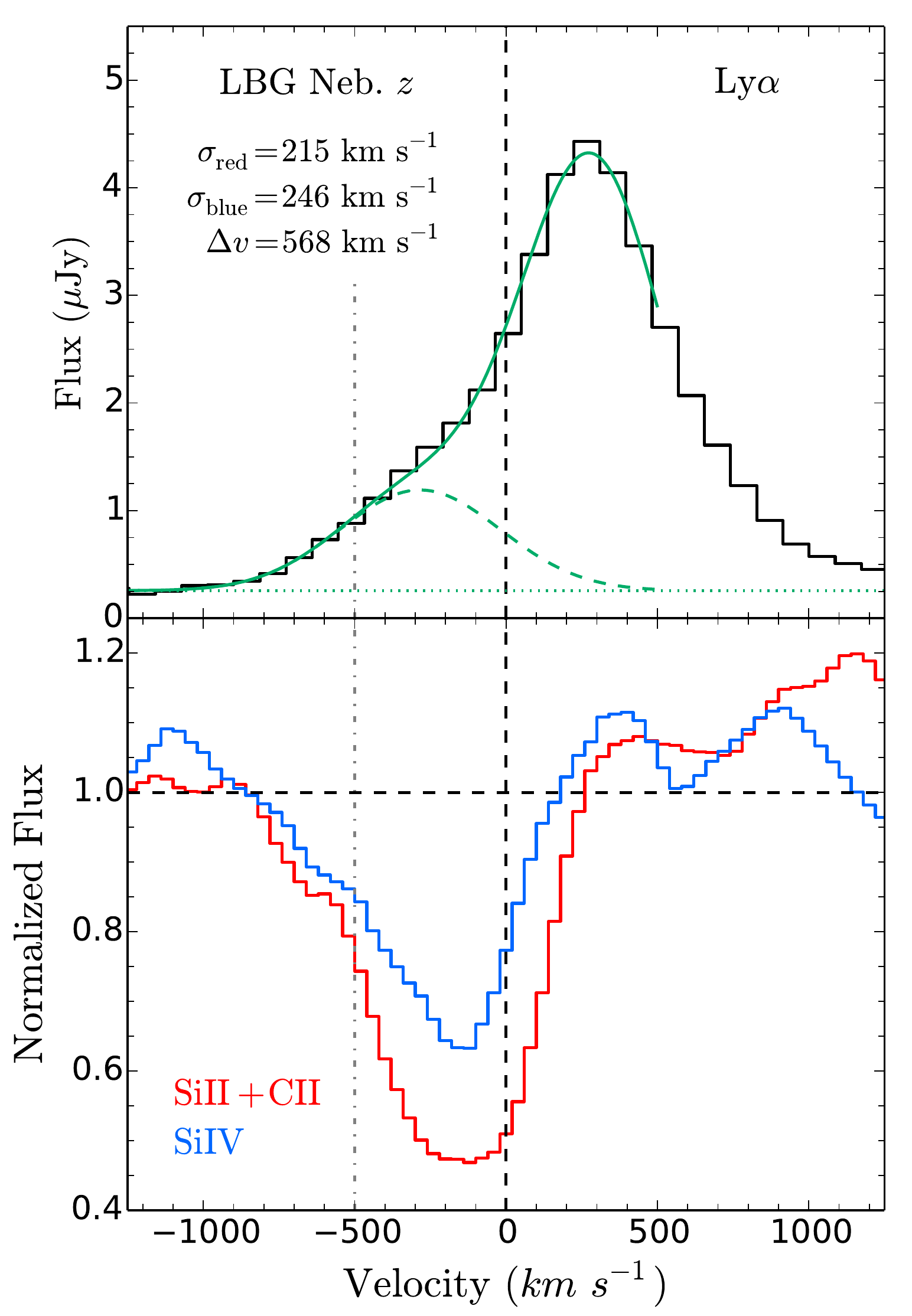}
\includegraphics[width=0.33\textwidth]{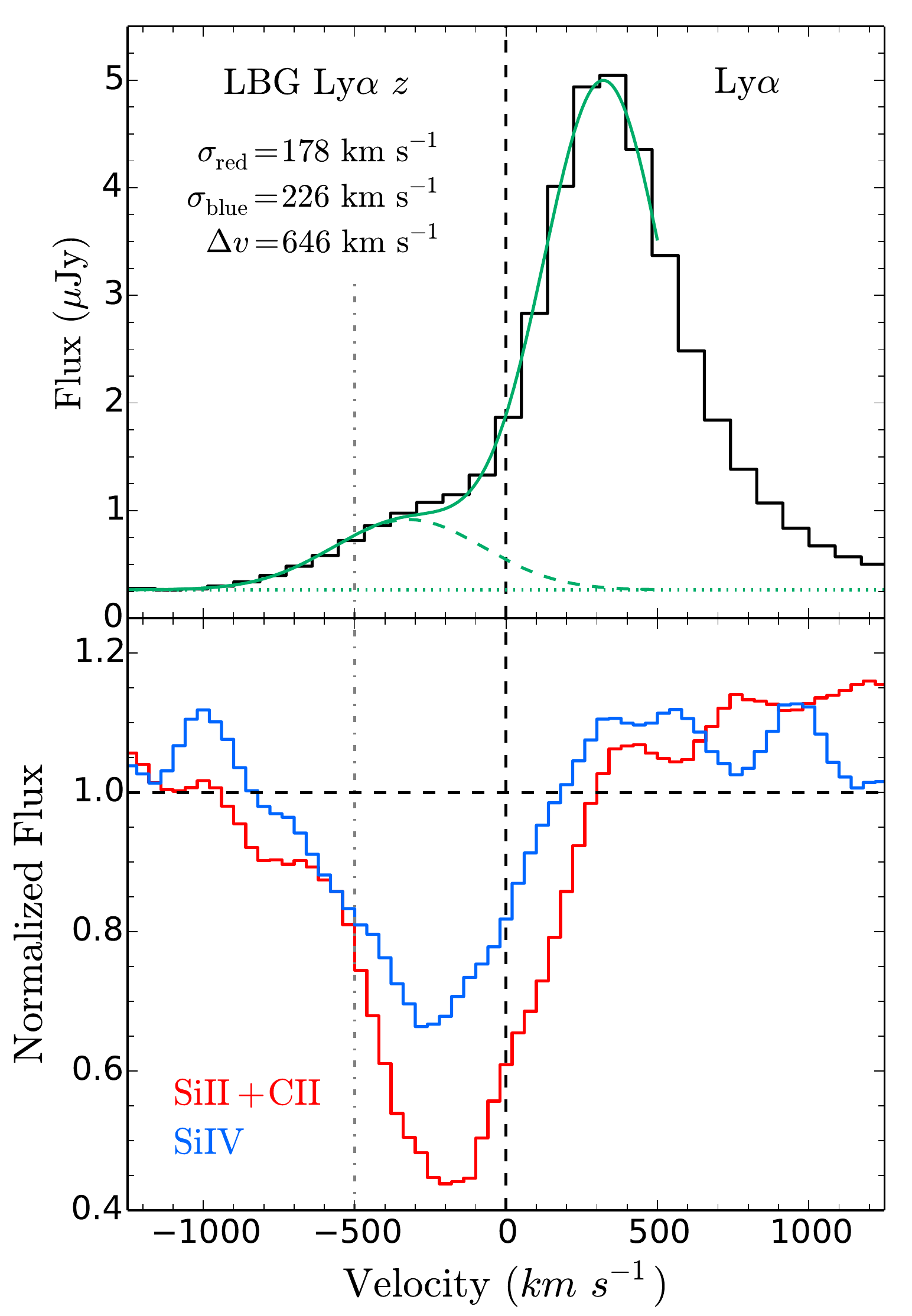}
\caption[Stacked \lya\ emission and absorption for LBG sample]{The
  composite \lya\ emission and low/high-ionization absorption line
  profiles (as in Fig.~\ref{fig:abs_ions}) for the KBSS LBG
  sample. {\it Left:} The composite spectrum is constructed using the
  nebular (systemic) redshift measurements. {\it Right:} The composite
  is constructed by stacking spectra according to their \lya\ emission
  peak, in the same way as the LAE samples of
  Fig.~\ref{fig:abs_ions}. As in Fig.~\ref{fig:lya_stack}, the
  composite \lya\ profile is distorted with respect to the composite
  constructed using systemic redshift measurements. However, the
  composite absorption profile is not significantly affected by the
  choice of nebular redshifts or calibrated \lya\ redshifts.}
\label{fig:abs_ions_lbg}
\end{figure*}

Spectra combined in this way are displayed in Fig.~\ref{fig:abs_ions}
for three subsamples divided according to $W\sub{\lya}$, with median
$W\sub{\lya} \approx 11$\AA, 37\AA, and 110\AA\ respectively. Because these 
stacks include features at multiple wavelengths, we 
subsampled them by a factor of two with respect to the original
wavelength scale before stacking in order to maximize the velocity
resolution of the stacks. Details of these subsamples are in Table~\ref{table:eqwsamples}.

Several key features are apparent in the bottom panels of Fig.~\ref{fig:abs_ions}. Firstly,
there is substantial absorption by low-ionization species in all three
panels at the systemic redshift of the galaxy ($v=0$), but there is
little or no corresponding high-ionization absorption. At the
negative edge of the absorption profiles ($v\sim - 500$ km s$^{-1}$),
however, the high-ionization and low-ionization absorption are well
aligned in each panel. This pattern may indicate that both ionized and
neutral gas are present in the material most likely associated with an
outflow, but that the interstellar medium is dominated by neutral \ion{H}{1} and
low-ionization metals at $v=0$. As noted above, however, the
\ion{Si}{4} transition is particularly sensitive to emission-filling, which will
tend to systematically reduce the observed absorption at $v\approx 0$.

Secondly, if the absorption lines are assumed to be optically-thick, there is a
slight variation in covering fraction as a function of $W\sub{\lya}$:
$f_c\sim30$\% for the $W\sub{\lya}\sim11$\AA\ sample, $f_c\sim20$\% for the 
intermediate sample, and $f_c\sim25$\% for the highest-$W\sub{\lya}$
sample. The decline in $f_c$ from $W\sub{\lya}\sim11$\AA\ to $W\sub{\lya}\sim37$\AA\
is consistent with expectations for \lya-emitting galaxies:
\citet{ste10} demonstrate that observed \lya-emission is closely
correlated with the covering fraction of the gas through which these
photons escape. The increase in apparent $f_c$ between the $\langle
W\sub{\lya}\rangle=37$\AA\ and $\langle W\sub{\lya}\rangle =110$\AA\
samples may not be significant because the continuum
emission is extremely faint in the latter stack; not only is
the statistical error large, but small systematics in the
spectroscopic background subtraction could be significant in the stack of these
faint objects, possibly biasing the measured continuum level. However, the
consistency between the photometric and spectroscopic continuum
estimates suggests this effect should be quite small. In any case, it
seems that the apparent depth of the interstellar absorption features
correlates only weakly with $W\sub{\lya}$ for the faintest, 
highest-$W\sub{\lya}$ bin of galaxies.

Thirdly, there is some dependence in the maximum absorption velocity as
a function of $W\sub{\lya}$. While the absorption extends to
$v\sub{abs}\sim-500$ km s$^{-1}$ in all three panels, the
$W\sub{\lya}\sim11$\AA\ stack shows absorption in \ion{Si}{4} extending to
higher velocities at low optical depths. Conversely, the
most-blueshifted edge of absorption in the $W\sub{\lya}\sim110$\AA\ stack extends no
further than $v\sub{abs}\sim-400$ km s$^{-1}$.\footnote{Note that the
 maximum absorption velocity is insensitive to any bias in the
 continuum level due to systematics in the background subtraction.} It
is important to note that the systematic velocities used here are
derived from a constant correction to the \lya\ redshift, and thus the
evolution of the \lya\ offset with $W\sub{\lya}$ could theoretically mimic an
evolution in absorption velocity with $W\sub{\lya}$. However, despite
previous measurements of the
significantly smaller \lya\ offsets among LAEs with respect to LBGs,
we find that the  {\it peak} of the \lya\ profile has a consistent 200 km
s$^{-1}$ redshift with respect to systemic at our observed resolution for both LBGs and
LAEs (Sec.~\ref{sublaes:lyashift}). Furthermore, no difference with 
$W\sub{\lya}$ is observed in the most-redshifted edge of the low-ionization absorption
(presumably corresponding to neutral interstellar gas at the systemic
redshift) across the three panels of
Fig.~\ref{fig:abs_ions}; such a shift would be expected from a
$W\sub{\lya}$-dependent estimator of the systemic redshift. Lastly, the
total velocity width of absorption decreases significantly with
increasing $W\sub{\lya}$ across the three panels, suggesting the presence of a
real anti-correlation of \lya\ equivalent width with outflow velocity among
these faint LAEs.

Given that $W\sub{\lya}$ is likely to be anti-correlated with
star-formation rate and stellar mass among these LAEs (as discussed in
Sec.~\ref{sublaes:lyaescape}), we interpret the above trends as a
positive correlation between star-formation rate and outflow
velocity. A similar analysis of the KBSS LBG sample
(Fig.~\ref{fig:abs_ions_lbg}, discussed further in
Sec.~\ref{sublaes:lbgoutflows}) suggests a continuation of this trend
to higher star-formation rates.


\subsection{Ly$\alpha$ emission signatures of outflows}\label{sublaes:emission}

The composite \lya\ emission profiles for each subsample (top panels of
Fig.~\ref{fig:abs_ions}) show evidence for changes in outflow velocity
with $W\sub{\lya}$ as well. Each \lya\ profile exhibits a narrow peak redward
of the systemic redshift (by construction, as they are stacked on the
basis of their peak \lya\ redshifts) as well as an extended blue
wing, which broadens with decreasing
$W\sub{\lya}$ (Fig.~\ref{fig:abs_ions}). To quantify this
effect, a two-component Gaussian fit was performed to each
stacked \lya\ profile with the following form:

\begin{eqnarray}
  f_\lambda(v) &=& A\sub{blue} e^{-(v+\Delta v/2)/2
    \sigma\sub{blue}}+ \nonumber \\
  & & A\sub{red} e^{-(v-\Delta v/2)/2
    \sigma\sub{red}} +f\sub{0,cont} \, \, ,
\label{eq:twogauss}
\end{eqnarray}

\noindent such that the fit profile consists of two Gaussian peaks of arbitrary
height and width, but with equal and opposite shifts with respect to
systemic ($\pm\Delta v/2$). The model intentionally evokes an idealized outflow
scenario in which the average LAE is surrounded by outflowing gas in
both directions along the line of sight. In all three stacks, the red
side of the \lya\ profile diverges strongly from a Gaussian shape at
large velocities, forming a broad red tail similar to that seen at $v\ll
0$. In order to maintain the simplicity of the model, the spectrum at
$v>+500$ km s$^{-1}$ was omitted from the fitting.

The resulting fits are displayed in Fig.~\ref{fig:abs_ions}, with fit
parameters given in Table~\ref{table:eqwsamples}. Both the width of
the blue component and the separation of the blue and red peaks
increases with decreasing $W\sub{\lya}$. While the \lya\ peak width and
separation may be increased by raising the \ion{H}{1} column
density, such a scenario would not predict a corresponding increase in 
the velocity width of metal absorption transitions, which have significantly smaller
optical depths than \lya. The similarity of the extended, blueshifted
\lya\ emission to the metal absorption profile therefore indicates
that emission from outflowing gas likely populates the wings of the
\lya\ emission profile. That is, we suggest that the extended \lya\
emission observed at large redshifts and blueshifts with respect to
systemic depends sensitively on the outflow velocity, rather than
merely the optical depth of the scattering medium. Models for this
behavior, including the dependence of \lya\ transmissivity on both the
amplitude and velocity gradient of outflowing gas, are discussed in
detail by \citet{ste10,ste11}.

Radiative transfer simulations of \lya\ transmission suggest that
  these processes can generate double-peaked emission profiles with
  asymmetric offsets with respect to the systemic redshift. In
  particular, \citet{ver06} and others find that a spherical shell of
  gas expanding with velocity $V\sub{exp}$ can in many cases produce
  a blue peak at $v=-V\sub{exp}$ and a red peak at $v=2V\sub{exp}$. 
  The stacked \lya\ profiles in Fig.~\ref{fig:abs_ions} (or
  Fig.~\ref{fig:abs_ions_lbg}, discussed below) are insensitive to the
  peak of the blue component and cannot discriminate between the two
  models (that is, whether the offset of the blue peak is equal to
  that of the red peak or half that of the red peak). However, the same
  qualitative relations hold for either model: the separation of the
  two peaks and the breadth of the extended blue and red components
  increases with increasing continuum luminosity.

\subsection{Comparison to continuum-bright galaxies}\label{sublaes:lbgoutflows}

While the trend of the inferred outflow properties with $W\sub{\lya}$ is weak
among the LAEs in this sample, we can probe a much larger range in
galaxy properties by extending the sample to include continuum-bright
galaxies. In particular, the composite \lya\ profile of the
KBSS LBGs (Fig.~\ref{fig:abs_ions_lbg}) displays a
continuation of several trends identified among the LAE subsamples in
Fig~\ref{fig:abs_ions}. The KBSS LBGs all have systemic redshift
measurements, so we can compare the results of stacking them by either
their nebular redshifts (Fig.~\ref{fig:abs_ions_lbg}, left panel) or their
\lya-derived redshifts (right panel, as in Fig.~\ref{fig:abs_ions}). While the detailed 
profile depends on which redshift is employed, the LBG composites have broader and more
widely-separated peaks than any of the LAE subsamples, and the
velocity extent of the broad tails of the \lya\ emission again
corresponds closely to the velocity extent of blueshifted absorption
from the higher- and lower-ionization species. Note that the
absorption profiles (bottom panels of Fig.~\ref{fig:abs_ions_lbg}) are
insensitive to which redshift indicator is used: the \ion{Si}{2}
$\lambda$1260, \ion{C}{2} $\lambda$1334, and \ion{Si}{4} doublet
absorption centroids each change by less than $\sim$25 km s$^{-1}$
when stacked according to the nebular or calibrated \lya\
redshifts. This consistency suggests that our measurements of
interstellar absorption velocities in the LAE composite are unaffected
by the scarcity of systemic redshifts among the LAE sample.

%

\begin{figure}
\center
\includegraphics[width=\linewidth]{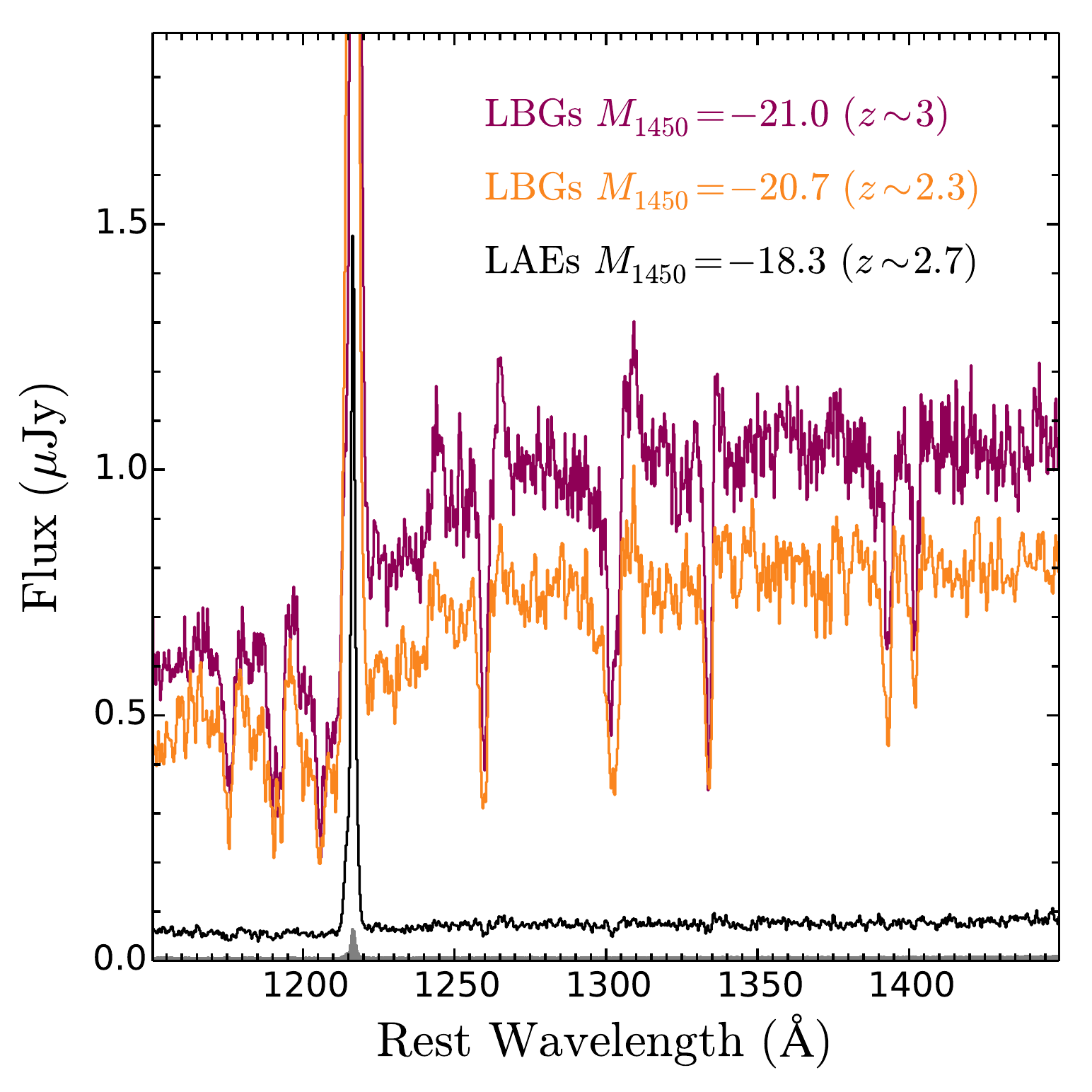}
\caption[Comparison of LAE and LBG spectra]{Composite UV
  continuum spectra for stacks of KBSS-\lya\ LAEs and two comparison
  samples of LBGs. Details of the LBG samples are given in
  Table~\ref{table:lbgsamples}. The LAE stack
   (black, bottom) is the same as 
   that in Fig.~\ref{fig:metal_abs}, and the bootstrapped error
   spectrum is again shown in grey (primarily visible near the
   wavelength of the \lya\ line). The KBSS LBG stack (orange,
   middle) has redshifts measured via nebular emission lines. The
   ``LyC LBG'' 
  stack (purple, top) has 
  redshifts measured by a calibrated combination of \lya\ emission
  and metal absorption redshifts. All three composite spectra have
  $R\sim1300$ for 1250\AA $<\lambda<$ 1400\AA. The challenge of measuring
  outflow signatures in faint LAEs is clear: not only is the continuum
  emission $\sim$10$\times$ fainter than typical LBGs, but the
  underlying absorption signatures are significantly weaker (Fig.~\ref{fig:laevslbg_abs}).}
\label{fig:laevslbg_wide}
\end{figure}

\begin{figure*}
\includegraphics[width=\textwidth]{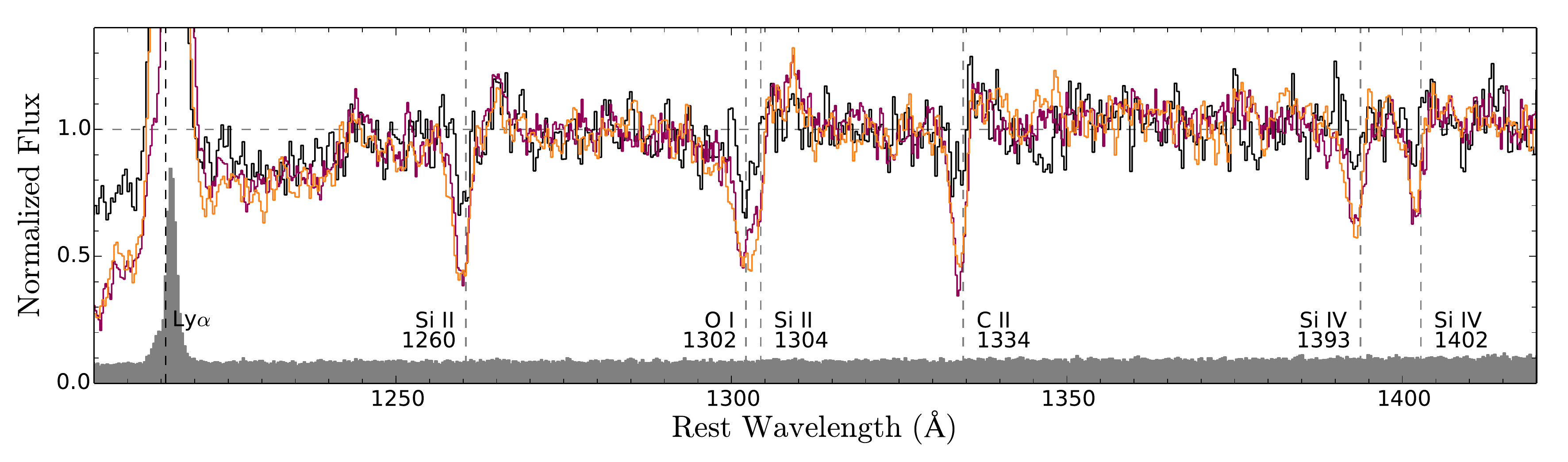}
\caption[Comparison of LAE and LBG continuum absorption]{Composite rest-FUV
  continuum spectra (as in Fig.~\ref{fig:laevslbg_wide}) normalized to
  a common $f\sub{1325}$ continuum level (1250\AA
  $<\lambda<$ 1400\AA). As in the preceding figures, the
  bootstrapped LAE error spectrum is in grey. The absorption lines displayed in
  Fig.~\ref{fig:metal_abs} are here marked by vertical dashed
  lines. The two
  LBG samples (orange, purple) are highly consistent with each other and exhibit clear
  contrasts with the LAE sample, including significantly broader and deeper
  absorption profiles.}
\label{fig:laevslbg_abs}
\end{figure*}

\begin{figure*}
\includegraphics[width=\textwidth]{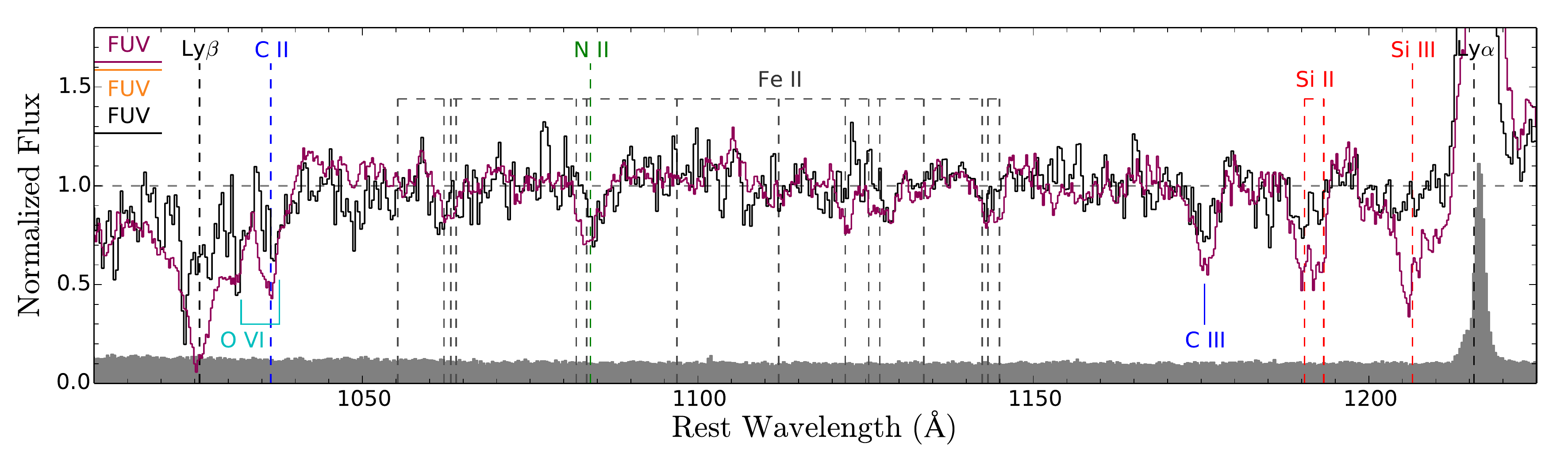}
\caption[Comparison of LAE and LBG EUV continuum absorption]{Composite rest-EUV
  continuum spectra (as in Fig.~\ref{fig:laevslbg_wide}) normalized to
  a common $f\sub{1125}$ continuum level (1050\AA
  $<\lambda<$ 1200\AA). Note that the normalizations differ from
  Fig.~\ref{fig:laevslbg_abs}: the continuum ratio 
  $f\sub{1325}/f\sub{1125}$ for each stack 
  is shown here by the horizontal lines labeled ``FUV'' on the far
  left of the figure. The KBSS LBG composite is omitted due to
  the low S/N in this region, but the continuum ratio for that
  spectrum is marked with the orange horizontal line. As in the preceding figures, the
  bootstrapped LAE error spectrum is in grey. Insterstellar 
  absorption lines are marked with vertical dashed lines, color-coded
  by atomic species and labeled from above. Horizontal dashed lines
  connect transitions of the same ion. Vertical solid lines labeled
  from below mark features with a photospheric component. Significant
  blending is visible due to both the absorption features and the many
  excited-state emission features throughout this region of the spectrum (not marked).}
\label{fig:laevslbg_euv}
\end{figure*}

In Figs.~\ref{fig:laevslbg_wide}$-$\ref{fig:laevslbg_euv}, we  
compare the stacked spectrum of all 318 LAEs in this sample with two
stacks of LBGs: the KBSS LBG \citep{ste14} sample, and another sample
from a survey for Lyman-continuum (LyC) emission in $z\sim 3$ galaxies
(C. Steidel et al., in prep.). A comparison of these two samples to
the KBSS-\lya\ LAE sample is given in Table \ref{table:lbgsamples}.

\begin{deluxetable*}{lcccccccccccc}
\tablecaption{Comparison of LAE and LBG Continuum and Absorption Properties}
\tablewidth{0pt}
\tablehead{
Sample & $\langle z \rangle$ & $m\sub{1450}$\tablenotemark{a} &
$M\sub{1450}$\tablenotemark{b} & 
$W\sub{\lya}$\tablenotemark{c} & $N\sub{obj}$ &
$W\sub{SiII}$\tablenotemark{d} & $v\sub{SiII}$\tablenotemark{e} &
$W\sub{CII}$\tablenotemark{d} & $v\sub{CII}$\tablenotemark{e} &
$W\sub{SiIV}$\tablenotemark{d} & $v\sub{SiIV}$\tablenotemark{e} &
$f_c$\tablenotemark{f}\\
& & (mag) & (mag) & (\AA) & & (\AA) & (km s$^{-1}$) &
(\AA) & (km s$^{-1}$) & (\AA) & (km s$^{-1}$) &
}

\startdata
KBSS-\lya & 2.70 & 27.0 & $-$18.3 & 42.7~ & 318 & 0.53 &
\pp$-$98 & 0.45 & $-$185 & 0.51 & $-$190 &0.3 \\  
KBSS LBGs & 2.36 & 24.4 & $-$20.7 & 27.2~ & \pp65 & 
1.67 & $-$224 & 1.35 & $-$201 & 1.98 & $-$276 & 0.6 \\ 
LyC LBGs & 3.07 & 24.6 & $-$21.0 & 21.5~ & \pp74 &
1.50 & $-$172 & 1.44 & $-$176 & 2.41 & $-$288 & 0.6
\enddata
\tablenotetext{a}{$m\sub{1450}$ is the observed AB magnitude
  corresponding to the rest-frame $\lambda \sim 1450$\AA\ UV
  continuum. At $z\sim 2.7$ (KBSS-\lya\ and KBSS LBGs), this
  correponds to the $G$ band, whereas the $\mathcal{R}$ magnitude is
  used for the LyC LBGs at $z\sim 3$. For consistency with the
  other samples, the LAE $m\sub{1450}$ value is not \lya-subtracted,
  though the \lya\ emission dominates the broadband flux in many cases.}
\tablenotetext{a}{$M\sub{1450}$ is the absolute magnitude
  corresponding to $m\sub{1450}$ at the typical redshift of the sample
  $\langle z \rangle$.}
\tablenotetext{c}{Because the LBG samples have no narrowband \lya\
  images, all three values of $W\sub{\lya}$ here were measured
  spectroscopically for consistency. Note that the spectroscopic LAE
  sample has a mean photometric equivalent width $W\sub{\lya}=85.6$\AA.} 
\tablenotetext{d}{$W\sub{SiII}$, $W\sub{CII}$, and $W\sub{SiIV}$ are
  the equivalent widths in absorption of \ion{Si}{2} $\lambda$1260,
  \ion{C}{2} $\lambda$1334, and \ion{Si}{4} $\lambda$1393+$\lambda$1402
  (combined), respectively. See Table~\ref{table:lines} and Fig.~\ref{fig:metal_abs}.} 
\tablenotetext{e}{$v\sub{SiII}$, $v\sub{CII}$, and $v\sub{SiIV}$ are
  the absorption-weighted velocity of \ion{Si}{2} $\lambda$1260,
  \ion{C}{2} $\lambda$1334, and \ion{Si}{4} $\lambda$1393+$\lambda$1402.}
\tablenotetext{f}{$f_c$ is the covering fraction of low-ionization gas implied
  by the depth of the saturated \ion{Si}{2} $\lambda$1260 absorption
  trough, which is consistent with the covering fraction
  implied by the depths of the \ion{O}{1} $\lambda$1304 and \ion{C}{2}
  $\lambda$1334 transitions in each composite spectrum.}
\label{table:lbgsamples}
\end{deluxetable*}

As discussed in Secs.~\ref{sublaes:specobs}~\&~\ref{sublaes:lyashift},
the KBSS LBG sample consists of
65 galaxies that each exhibit \lya\ emission in their UV spectra and
have accurate systemic redshifts measured from 
rest-frame optical emission lines. These rest-UV spectra were observed using
Keck/LRIS-B with the 600-line grism (the same used for the KBSS-\lya\
LAEs), providing a spectral resolution 
$R\sim1300$, or $\sigma\sub{inst} \approx 100$ km s$^{-1}$. There is significantly
stronger metal absorption in each of the spectral lines shown in
Figs.~\ref{fig:metal_abs}~\&~\ref{fig:abs_ions_lbg}, and 
the absorption profiles are both broader (extending to $v\approx
-1000$ km s$^{-1}$) and deeper ($\lesssim$0.5$\times$ the continuum
value) than the corresponding features in the LAE stack. The excited fine-structure and
nebular emission lines (Fig.~\ref{fig:laevslbg_abs}), however, are
well-matched to the corresponding LAE features, further validating
the redshift offset employed for the KBSS-\lya\ stack.

The LyC LBG sample consists of 74 galaxies, most of which do not
have measured nebular redshifts; the redshifts for this sample were estimated via a
combination of their \lya\ emission and interstellar absorption redshifts
with a  calibration based on the results of \citet{ste10} and
\citet{rak11}. Furthermore, the composite spectrum was checked after
stacking using stellar photospheric features. About 15\% of these galaxies
have redshifts derived solely from the \lya\ line and assume a 300 km
s$^{-1}$ offset (with respect to the \lya\ centroid) based on stacks
of the subset of objects with nebular emission redshifts. Like the LAE
and KBSS LBG comparison samples, only objects with
spectroscopically-detected \lya\ emission were included in the
sample.  The correspondence of the absorption and emission 
signatures of the stack to those of the KBSS LBG stack again
demonstrates the efficacy of the calibration based on UV
features. Being at higher redshifts than the other samples, these spectra were observed using
a combination of the LRIS-R 600-line grating (covering the spectrum
$\lambda\sub{rest}\gtrsim1200$\AA) and the LRIS-B 400-line grism (for
$\lambda\sub{rest}\lesssim1200$\AA). This lower resolution was
employed to maximize the S/N of the spectra shortward of \lya\ (and
the Lyman limit), where the observed spectrum becomes particularly
faint. These spectra also have significantly larger integration times
compared to the KBSS LBGs and KBSS-\lya\ LAEs (8$-$10 hours
vs. 1.5$-$2 hours). Additionally, the higher redshifts of this sample ($z\sim3$)
with repect to the KBSS LBG sample ($z\sim2.3$) cause the $1000\AA <
\lambda\sub{rest} < 1200\AA$ continuum to lie at observed wavelengths
with significantly higher LRIS-B throughput, further boosting the S/N
of the LyC LBG composite spectrum at these wavelengths with respect to that of the
KBSS LBGs. This stack is thus particularly useful for
comparing the absorption due to Ly$\beta$ and other EUV transitions
among the LBG and LAE samples (Fig.~\ref{fig:laevslbg_euv}).

The absorption profiles of the two LBG composite spectra
are quite similar (Fig.~\ref{fig:laevslbg_abs}). The velocity extent of the absorption is almost
identical in both samples ($v\sub{max} \approx -1000$ km
s$^{-1}$), though the LyC LBG stack shows slightly deeper absorption across all the lines.
This difference may signify a difference in the
galaxy properties: at $z\sim 3$, the LyC LBGs have a brighter
average rest-UV luminosity than those at $z\sim2.3$ (which have
similar apparent magnitudes). The total equivalent width in
interstellar absorption (Table \ref{table:lbgsamples}), which is
insensitive to the spectral resolution, is similar for the LyC and
KBSS LBG stacks. 

While the blending of \ion{Si}{2} $\lambda$1304 with \ion{O}{1}
$\lambda$1302 inhibits a direct measurement of the \ion{Si}{2} optical
depth, \citet{sha03} demonstrated that the \ion{Si}{2} $\lambda$1260 transition is
typically saturated in similar samples of bright LBGs, and the
absorption of the \ion{Si}{2} $\lambda$1190, $\lambda$1193, and $\lambda$1526
transitions in the LBG stacks are consistent with this
interpretation. Furthermore, the consistency of 
the absorption profile depths among the low-ionization species in the LBG
stacks suggests a gas covering fraction $f_c\sim 0.6$ for enriched,
neutral/low-ionization material for the two LBG samples. The observed line ratio of the
\ion{Si}{4} doublet is $W\sub{1393}/W\sub{1402}\sim 2$ for both LBG
stacks (as in \citealt{sha03}), suggesting optically-thin
absorption and $f_c\gtrsim 0.4$ for the ionized \ion{H}{2} gas. 

The difference in absorption line strength between the LBG and LAE samples
can likewise be seen in the Ly$\beta$ transition
(Fig.~\ref{fig:laevslbg_euv}). While the LBG composite is almost
completely absorbed at the Ly$\beta$ line, the LAE
spectrum has significantly higher transmission ($\gtrsim50$\%). The
continuum ratio $f\sub{1325}/f\sub{1125}$ (displayed as horizontal lines on
the left side of Fig.~\ref{fig:laevslbg_euv}) also varies
significantly between the LBG and LAE samples: the LBG composites have a
flux ratio $f\sub{1325}/f\sub{1125}\approx1.6$, while the LAE composite has
$f\sub{1325}/f\sub{1125}\approx1.3$. The flux ratio for the $z\sim 3$
LyC LBG composite is uncertain because the spectra at $\lambda\sub{rest}\gtrsim1200$\AA\ and $\lambda\sub{rest}\lesssim1200$\AA\ fall on
different detectors (LRIS-R and LRIS-B) within the spectrograph, so it
is not entirely clear how much of the observed variation in continuum
absorption is due to physical differences in the LAE and LBG galaxy populations,
rather than the changing transmissivity of the IGM with redshift
(e.g., \citealt{mad95}). Both effects are likely to be present, as the
UV continuum level blueward of \lya\ is particularly 
sensitive to the dust content and metallicity of the ISM (the latter
due to metal-line blanketing) and the age of the stellar population,
all of which are observed to be lower among LAEs than LBGs.

Through comparison of the LAE and LBG stacks, it seems clear that the
qualitative links between \lya\ emission and and interstellar
absorption discussed in \citet{sha03} persist down to far fainter
continuum luminosities than those probed by samples of LBGs. The outflow
velocity (represented by either the absorption-weighted mean or the
highest-velocity edge of absorption) decreases with increasing $W\sub{\lya}$
across the LBG and LAE subsamples, and the covering fraction of gas
drops steadily with increasing $W\sub{\lya}$ to at least the intermediate
subsample of LAEs ($W\sub{\lya}\sim 37$\AA). The offset of the \lya\ emission
centroid likewise decreases from the LBG samples (+300 km s$^{-1}$) to the
LAE stacks (+200 km s$^{-1}$). While further
observations and modeling are required to determine the physics of
star-formation driven outflows, we suggest that our results provide an
important link between the physical properties of these galaxies and those of
the gas in and around them. We provide some initial interpretation
of these results in the context of previous work below.

\section{Discussion}\label{laes:discussion}

As described above, the UV continuum
luminosities of these galaxies are closely linked to their
star-formation rates, and our brighter samples of galaxies (e.g., LBGs
and low-$W\sub{\lya}$ LAEs) are associated with higher velocity
dispersions and thus larger dynamical masses. Assuming that the
star-formation rate (SFR) and stellar mass ($M_*$) are roughly proportional to the
UV luminosity (i.e., the luminosity corresponding to $M\sub{1450}$ in
Table~\ref{table:lbgsamples}), the masses and star formation rates of
the KBSS LBGs likely exceed those of the LAEs by a typical factor
SFR\sub{LBG}$/$SFR\sub{LAE} $\sim$
$M_{*\mathrm{,LBG}}/M_{*\mathrm{,LAE}}$ $\sim$ 10. Similarly, the
measured nebular velocity dispersions $\sigma\sub{neb}$ are
$\gtrsim$2$\times$ larger for the LBG sample on average
(Sec.~\ref{sublaes:lbgemission}; Fig.~\ref{fig:signeb}), which likewise suggests
$M\sub{dyn,LBG}/M\sub{dyn,LAE} \approx \sigma\sub{LBG}^2
a\sub{LBG}/\sigma\sub{LAE}^2 a\sub{LAE}\sim 10$, given the slightly
larger physical sizes ($a$) of the typical KBSS LBGs \citep{erb06b}.

In comparison, various indicators of outflow velocity differ by a
factor $\sim$1.5$-$2 between the LAE and LBG samples: the \lya\ emission
centroid ($v\sub{LAE}\approx 200$ \kms, $v\sub{LBG}\approx 300$ \kms;
Figs.~\ref{fig:abs_ions}~\&~\ref{fig:abs_ions_lbg}), the absorption
centroid of low-ionization lines ($v\sub{LAE}\approx -100$ \kms,
$v\sub{LBG}\approx -200$ \kms; Table~\ref{table:lbgsamples}), and the
maximum velocity of blueshifted absorption ($v\sub{LAE}\approx 500$
\kms, $v\sub{LBG}\approx 800$ \kms;
Figs.~\ref{fig:abs_ions}~\&~\ref{fig:abs_ions_lbg}). This factor of
$\sim$1.5$-$2$\times$ in velocity over a $10\times$ range in
mass and star-formation rate is broadly consistent with the scaling seen in
samples of galaxies at at low and intermediate redshifts
\citep{mar05,rup05,wei09}, in which $v\sub{outflow}$ is measured to
scale as SFR$^{0.2-0.35}$. This positive scaling stands in contrast
to the lack of outflow velocity variation with either star-formation
rate or velocity dispersion observed by \citet{ste10}. In that paper,
the authors note that their LBG samples have star-formation rates
similar to the highest-SFR galaxies of \citet{mar05} and
\citet{rup05}, where the relationship between SFR and $v\sub{outflow}$
flattens off. Our LAEs are likely to have similar star-formation rates
to the low-SFR galaxies of the low-redshift samples (SFR $\lesssim 10$
\msun\ yr$^{-1}$), so our results may indicate that the relationship between
star-formation rate and outflow velocity (i.e., positive scaling at low SFR,
saturation at SFR $\gtrsim 10$\msun\ yr$^{-1}$) is similar at low
and high redshifts.

Analogously, the doubling of $\sigma\sub{neb}$ between the LAE and LBG
samples here suggests that outflow velocity grows in a roughly linear
(or marginally sub-linear) fashion with the stellar dynamics of the
galaxies. Similar scaling is observed in the lower-redshift samples
noted above. Furthermore, if the escape speed scales as
$v\sub{esc}\approx 5-6\sigma\sub{neb}$ (as discussed in
\citealt{wei09}), then the escape velocities of the LAEs are
$v\sub{esc} \approx 200$. This velocity is of order the typical
absorption-weighted centroid velocity of the outflows (and significantly less
than the maximum velocity of absorption), suggesting that a
significant fraction of outflowing gas is permanantly ejected from the
galaxy halos (similar to the lower-redshift results of \citealt{wei09}). More
detailed measurements of the masses and 
star-formation rates of individual LAEs are required to probe the
scaling of these feedback processes within the sample of low-mass, high-redshift galaxies.



It may seem
paradoxical that the centroid or width of the \lya\ line 
is linked to outflow velocity and/or the star-formation rate in a
galaxy while the \lya\ peak velocity at fixed resolution shows no
dependence on continuum luminosity. In particular,
Figs.~\ref{fig:abs_ions}~\&~\ref{fig:abs_ions_lbg} seem to show a
decoupling between the position of the narrow \lya\ peak and the
breadth of the extended \lya\ emission tails to positive and
negative velocities. We suggest that previous studies of \lya\ emission
actually show a similar decoupling between the \lya\ peak and the
width of the overall velocity distribution of \lya\ flux. The
radiative-transfer models employed by \citet{kul12} and \citet{cho13}
seem to show that the dominant red peak of \lya\ transmission through an
expanding spherical shell of gas does not sensitively depend on gas
velocity (at least for $v\sub{gas}\lesssim 300$ km s$^{-1}$), but the
distribution of flux in secondary and tertiary red 
and blue peaks becomes more extended with increasing gas
velocity. These narrow emission line features are observed
in higher-resolution spectra of bright and nearby galaxies
\citep{ver08,cho13,mar15} and may be useful for modeling the effects of
radiative transfer and gas velocity directly, but they are not
feasibly observable in $L\sim0.1L_*$ galaxies at $z\gtrsim 2$. We
therefore suggest that the observed uniformity of the \lya\ peak offset
is due to the absorbing presence of {\it some} neutral gas at $v=0$ in all of
these star-forming galaxies (which sets the minimum velocity shift for
an escaping \lya\ photon), whereas the emission at larger (positive
or negative) velocities reflects those photons scattered off of the
high-velocity outflowing gas (which depends more strongly on galaxy properties).
The broad wings of the \lya\ emission line (e.g.,
Fig.~\ref{fig:abs_ions}) are therefore likely to
be populated by scattering off of outflowing gas, and thus are
sensitive to the gas velocity and show a velocity distribution similar
to that of the extended metal absorption. 


However, the
relatively low measured covering fractions of these outflows and the
consistency of the \lya\ peak offset may suggest 
that the majority of \lya\ scatterings in LAEs occur without incurring
any single large kinematic ``kick'' from an interaction with the
highest-velocity gas. This tendency
would be likely in a scenario where galaxy-scale outflows are highly
anisotropic and thus create an orientation-dependent model for \lya\
scattering and escape. The anisotropy of galaxy-scale outflows is highly
favored by simulations (e.g., \citealt{hop12}) and lower-redshift
observations (e.g., \citealt{kor12} at $z\sim1$), both of which suggest that 
outflows dominate along the axis perpindicular to galactic disks, but
it is not clear whether the anisotropy of the outflows and that of \lya\ emissivity
coincide. \citet{ver12} suggest that the \lya\ flux and equivalent
width are maximized when a galaxy is observed face-on. However
this model would naively predict a positive correlation between \lya\
velocity offset and equivalent width (rather than the anti- or absence of
correlation observed) because both the \lya\ emission and the
line-of-sight outflow velocity would be maximized for face-on galaxy disks.

Local star-forming galaxies can help probe the geometry of \lya\
escape. In particular, the \lya\ Reference Sample (LARS;
\citealt{hay13,ost14}) consists of 14 local galaxies with {\it
HST} imaging and spectroscopy of their \lya\ emission. This sample was
selected to provide a set of local analogs to high-redshift LAEs that
can be observed at much higher spatial and spectral
resolution. \citet{riv15} use {\it HST}/COS spectra to analyze \lya\ emission
along with absorption due to \ion{Si}{2} and \ion{Si}{4} along the
line of sight to the brightest star-forming regions within these
galaxies, finding several results broadly consistent with those of this
paper. \citeauthor{riv15} use four spectrally-resolved \ion{Si}{2} transitions to 
show that the typical absorbing medium is optically thick, with a
covering fraction that depends on both position and velocity. As in
this paper, they find that higher local
(and global) \lya\ transmissivity is associated with low covering fractions
of \ion{Si}{2}, the \ion{Si}{2}/\ion{Si}{4} absorption ratio is high at $v=0$,
and the resolved \lya\ emission peak shows little dependence on
outflow velocity. \footnote{In contrast, \citet{hen15} argue that
 the covering fraction of neutral gas does not dominate \lya\ escape
 in low-redshift, lower-mass galaxies, basing their claim on the
 appearance of Lyman-series absorption lines with zero residual
 transmission in COS spectra of \lya-emitting Green Pea galaxies
 (GPs; \citealt{car09}). However, the KBSS LAEs show significant transmission
 at Ly$\beta$ (Fig.~\ref{fig:laevslbg_euv}), suggesting that the
 gaseous environments of GPs are qualitatively different from those of
$z\sim2-3$ LAEs despite their similarly low masses and compact
sizes.} The authors interpret these results in light of a 
model where \lya\ escapes through an expanding medium that is
both porous and non-uniform in velocity, with inhomogeneities
potentially driven by Rayleigh-Taylor instabilities in expanding
bubbles. \citet{hay14} find that the \lya\ emission is spatially 
extended with respect to the FUV continuum emission (in agreement with
high-redshift studies); the local \lya/\ha\ ratio is higher at larger
galacto-centric distances. In some edge-on disks, \lya/\ha\ is seen to
peak above and below the disk plane, but in most LARS galaxies the
ratio appears to vary with no simple dependence on galaxy geometry or
structure. 

Furthermore, \citet{law09} find that most galaxies at $z\sim2-3$ are 
not dynamically dominated by rotation and thus are less likely than low-redshift
galaxies to generate coherent, collimated outflows. The young, faint
populations of LAEs considered in this paper may be even less likely to maintain
such structural coherence. Therefore, the KBSS-\lya\ LAEs may be best
modeled by irregular morphologies in which \lya\ photons escape
through a patchy ISM with minimal planar symmetry. Such a model, in
which \lya\ photons escape through ``holes'' in the neutral gas
distribution, would similarly explain the possible association between
\lya\ and LyC emissivity (e.g., \citealt{mos13}).

LBGs and
other bright star-forming galaxies can have extremely low \lya\ and
LyC escape fractions even when low-resolution spectroscopy suggests
that the gas covering fractions are $f_c<1$; in many cases, there may be
extremely narrow, unresolved absorption components 
covering the line of sight that are still capable of absorbing all of
the incident LyC flux (e.g.,the galaxy Q0000-D6, as discussed by
\citealt{sha03} and \citealt{gia02}). Opening up lines of sight
through the neutral ISM that are transparent to both \lya\ and LyC photons
may be easier for low-mass galaxies.  


Other mechanisms of \lya\ escape could lead to decoupling between the
\lya\ and LyC emissivity of galaxies. While the scattering of \lya\
photons is driven by \ion{H}{1}, 
the eventual absorption and destruction of these photons is dominated
by dust. If the LAEs contain little interstellar dust and have velocity fields
capable of quickly scattering \lya\ photons out of resonance, then a
large fraction of \lya\ photons may escape before encountering a
dust particle (without requiring a ``hole'' in the \ion{H}{1}
distribution). If \lya\ escape is dominated by this channel, then
the \lya\ and LyC escape fractions may be less strongly correlated, given
that LyC photons are absorbed by \ion{H}{1} rather than merely
scattered.  The apparent breakdown of the correlations of galaxy 
properties with $W\sub{\lya}$ for the highest-$W\sub{\lya}$ bin may
indicate the transition to such a regime where 
dust formation and mixing into the ISM has not fully progressed. The
faintest LAEs at $z\sim2-3$ may be most analogous to those observed at
$z\sim6$, where the extremely blue UV slopes suggest that escaping
photons encounter very little dust while exiting the galaxy. More
robust fitting of the LAE spectral energy distributions, including
deep NIR images from MOSFIRE, could strongly constrain the dust
content of these young galaxies. Similarly, deep NIR spectra may
constrain the metallicity (and thus the chemistry) of their ISM. 

Another possibility is that the decoupling between \lya\ emission and galaxy properties
in this sample at the highest values of $W\sub{\lya}$ may result from
the contribution of the QSO ionizing radiation field to their \lya\
flux. For fluorescently-illuminated 
systems, the total \lya\ flux is proportional to the surface area of
optically-thick \ion{H}{1} and the strength of the local QSO-generated
ionizing field \citep{can05,kol10}, with minimal dependence on the
kinematic or chemical properties of the galaxies. Our forthcoming
analysis of the {\it HST} NIR and optical images of these fields will
shed more light on the physical sizes, morphologies, and stellar
components of the \lya-emitting regions of these galaxies.




%

\section{Conclusions} \label{laes:conclusions}

This paper has presented the rest-frame UV spectral properties
of \lya-emitters near hyperluminous QSOs at $z\sim 2.7$ from the
KBSS-\lya\ survey. These LAEs 
are significantly fainter ($L\sim 0.1 L_*$) than previous
spectroscopic samples of high-redshift star-forming galaxies, which
enables us to extend observations of interstellar enrichment and
kinematics to a new galaxy regime. Our primary results are as follows:

\begin{itemize}[leftmargin=*]
\item The faint LAEs have \lya\ escape fractions
  $f\sub{esc,\lya}\approx 30\%$ (based on the observed \lya\ to \ha\
  flux ratio), significantly higher than previous observations of brighter
  star-forming galaxies at high-redshift
  (Sec.~\ref{sublaes:lyaescape}). This result is consistent with the
  conclusions of \citet{ste11}.
\item Like continuum-bright galaxies at similar redshifts, LAEs
  produce \lya\ emission that is significantly shifted from the
  systemic redshift. While the flux-weighted \lya\ emission centroid
  varies with galaxy luminosity across our combined sample of LAEs and
  LBGs, the \lya\ peak (when observed with $R\approx 1300$)
  displays a highly consistent shift of $\sim$200 km s$^{-1}$
  irrespective of continuum luminosity. This offset may be used to
  estimate redshifts of diverse \lya-emitting galaxies observed at
  comparable resolution when rest-frame
  optical spectroscopy is unavailable (Sec.~\ref{sublaes:lyashift}).
\item Faint LAEs display a range of diverse \lya\ morphologies and are associated with
  narrower \lya\ peaks with smaller peak separations than
  those of LBGs. These results are consistent with smaller velocities and/or
  column densities of neutral gas in the circumgalactic medium of LAEs
  compared to
  LBGs. While most of the \lya\ emission lines have a characteristic
  red-dominant morphology, there is a minority (13\%, 4 objects) with
  blueshifted primary peaks (Secs.~\ref{sublaes:lyashift}$-$\ref{sublaes:lbgemission}).
\item By combining the UV continuum spectra of all 318 LAEs in our
  spectroscopic sample, we find
  absorption signatures of several metal ions that have been shown to trace
  outflowing material in more luminous star-forming galaxies. We
  believe these
  are the first measurements of metal-enriched outflowing gas in such
  faint high-redshift galaxies. We
  measure absorption-weighted velocities of $\sim100-200$ km s$^{-1}$
  and maximum velocities of $\sim500$ km s$^{-1}$, both significantly
  smaller than those seen in comparison samples of LBGs (Sec.~\ref{laes:stacks}).
\item Using multiple transitions of \ion{Si}{2}, we infer the
  saturation of the \ion{Si}{2} $\lambda$1260 transition, thereby
  constraining the covering fraction of low-ionization material in
  front of the LAEs to $f_c\approx0.3$. This covering fraction is
  significantly smaller than that observed in LBGs, with $f_c\sim
  0.5-0.7$ in the same ions. This low covering fraction of obscuring
  gas likely drives the observed high escape fractions of \lya\ and
  Ly-continuum photons from faint galaxies (Sec.~\ref{sublaes:abs}).
\item We find that outflow velocity -- whether probed by maximum velocity of
  metal absorption, absorption-weighted mean velocity, or \lya\
  emission morphology -- increases with continuum luminosity
  (decreases with \lya\ equivalent width) among LAEs and with respect
  to LBGs. This result suggests that these gas outflows are tied to
  physical properties of their associated galaxies -- e.g., the
  mass and star-formation rate, for which the nebular velocity dispersion and
  continuum luminosity are proxies. In particular, our results are
  consistent with weak dependence of outflow velocity on
  star-formation rate ($v\sub{outflow}\sim$ SFR$^{0.25}$), and we find
  that the outflow velocities (up to 500 \kms) are large compared to
  the probable $\sim$200 \kms\ escape velocities of the LAEs
  (Sec.~\ref{laes:discussion}). 
\end{itemize}

In sum, these observations demonstrate that stellar feedback is an
essential ingredient in realistic simulations of even the youngest,
least massive galaxies at high redshift. In particular, these feedback
processes are key to understanding the escape of line and continuum
photons (particularly ionizing photons) from faint galaxies and their
effect on their surrounding IGM. Further work is necessary to 
analyze the apparent absence of correlations between $W\sub{\lya}$
and covering fraction of ISM absorption at the highest equivalent
widths. This effect may be the result of a qualitative transition in
the properties of galaxies to those with minimal enrichment and dust,
and could also point to the effect of the QSO in picking out galaxies
independently of their star-formation rates.

In each of these topics, further progress will be driven by 
continued near-IR observations of faint galaxy populations. Rest-frame
optical spectra (in combination with observations in the rest-UV) are
the only way to effectively constrain the star-formation rates, metal
enrichment, and gas kinematics of these newborn galaxies, thereby
establishing the relationships between their stellar and gaseous 
properties and characterizing evolution of the baryon cycle over
cosmic time. 

\acknowledgments

\noindent We are indebted to the staff 
of the W.M. Keck Observatory who keep the 
instruments and telescopes running effectively. We also wish to extend
thanks to those of Hawaiian ancestry on whose sacred mountain we are
privileged to be guests. This work has been supported in part by the US
National Science Foundation through grants AST-0908805 and AST-1313472. We
thank Mariska Kriek and Eliot Quataert for many useful discussions
as well as the anonymous referee for comments and suggestions that
significantly improved this work. RFT also acknowledges support from
Dennis and Carol Troesh and from the Miller Institute for Basic
Research in Science at the University of California, Berkeley.

\bibliographystyle{apj}
\bibliography{thesis}

\begin{thebibliography}{95}
\expandafter\ifx\csname natexlab\endcsname\relax\def\natexlab#1{#1}\fi

\bibitem[{{Adelberger} {et~al.}(2005){Adelberger}, {Shapley}, {Steidel},
  {Pettini}, {Erb}, \& {Reddy}}]{ade05d}
{Adelberger}, K.~L., {Shapley}, A.~E., {Steidel}, C.~C., {Pettini}, M., {Erb},
  D.~K., \& {Reddy}, N.~A. 2005, \apj, 629, 636

\bibitem[{{Adelberger} {et~al.}(2006){Adelberger}, {Steidel}, {Kollmeier}, \&
  {Reddy}}]{ade06}
{Adelberger}, K.~L., {Steidel}, C.~C., {Kollmeier}, J.~A., \& {Reddy}, N.~A.
  2006, \apj, 637, 74

\bibitem[{{Adelberger} {et~al.}(2003){Adelberger}, {Steidel}, {Shapley}, \&
  {Pettini}}]{ade03}
{Adelberger}, K.~L., {Steidel}, C.~C., {Shapley}, A.~E., \& {Pettini}, M. 2003,
  \apj, 584, 45

\bibitem[{{Aguirre} {et~al.}(2001){Aguirre}, {Hernquist}, {Schaye}, {Katz},
  {Weinberg}, \& {Gardner}}]{agu01}
{Aguirre}, A., {Hernquist}, L., {Schaye}, J., {Katz}, N., {Weinberg}, D.~H., \&
  {Gardner}, J. 2001, \apj, 561, 521

\bibitem[{{Ando} {et~al.}(2006){Ando}, {Ohta}, {Iwata}, {Akiyama}, {Aoki}, \&
  {Tamura}}]{and06}
{Ando}, M., {Ohta}, K., {Iwata}, I., {Akiyama}, M., {Aoki}, K., \& {Tamura}, N.
  2006, \apjl, 645, L9

\bibitem[{{Andrews} \& {Martini}(2013)}]{and13}
{Andrews}, B.~H. \& {Martini}, P. 2013, \apj, 765, 140

\bibitem[{{Atek} {et~al.}(2014){Atek}, {Kunth}, {Schaerer}, {Mas-Hesse},
  {Hayes}, {{\"O}stlin}, \& {Kneib}}]{ate14}
{Atek}, H., {Kunth}, D., {Schaerer}, D., {Mas-Hesse}, J.~M., {Hayes}, M.,
  {{\"O}stlin}, G., \& {Kneib}, J.-P. 2014, \aap, 561, A89

\bibitem[{{Birnboim} \& {Dekel}(2003)}]{bir03}
{Birnboim}, Y. \& {Dekel}, A. 2003, \mnras, 345, 349

\bibitem[{{Brocklehurst}(1971)}]{bro71}
{Brocklehurst}, M. 1971, \mnras, 153, 471

\bibitem[{{Cantalupo} {et~al.}(2012){Cantalupo}, {Lilly}, \&
  {Haehnelt}}]{can12}
{Cantalupo}, S., {Lilly}, S.~J., \& {Haehnelt}, M.~G. 2012, \mnras, 425, 1992

\bibitem[{{Cantalupo} {et~al.}(2007){Cantalupo}, {Lilly}, \&
  {Porciani}}]{can07}
{Cantalupo}, S., {Lilly}, S.~J., \& {Porciani}, C. 2007, \apj, 657, 135

\bibitem[{{Cantalupo} {et~al.}(2005){Cantalupo}, {Porciani}, {Lilly}, \&
  {Miniati}}]{can05}
{Cantalupo}, S., {Porciani}, C., {Lilly}, S.~J., \& {Miniati}, F. 2005, \apj,
  628, 61

\bibitem[{{Cardamone} {et~al.}(2009){Cardamone}, {Schawinski}, {Sarzi},
  {Bamford}, {Bennert}, {Urry}, {Lintott}, {Keel}, {Parejko}, {Nichol},
  {Thomas}, {Andreescu}, {Murray}, {Raddick}, {Slosar}, {Szalay}, \&
  {Vandenberg}}]{car09}
{Cardamone}, C., {Schawinski}, K., {Sarzi}, M., {Bamford}, S.~P., {Bennert},
  N., {Urry}, C.~M., {Lintott}, C., {Keel}, W.~C., {Parejko}, J., {Nichol},
  R.~C., {Thomas}, D., {Andreescu}, D., {Murray}, P., {Raddick}, M.~J.,
  {Slosar}, A., {Szalay}, A., \& {Vandenberg}, J. 2009, \mnras, 399, 1191

\bibitem[{{Chonis} {et~al.}(2013){Chonis}, {Blanc}, {Hill}, {Adams},
  {Finkelstein}, {Gebhardt}, {Kollmeier}, {Ciardullo}, {Drory}, {Gronwall},
  {Hagen}, {Overzier}, {Song}, \& {Zeimann}}]{cho13}
{Chonis}, T.~S., {Blanc}, G.~A., {Hill}, G.~J., {Adams}, J.~J., {Finkelstein},
  S.~L., {Gebhardt}, K., {Kollmeier}, J.~A., {Ciardullo}, R., {Drory}, N.,
  {Gronwall}, C., {Hagen}, A., {Overzier}, R.~A., {Song}, M., \& {Zeimann},
  G.~R. 2013, \apj, 775, 99

\bibitem[{{Ciardullo} {et~al.}(2013){Ciardullo}, {Gronwall}, {Adams}, {Blanc},
  {Gebhardt}, {Finkelstein}, {Jogee}, {Hill}, {Drory}, {Hopp}, {Schneider},
  {Zeimann}, \& {Dalton}}]{cia13}
{Ciardullo}, R., {Gronwall}, C., {Adams}, J.~J., {Blanc}, G.~A., {Gebhardt},
  K., {Finkelstein}, S.~L., {Jogee}, S., {Hill}, G.~J., {Drory}, N., {Hopp},
  U., {Schneider}, D.~P., {Zeimann}, G.~R., \& {Dalton}, G.~B. 2013, \apj, 769,
  83

\bibitem[{{Ciardullo} {et~al.}(2014){Ciardullo}, {Zeimann}, {Gronwall},
  {Gebhardt}, {Schneider}, {Hagen}, {Malz}, {Blanc}, {Hill}, {Drory}, \&
  {Gawiser}}]{cia14}
{Ciardullo}, R., {Zeimann}, G.~R., {Gronwall}, C., {Gebhardt}, H., {Schneider},
  D.~P., {Hagen}, A., {Malz}, A.~I., {Blanc}, G.~A., {Hill}, G.~J., {Drory},
  N., \& {Gawiser}, E. 2014, \apj, 796, 64

\bibitem[{{Dijkstra} {et~al.}(2006){Dijkstra}, {Haiman}, \& {Spaans}}]{dij06}
{Dijkstra}, M., {Haiman}, Z., \& {Spaans}, M. 2006, \apj, 649, 14

\bibitem[{{Dopita} \& {Sutherland}(2003)}]{dop03}
{Dopita}, M.~A. \& {Sutherland}, R.~S. 2003, Astrophysics of the diffuse
  universe, Berlin, New York: Springer, 2003.~Astronomy and astrophysics
  library, ISBN 3540433627

\bibitem[{{Dressler} {et~al.}(2014){Dressler}, {Henry}, {Martin}, {Sawicki},
  {McCarthy}, \& {Villaneuva}}]{dre14}
{Dressler}, A., {Henry}, A., {Martin}, C.~L., {Sawicki}, M., {McCarthy}, P., \&
  {Villaneuva}, E. 2014, ArXiv e-prints

\bibitem[{{Erb} {et~al.}(2006{\natexlab{a}}){Erb}, {Shapley}, {Pettini},
  {Steidel}, {Reddy}, \& {Adelberger}}]{erb06a}
{Erb}, D.~K., {Shapley}, A.~E., {Pettini}, M., {Steidel}, C.~C., {Reddy},
  N.~A., \& {Adelberger}, K.~L. 2006{\natexlab{a}}, \apj, 644, 813

\bibitem[{{Erb} {et~al.}(2006{\natexlab{b}}){Erb}, {Steidel}, {Shapley},
  {Pettini}, {Reddy}, \& {Adelberger}}]{erb06b}
{Erb}, D.~K., {Steidel}, C.~C., {Shapley}, A.~E., {Pettini}, M., {Reddy},
  N.~A., \& {Adelberger}, K.~L. 2006{\natexlab{b}}, \apj, 646, 107

\bibitem[{{Erb} {et~al.}(2014){Erb}, {Steidel}, {Trainor}, {Bogosavljevi{\'c}},
  {Shapley}, {Nestor}, {Kulas}, {Law}, {Strom}, {Rudie}, {Reddy}, {Pettini},
  {Konidaris}, {Mace}, {Matthews}, \& {McLean}}]{erb14}
{Erb}, D.~K., {Steidel}, C.~C., {Trainor}, R.~F., {Bogosavljevi{\'c}}, M.,
  {Shapley}, A.~E., {Nestor}, D.~B., {Kulas}, K.~R., {Law}, D.~R., {Strom},
  A.~L., {Rudie}, G.~C., {Reddy}, N.~A., {Pettini}, M., {Konidaris}, N.~P.,
  {Mace}, G., {Matthews}, K., \& {McLean}, I.~S. 2014, \apj, 795, 33

\bibitem[{{Gawiser} {et~al.}(2006){Gawiser}, {van Dokkum}, {Gronwall},
  {Ciardullo}, {Blanc}, {Castander}, {Feldmeier}, {Francke}, {Franx},
  {Haberzettl}, {Herrera}, {Hickey}, {Infante}, {Lira}, {Maza}, {Quadri},
  {Richardson}, {Schawinski}, {Schirmer}, {Taylor}, {Treister}, {Urry}, \&
  {Virani}}]{gaw06}
{Gawiser}, E., {van Dokkum}, P.~G., {Gronwall}, C., {Ciardullo}, R., {Blanc},
  G.~A., {Castander}, F.~J., {Feldmeier}, J., {Francke}, H., {Franx}, M.,
  {Haberzettl}, L., {Herrera}, D., {Hickey}, T., {Infante}, L., {Lira}, P.,
  {Maza}, J., {Quadri}, R., {Richardson}, A., {Schawinski}, K., {Schirmer}, M.,
  {Taylor}, E.~N., {Treister}, E., {Urry}, C.~M., \& {Virani}, S.~N. 2006,
  \apjl, 642, L13

\bibitem[{{Giallongo} {et~al.}(2002){Giallongo}, {Cristiani}, {D'Odorico}, \&
  {Fontana}}]{gia02}
{Giallongo}, E., {Cristiani}, S., {D'Odorico}, S., \& {Fontana}, A. 2002,
  \apjl, 568, L9

\bibitem[{{Giavalisco} {et~al.}(1996){Giavalisco}, {Koratkar}, \&
  {Calzetti}}]{gia96}
{Giavalisco}, M., {Koratkar}, A., \& {Calzetti}, D. 1996, \apj, 466, 831

\bibitem[{{Hashimoto} {et~al.}(2013){Hashimoto}, {Ouchi}, {Shimasaku}, {Ono},
  {Nakajima}, {Rauch}, {Lee}, \& {Okamura}}]{has13}
{Hashimoto}, T., {Ouchi}, M., {Shimasaku}, K., {Ono}, Y., {Nakajima}, K.,
  {Rauch}, M., {Lee}, J., \& {Okamura}, S. 2013, \apj, 765, 70

\bibitem[{{Hayes} {et~al.}(2014){Hayes}, {{\"O}stlin}, {Duval}, {Sandberg},
  {Guaita}, {Melinder}, {Adamo}, {Schaerer}, {Verhamme}, {Orlitov{\'a}},
  {Mas-Hesse}, {Cannon}, {Atek}, {Kunth}, {Laursen}, {Ot{\'{\i}}-Floranes},
  {Pardy}, {Rivera-Thorsen}, \& {Herenz}}]{hay14}
{Hayes}, M., {{\"O}stlin}, G., {Duval}, F., {Sandberg}, A., {Guaita}, L.,
  {Melinder}, J., {Adamo}, A., {Schaerer}, D., {Verhamme}, A., {Orlitov{\'a}},
  I., {Mas-Hesse}, J.~M., {Cannon}, J.~M., {Atek}, H., {Kunth}, D., {Laursen},
  P., {Ot{\'{\i}}-Floranes}, H., {Pardy}, S., {Rivera-Thorsen}, T., \&
  {Herenz}, E.~C. 2014, \apj, 782, 6

\bibitem[{{Hayes} {et~al.}(2010){Hayes}, {{\"O}stlin}, {Schaerer}, {Mas-Hesse},
  {Leitherer}, {Atek}, {Kunth}, {Verhamme}, {de Barros}, \& {Melinder}}]{hay10}
{Hayes}, M., {{\"O}stlin}, G., {Schaerer}, D., {Mas-Hesse}, J.~M., {Leitherer},
  C., {Atek}, H., {Kunth}, D., {Verhamme}, A., {de Barros}, S., \& {Melinder},
  J. 2010, \nat, 464, 562

\bibitem[{{Hayes} {et~al.}(2013){Hayes}, {{\"O}stlin}, {Schaerer}, {Verhamme},
  {Mas-Hesse}, {Adamo}, {Atek}, {Cannon}, {Duval}, {Guaita}, {Herenz}, {Kunth},
  {Laursen}, {Melinder}, {Orlitov{\'a}}, {Ot{\'{\i}}-Floranes}, \&
  {Sandberg}}]{hay13}
{Hayes}, M., {{\"O}stlin}, G., {Schaerer}, D., {Verhamme}, A., {Mas-Hesse},
  J.~M., {Adamo}, A., {Atek}, H., {Cannon}, J.~M., {Duval}, F., {Guaita}, L.,
  {Herenz}, E.~C., {Kunth}, D., {Laursen}, P., {Melinder}, J., {Orlitov{\'a}},
  I., {Ot{\'{\i}}-Floranes}, H., \& {Sandberg}, A. 2013, \apjl, 765, L27

\bibitem[{{Henry} {et~al.}(2015){Henry}, {Scarlata}, {Martin}, \&
  {Erb}}]{hen15}
{Henry}, A., {Scarlata}, C., {Martin}, C.~L., \& {Erb}, D. 2015, ArXiv e-prints

\bibitem[{{Hogan} \& {Weymann}(1987)}]{hog87}
{Hogan}, C.~J. \& {Weymann}, R.~J. 1987, \mnras, 225, 1P

\bibitem[{{Hogg} {et~al.}(1998){Hogg}, {Cohen}, {Blandford}, \&
  {Pahre}}]{hog98}
{Hogg}, D.~W., {Cohen}, J.~G., {Blandford}, R., \& {Pahre}, M.~A. 1998, \apj,
  504, 622

\bibitem[{{Hopkins} {et~al.}(2014){Hopkins}, {Kere{\v s}}, {O{\~n}orbe},
  {Faucher-Gigu{\`e}re}, {Quataert}, {Murray}, \& {Bullock}}]{hop14}
{Hopkins}, P.~F., {Kere{\v s}}, D., {O{\~n}orbe}, J., {Faucher-Gigu{\`e}re},
  C.-A., {Quataert}, E., {Murray}, N., \& {Bullock}, J.~S. 2014, \mnras, 445,
  581

\bibitem[{{Hopkins} {et~al.}(2012){Hopkins}, {Quataert}, \& {Murray}}]{hop12}
{Hopkins}, P.~F., {Quataert}, E., \& {Murray}, N. 2012, \mnras, 421, 3522

\bibitem[{{Hummels} {et~al.}(2013){Hummels}, {Bryan}, {Smith}, \&
  {Turk}}]{hum13}
{Hummels}, C.~B., {Bryan}, G.~L., {Smith}, B.~D., \& {Turk}, M.~J. 2013,
  \mnras, 430, 1548

\bibitem[{{Jones} {et~al.}(2012){Jones}, {Stark}, \& {Ellis}}]{jon12}
{Jones}, T., {Stark}, D.~P., \& {Ellis}, R.~S. 2012, \apj, 751, 51

\bibitem[{{Jones} {et~al.}(2013){Jones}, {Ellis}, {Schenker}, \&
  {Stark}}]{jon13}
{Jones}, T.~A., {Ellis}, R.~S., {Schenker}, M.~A., \& {Stark}, D.~P. 2013,
  \apj, 779, 52

\bibitem[{{Kennicutt}(1998)}]{ken98}
{Kennicutt}, Jr., R.~C. 1998, \araa, 36, 189

\bibitem[{{Kere{\v s}} {et~al.}(2009){Kere{\v s}}, {Katz}, {Dav{\'e}},
  {Fardal}, \& {Weinberg}}]{ker09}
{Kere{\v s}}, D., {Katz}, N., {Dav{\'e}}, R., {Fardal}, M., \& {Weinberg},
  D.~H. 2009, \mnras, 396, 2332

\bibitem[{{Kere{\v s}} {et~al.}(2005){Kere{\v s}}, {Katz}, {Weinberg}, \&
  {Dav{\'e}}}]{ker05}
{Kere{\v s}}, D., {Katz}, N., {Weinberg}, D.~H., \& {Dav{\'e}}, R. 2005,
  \mnras, 363, 2

\bibitem[{{Kollmeier} {et~al.}(2010){Kollmeier}, {Zheng}, {Dav{\'e}}, {Gould},
  {Katz}, {Miralda-Escud{\'e}}, \& {Weinberg}}]{kol10}
{Kollmeier}, J.~A., {Zheng}, Z., {Dav{\'e}}, R., {Gould}, A., {Katz}, N.,
  {Miralda-Escud{\'e}}, J., \& {Weinberg}, D.~H. 2010, \apj, 708, 1048

\bibitem[{{Kornei} {et~al.}(2010){Kornei}, {Shapley}, {Erb}, {Steidel},
  {Reddy}, {Pettini}, \& {Bogosavljevi{\'c}}}]{kor10}
{Kornei}, K.~A., {Shapley}, A.~E., {Erb}, D.~K., {Steidel}, C.~C., {Reddy},
  N.~A., {Pettini}, M., \& {Bogosavljevi{\'c}}, M. 2010, \apj, 711, 693

\bibitem[{{Kornei} {et~al.}(2012){Kornei}, {Shapley}, {Martin}, {Coil}, {Lotz},
  {Schiminovich}, {Bundy}, \& {Noeske}}]{kor12}
{Kornei}, K.~A., {Shapley}, A.~E., {Martin}, C.~L., {Coil}, A.~L., {Lotz},
  J.~M., {Schiminovich}, D., {Bundy}, K., \& {Noeske}, K.~G. 2012, \apj, 758,
  135

\bibitem[{{Kulas} {et~al.}(2012){Kulas}, {Shapley}, {Kollmeier}, {Zheng},
  {Steidel}, \& {Hainline}}]{kul12}
{Kulas}, K.~R., {Shapley}, A.~E., {Kollmeier}, J.~A., {Zheng}, Z., {Steidel},
  C.~C., \& {Hainline}, K.~N. 2012, \apj, 745, 33

\bibitem[{{Kunth} {et~al.}(1998){Kunth}, {Mas-Hesse}, {Terlevich}, {Terlevich},
  {Lequeux}, \& {Fall}}]{kun98}
{Kunth}, D., {Mas-Hesse}, J.~M., {Terlevich}, E., {Terlevich}, R., {Lequeux},
  J., \& {Fall}, S.~M. 1998, \aap, 334, 11

\bibitem[{{Law} {et~al.}(2009){Law}, {Steidel}, {Erb}, {Larkin}, {Pettini},
  {Shapley}, \& {Wright}}]{law09}
{Law}, D.~R., {Steidel}, C.~C., {Erb}, D.~K., {Larkin}, J.~E., {Pettini}, M.,
  {Shapley}, A.~E., \& {Wright}, S.~A. 2009, \apj, 697, 2057

\bibitem[{{Madau}(1995)}]{mad95}
{Madau}, P. 1995, \apj, 441, 18

\bibitem[{{Martin}(2005)}]{mar05}
{Martin}, C.~L. 2005, \apj, 621, 227

\bibitem[{{Martin} {et~al.}(2015){Martin}, {Dijkstra}, {Henry}, {Soto},
  {Danforth}, \& {Wong}}]{mar15}
{Martin}, C.~L., {Dijkstra}, M., {Henry}, A., {Soto}, K.~T., {Danforth}, C.~W.,
  \& {Wong}, J. 2015, \apj, 803, 6

\bibitem[{{Martin} {et~al.}(2010){Martin}, {Scannapieco}, {Ellison}, {Hennawi},
  {Djorgovski}, \& {Fournier}}]{mar10}
{Martin}, C.~L., {Scannapieco}, E., {Ellison}, S.~L., {Hennawi}, J.~F.,
  {Djorgovski}, S.~G., \& {Fournier}, A.~P. 2010, \apj, 721, 174

\bibitem[{{Martin} {et~al.}(2012){Martin}, {Shapley}, {Coil}, {Kornei},
  {Bundy}, {Weiner}, {Noeske}, \& {Schiminovich}}]{mar12}
{Martin}, C.~L., {Shapley}, A.~E., {Coil}, A.~L., {Kornei}, K.~A., {Bundy}, K.,
  {Weiner}, B.~J., {Noeske}, K.~G., \& {Schiminovich}, D. 2012, \apj, 760, 127

\bibitem[{{Mas-Hesse} {et~al.}(2003){Mas-Hesse}, {Kunth}, {Tenorio-Tagle},
  {Leitherer}, {Terlevich}, \& {Terlevich}}]{mas03}
{Mas-Hesse}, J.~M., {Kunth}, D., {Tenorio-Tagle}, G., {Leitherer}, C.,
  {Terlevich}, R.~J., \& {Terlevich}, E. 2003, \apj, 598, 858

\bibitem[{{Matsuda} {et~al.}(2004){Matsuda}, {Yamada}, {Hayashino}, {Tamura},
  {Yamauchi}, {Ajiki}, {Fujita}, {Murayama}, {Nagao}, {Ohta}, {Okamura},
  {Ouchi}, {Shimasaku}, {Shioya}, \& {Taniguchi}}]{mat04}
{Matsuda}, Y., {Yamada}, T., {Hayashino}, T., {Tamura}, H., {Yamauchi}, R.,
  {Ajiki}, M., {Fujita}, S.~S., {Murayama}, T., {Nagao}, T., {Ohta}, K.,
  {Okamura}, S., {Ouchi}, M., {Shimasaku}, K., {Shioya}, Y., \& {Taniguchi}, Y.
  2004, \aj, 128, 569

\bibitem[{{McLean} {et~al.}(2010){McLean}, {Steidel}, {Epps}, {Matthews},
  {Adkins}, {Konidaris}, {Weber}, {Aliado}, {Brims}, {Canfield}, {Cromer},
  {Fucik}, {Kulas}, {Mace}, {Magnone}, {Rodriguez}, {Wang}, \& {Weiss}}]{mcl10}
{McLean}, I.~S., {Steidel}, C.~C., {Epps}, H., {Matthews}, K., {Adkins}, S.,
  {Konidaris}, N., {Weber}, B., {Aliado}, T., {Brims}, G., {Canfield}, J.,
  {Cromer}, J., {Fucik}, J., {Kulas}, K., {Mace}, G., {Magnone}, K.,
  {Rodriguez}, H., {Wang}, E., \& {Weiss}, J. 2010, in Society of Photo-Optical
  Instrumentation Engineers (SPIE) Conference Series, Vol. 7735, Society of
  Photo-Optical Instrumentation Engineers (SPIE) Conference Series

\bibitem[{{McLean} {et~al.}(2012){McLean}, {Steidel}, {Epps}, {Konidaris},
  {Matthews}, {Adkins}, {Aliado}, {Brims}, {Canfield}, {Cromer}, {Fucik},
  {Kulas}, {Mace}, {Magnone}, {Rodriguez}, {Rudie}, {Trainor}, {Wang}, {Weber},
  \& {Weiss}}]{mcl12}
{McLean}, I.~S., {Steidel}, C.~C., {Epps}, H.~W., {Konidaris}, N., {Matthews},
  K.~Y., {Adkins}, S., {Aliado}, T., {Brims}, G., {Canfield}, J.~M., {Cromer},
  J.~L., {Fucik}, J., {Kulas}, K., {Mace}, G., {Magnone}, K., {Rodriguez}, H.,
  {Rudie}, G., {Trainor}, R., {Wang}, E., {Weber}, B., \& {Weiss}, J. 2012, in
  Society of Photo-Optical Instrumentation Engineers (SPIE) Conference Series,
  Vol. 8446, Society of Photo-Optical Instrumentation Engineers (SPIE)
  Conference Series

\bibitem[{{McLinden} {et~al.}(2011){McLinden}, {Finkelstein}, {Rhoads},
  {Malhotra}, {Hibon}, {Richardson}, {Cresci}, {Quirrenbach}, {Pasquali},
  {Bian}, {Fan}, \& {Woodward}}]{mcl11}
{McLinden}, E.~M., {Finkelstein}, S.~L., {Rhoads}, J.~E., {Malhotra}, S.,
  {Hibon}, P., {Richardson}, M.~L.~A., {Cresci}, G., {Quirrenbach}, A.,
  {Pasquali}, A., {Bian}, F., {Fan}, X., \& {Woodward}, C.~E. 2011, \apj, 730,
  136

\bibitem[{{Mostardi} {et~al.}(2013){Mostardi}, {Shapley}, {Nestor}, {Steidel},
  {Reddy}, \& {Trainor}}]{mos13}
{Mostardi}, R.~E., {Shapley}, A.~E., {Nestor}, D.~B., {Steidel}, C.~C.,
  {Reddy}, N.~A., \& {Trainor}, R.~F. 2013, \apj, 779, 65

\bibitem[{{Nestor} {et~al.}(2013){Nestor}, {Shapley}, {Kornei}, {Steidel}, \&
  {Siana}}]{nes13}
{Nestor}, D.~B., {Shapley}, A.~E., {Kornei}, K.~A., {Steidel}, C.~C., \&
  {Siana}, B. 2013, \apj, 765, 47

\bibitem[{{{\"O}stlin} {et~al.}(2014){{\"O}stlin}, {Hayes}, {Duval},
  {Sandberg}, {Rivera-Thorsen}, {Marquart}, {Orlitov{\'a}}, {Adamo},
  {Melinder}, {Guaita}, {Atek}, {Cannon}, {Gruyters}, {Herenz}, {Kunth},
  {Laursen}, {Mas-Hesse}, {Micheva}, {Ot{\'{\i}}-Floranes}, {Pardy}, {Roth},
  {Schaerer}, \& {Verhamme}}]{ost14}
{{\"O}stlin}, G., {Hayes}, M., {Duval}, F., {Sandberg}, A., {Rivera-Thorsen},
  T., {Marquart}, T., {Orlitov{\'a}}, I., {Adamo}, A., {Melinder}, J.,
  {Guaita}, L., {Atek}, H., {Cannon}, J.~M., {Gruyters}, P., {Herenz}, E.~C.,
  {Kunth}, D., {Laursen}, P., {Mas-Hesse}, J.~M., {Micheva}, G.,
  {Ot{\'{\i}}-Floranes}, H., {Pardy}, S.~A., {Roth}, M.~M., {Schaerer}, D., \&
  {Verhamme}, A. 2014, \apj, 797, 11

\bibitem[{{Ouchi} {et~al.}(2008){Ouchi}, {Shimasaku}, {Akiyama}, {Simpson},
  {Saito}, {Ueda}, {Furusawa}, {Sekiguchi}, {Yamada}, {Kodama}, {Kashikawa},
  {Okamura}, {Iye}, {Takata}, {Yoshida}, \& {Yoshida}}]{ouc08}
{Ouchi}, M., {Shimasaku}, K., {Akiyama}, M., {Simpson}, C., {Saito}, T.,
  {Ueda}, Y., {Furusawa}, H., {Sekiguchi}, K., {Yamada}, T., {Kodama}, T.,
  {Kashikawa}, N., {Okamura}, S., {Iye}, M., {Takata}, T., {Yoshida}, M., \&
  {Yoshida}, M. 2008, \apjs, 176, 301

\bibitem[{{Partridge} \& {Peebles}(1967)}]{par67}
{Partridge}, R.~B. \& {Peebles}, P.~J.~E. 1967, \apj, 147, 868

\bibitem[{{Pettini} {et~al.}(2003){Pettini}, {Madau}, {Bolte}, {Prochaska},
  {Ellison}, \& {Fan}}]{pet03}
{Pettini}, M., {Madau}, P., {Bolte}, M., {Prochaska}, J.~X., {Ellison}, S.~L.,
  \& {Fan}, X. 2003, \apj, 594, 695

\bibitem[{{Pettini} {et~al.}(2002){Pettini}, {Rix}, {Steidel}, {Adelberger},
  {Hunt}, \& {Shapley}}]{pet02}
{Pettini}, M., {Rix}, S.~A., {Steidel}, C.~C., {Adelberger}, K.~L., {Hunt},
  M.~P., \& {Shapley}, A.~E. 2002, \apj, 569, 742

\bibitem[{{Pettini} {et~al.}(2001){Pettini}, {Shapley}, {Steidel}, {Cuby},
  {Dickinson}, {Moorwood}, {Adelberger}, \& {Giavalisco}}]{pet01}
{Pettini}, M., {Shapley}, A.~E., {Steidel}, C.~C., {Cuby}, J.-G., {Dickinson},
  M., {Moorwood}, A.~F.~M., {Adelberger}, K.~L., \& {Giavalisco}, M. 2001,
  \apj, 554, 981

\bibitem[{{Pettini} {et~al.}(2000){Pettini}, {Steidel}, {Adelberger},
  {Dickinson}, \& {Giavalisco}}]{pet00}
{Pettini}, M., {Steidel}, C.~C., {Adelberger}, K.~L., {Dickinson}, M., \&
  {Giavalisco}, M. 2000, \apj, 528, 96

\bibitem[{{Rakic} {et~al.}(2011){Rakic}, {Schaye}, {Steidel}, \&
  {Rudie}}]{rak11}
{Rakic}, O., {Schaye}, J., {Steidel}, C.~C., \& {Rudie}, G.~C. 2011, \mnras,
  414, 3265

\bibitem[{{Reddy} {et~al.}(2008){Reddy}, {Steidel}, {Pettini}, {Adelberger},
  {Shapley}, {Erb}, \& {Dickinson}}]{red08}
{Reddy}, N.~A., {Steidel}, C.~C., {Pettini}, M., {Adelberger}, K.~L.,
  {Shapley}, A.~E., {Erb}, D.~K., \& {Dickinson}, M. 2008, \apjs, 175, 48

\bibitem[{{Rivera-Thorsen} {et~al.}(2015){Rivera-Thorsen}, {Hayes},
  {{\"O}stlin}, {Duval}, {Orlitov{\'a}}, {Verhamme}, {Mas-Hesse}, {Schaerer},
  {Cannon}, {Ot{\'{\i}}-Floranes}, {Sandberg}, {Guaita}, {Adamo}, {Atek},
  {Herenz}, {Kunth}, {Laursen}, \& {Melinder}}]{riv15}
{Rivera-Thorsen}, T.~E., {Hayes}, M., {{\"O}stlin}, G., {Duval}, F.,
  {Orlitov{\'a}}, I., {Verhamme}, A., {Mas-Hesse}, J.~M., {Schaerer}, D.,
  {Cannon}, J.~M., {Ot{\'{\i}}-Floranes}, H., {Sandberg}, A., {Guaita}, L.,
  {Adamo}, A., {Atek}, H., {Herenz}, E.~C., {Kunth}, D., {Laursen}, P., \&
  {Melinder}, J. 2015, ArXiv e-prints

\bibitem[{{Robertson} {et~al.}(2013){Robertson}, {Furlanetto}, {Schneider},
  {Charlot}, {Ellis}, {Stark}, {McLure}, {Dunlop}, {Koekemoer}, {Schenker},
  {Ouchi}, {Ono}, {Curtis-Lake}, {Rogers}, {Bowler}, \& {Cirasuolo}}]{rob13}
{Robertson}, B.~E., {Furlanetto}, S.~R., {Schneider}, E., {Charlot}, S.,
  {Ellis}, R.~S., {Stark}, D.~P., {McLure}, R.~J., {Dunlop}, J.~S.,
  {Koekemoer}, A., {Schenker}, M.~A., {Ouchi}, M., {Ono}, Y., {Curtis-Lake},
  E., {Rogers}, A.~B., {Bowler}, R.~A.~A., \& {Cirasuolo}, M. 2013, \apj, 768,
  71

\bibitem[{{Rudie} {et~al.}(2013){Rudie}, {Steidel}, {Shapley}, \&
  {Pettini}}]{rud13}
{Rudie}, G.~C., {Steidel}, C.~C., {Shapley}, A.~E., \& {Pettini}, M. 2013,
  \apj, 769, 146

\bibitem[{{Rudie} {et~al.}(2012){Rudie}, {Steidel}, {Trainor}, {Rakic},
  {Bogosavljevi{\'c}}, {Pettini}, {Reddy}, {Shapley}, {Erb}, \& {Law}}]{rud12a}
{Rudie}, G.~C., {Steidel}, C.~C., {Trainor}, R.~F., {Rakic}, O.,
  {Bogosavljevi{\'c}}, M., {Pettini}, M., {Reddy}, N., {Shapley}, A.~E., {Erb},
  D.~K., \& {Law}, D.~R. 2012, \apj, 750, 67

\bibitem[{{Rupke} {et~al.}(2005){Rupke}, {Veilleux}, \& {Sanders}}]{rup05}
{Rupke}, D.~S., {Veilleux}, S., \& {Sanders}, D.~B. 2005, \apjs, 160, 115

\bibitem[{{Schenker} {et~al.}(2013){Schenker}, {Ellis}, {Konidaris}, \&
  {Stark}}]{sch13}
{Schenker}, M.~A., {Ellis}, R.~S., {Konidaris}, N.~P., \& {Stark}, D.~P. 2013,
  \apj, 777, 67

\bibitem[{{Shapley} {et~al.}(2003){Shapley}, {Steidel}, {Pettini}, \&
  {Adelberger}}]{sha03}
{Shapley}, A.~E., {Steidel}, C.~C., {Pettini}, M., \& {Adelberger}, K.~L. 2003,
  \apj, 588, 65

\bibitem[{{Shibuya} {et~al.}(2014){Shibuya}, {Ouchi}, {Nakajima}, {Hashimoto},
  {Ono}, {Rauch}, {Gauthier}, {Shimasaku}, {Goto}, {Mori}, \&
  {Umemura.}}]{shi14}
{Shibuya}, T., {Ouchi}, M., {Nakajima}, K., {Hashimoto}, T., {Ono}, Y.,
  {Rauch}, M., {Gauthier}, J.-R., {Shimasaku}, K., {Goto}, R., {Mori}, M., \&
  {Umemura.}, M. 2014, \apj, 788, 74

\bibitem[{{Stark} {et~al.}(2010){Stark}, {Ellis}, {Chiu}, {Ouchi}, \&
  {Bunker}}]{sta10}
{Stark}, D.~P., {Ellis}, R.~S., {Chiu}, K., {Ouchi}, M., \& {Bunker}, A. 2010,
  \mnras, 408, 1628

\bibitem[{{Steidel} {et~al.}(2003){Steidel}, {Adelberger}, {Shapley},
  {Pettini}, {Dickinson}, \& {Giavalisco}}]{ste03}
{Steidel}, C.~C., {Adelberger}, K.~L., {Shapley}, A.~E., {Pettini}, M.,
  {Dickinson}, M., \& {Giavalisco}, M. 2003, \apj, 592, 728

\bibitem[{{Steidel} {et~al.}(2011){Steidel}, {Bogosavljevi{\'c}}, {Shapley},
  {Kollmeier}, {Reddy}, {Erb}, \& {Pettini}}]{ste11}
{Steidel}, C.~C., {Bogosavljevi{\'c}}, M., {Shapley}, A.~E., {Kollmeier},
  J.~A., {Reddy}, N.~A., {Erb}, D.~K., \& {Pettini}, M. 2011, \apj, 736, 160

\bibitem[{{Steidel} {et~al.}(2010){Steidel}, {Erb}, {Shapley}, {Pettini},
  {Reddy}, {Bogosavljevi{\'c}}, {Rudie}, \& {Rakic}}]{ste10}
{Steidel}, C.~C., {Erb}, D.~K., {Shapley}, A.~E., {Pettini}, M., {Reddy}, N.,
  {Bogosavljevi{\'c}}, M., {Rudie}, G.~C., \& {Rakic}, O. 2010, \apj, 717, 289

\bibitem[{{Steidel} {et~al.}(1996){Steidel}, {Giavalisco}, {Pettini},
  {Dickinson}, \& {Adelberger}}]{ste96}
{Steidel}, C.~C., {Giavalisco}, M., {Pettini}, M., {Dickinson}, M., \&
  {Adelberger}, K.~L. 1996, \apjl, 462, L17

\bibitem[{{Steidel} {et~al.}(2014){Steidel}, {Rudie}, {Strom}, {Pettini},
  {Reddy}, {Shapley}, {Trainor}, {Erb}, {Turner}, {Konidaris}, {Kulas}, {Mace},
  {Matthews}, \& {McLean}}]{ste14}
{Steidel}, C.~C., {Rudie}, G.~C., {Strom}, A.~L., {Pettini}, M., {Reddy},
  N.~A., {Shapley}, A.~E., {Trainor}, R.~F., {Erb}, D.~K., {Turner}, M.~L.,
  {Konidaris}, N.~P., {Kulas}, K.~R., {Mace}, G., {Matthews}, K., \& {McLean},
  I.~S. 2014, \apj, 795, 165

\bibitem[{{Steidel} {et~al.}(2004){Steidel}, {Shapley}, {Pettini},
  {Adelberger}, {Erb}, {Reddy}, \& {Hunt}}]{ste04}
{Steidel}, C.~C., {Shapley}, A.~E., {Pettini}, M., {Adelberger}, K.~L., {Erb},
  D.~K., {Reddy}, N.~A., \& {Hunt}, M.~P. 2004, \apj, 604, 534

\bibitem[{{Trainor} \& {Steidel}(2013)}]{tra13}
{Trainor}, R. \& {Steidel}, C.~C. 2013, \apjl, 775, L3

\bibitem[{{Trainor} \& {Steidel}(2012)}]{tra12}
{Trainor}, R.~F. \& {Steidel}, C.~C. 2012, \apj, 752, 39

\bibitem[{{Turner} {et~al.}(2014){Turner}, {Schaye}, {Steidel}, {Rudie}, \&
  {Strom}}]{tur14}
{Turner}, M.~L., {Schaye}, J., {Steidel}, C.~C., {Rudie}, G.~C., \& {Strom},
  A.~L. 2014, \mnras, 445, 794

\bibitem[{{Verhamme} {et~al.}(2012){Verhamme}, {Dubois}, {Blaizot}, {Garel},
  {Bacon}, {Devriendt}, {Guiderdoni}, \& {Slyz}}]{ver12}
{Verhamme}, A., {Dubois}, Y., {Blaizot}, J., {Garel}, T., {Bacon}, R.,
  {Devriendt}, J., {Guiderdoni}, B., \& {Slyz}, A. 2012, \aap, 546, A111

\bibitem[{{Verhamme} {et~al.}(2008){Verhamme}, {Schaerer}, {Atek}, \&
  {Tapken}}]{ver08}
{Verhamme}, A., {Schaerer}, D., {Atek}, H., \& {Tapken}, C. 2008, \aap, 491, 89

\bibitem[{{Verhamme} {et~al.}(2006){Verhamme}, {Schaerer}, \&
  {Maselli}}]{ver06}
{Verhamme}, A., {Schaerer}, D., \& {Maselli}, A. 2006, \aap, 460, 397

\bibitem[{{Weiner} {et~al.}(2009){Weiner}, {Coil}, {Prochaska}, {Newman},
  {Cooper}, {Bundy}, {Conselice}, {Dutton}, {Faber}, {Koo}, {Lotz}, {Rieke}, \&
  {Rubin}}]{wei09}
{Weiner}, B.~J., {Coil}, A.~L., {Prochaska}, J.~X., {Newman}, J.~A., {Cooper},
  M.~C., {Bundy}, K., {Conselice}, C.~J., {Dutton}, A.~A., {Faber}, S.~M.,
  {Koo}, D.~C., {Lotz}, J.~M., {Rieke}, G.~H., \& {Rubin}, K.~H.~R. 2009, \apj,
  692, 187

\bibitem[{{White} \& {Frenk}(1991)}]{whi91}
{White}, S.~D.~M. \& {Frenk}, C.~S. 1991, \apj, 379, 52

\bibitem[{{Wofford} {et~al.}(2013){Wofford}, {Leitherer}, \& {Salzer}}]{wof13}
{Wofford}, A., {Leitherer}, C., \& {Salzer}, J. 2013, \apj, 765, 118

\bibitem[{{Yamada} {et~al.}(2012{\natexlab{a}}){Yamada}, {Matsuda}, {Kousai},
  {Hayashino}, {Morimoto}, \& {Umemura}}]{yam12b}
{Yamada}, T., {Matsuda}, Y., {Kousai}, K., {Hayashino}, T., {Morimoto}, N., \&
  {Umemura}, M. 2012{\natexlab{a}}, \apj, 751, 29

\bibitem[{{Yamada} {et~al.}(2012{\natexlab{b}}){Yamada}, {Nakamura}, {Matsuda},
  {Hayashino}, {Yamauchi}, {Morimoto}, {Kousai}, \& {Umemura}}]{yam12a}
{Yamada}, T., {Nakamura}, Y., {Matsuda}, Y., {Hayashino}, T., {Yamauchi}, R.,
  {Morimoto}, N., {Kousai}, K., \& {Umemura}, M. 2012{\natexlab{b}}, \aj, 143,
  79

\bibitem[{{Zheng} \& {Miralda-Escud{\'e}}(2002)}]{zhe02}
{Zheng}, Z. \& {Miralda-Escud{\'e}}, J. 2002, \apj, 578, 33

\bibitem[{{Zheng} \& {Wallace}(2013)}]{zhe13}
{Zheng}, Z. \& {Wallace}, J. 2013, ArXiv e-prints

\end{thebibliography}

\end{document}